\documentclass[12pt]{article}
\usepackage{graphicx}
\usepackage{pdflscape}
\usepackage{xcolor}
\usepackage{subcaption}
\usepackage[margin=0.8in]{geometry}
\usepackage{amsmath, amsfonts, amssymb, afterpage, bbm, tabularx}
\usepackage[hypertexnames=false, colorlinks=true, linkcolor=red, citecolor=blue, urlcolor=magenta]{hyperref}
\usepackage{longtable}
\usepackage{array}
\usepackage{siunitx}
\usepackage{booktabs}
\usepackage{appendix}
\usepackage{needspace}
\usepackage{bm}

\usepackage{placeins}

\usepackage[T1]{fontenc}
\usepackage{newtxtext}
\usepackage{newtxmath}

\usepackage{caption}
\usepackage{anyfontsize}

\usepackage{setspace}
\usepackage[round,authoryear]{natbib}
\usepackage{amsthm}
\newtheorem{proposition}{Proposition}

\usepackage{soul}
\usepackage{autonum}

\usepackage{microtype}
\microtypesetup{nopatch=footnote}

\usepackage[rightcaption]{sidecap}
\sidecaptionvpos{figure}{t}

\newcommand{\webautoref}[1]{\textcolor{red}{Web }\autoref{#1}}

\renewenvironment{abstract}{
  \vspace{1em}
  \noindent\textbf{Abstract}\vspace{0.3em}\\
  \noindent
}{
  \par\vspace{1em}
}

\makeatletter
\renewcommand{\maketitle}{
  \begin{center}
    {\LARGE\bfseries \@title \par}
    \vspace{1.0em}
    {\large Divan Aristo Burger$^{1,2}$\par}
    \vspace{0.9em}
    {$^{1}$ Cytel Inc., Waltham, MA, USA \\
    $^{2}$ Department of Mathematical Statistics and Actuarial Science, University of the Free State, Bloemfontein, Free State, South Africa\par}
    \vspace{0.9em}
    {\textbf{Corresponding author:} Divan Burger \\
    Cytel Inc., 1050 Winter Street, Waltham, MA 02451, USA \\
    \texttt{divanaburger@gmail.com}}
  \end{center}
}
\makeatother

\title{Reference-based sensitivity analysis for bounded longitudinal outcomes}

\date{}

\begin{document}

\maketitle

\begin{abstract}
Reference-based sensitivity analysis is widely used to assess departures from the assumption that data are missing at random in longitudinal randomized trials, but existing methods are typically formulated for approximately Gaussian continuous outcomes and do not directly accommodate bounded outcomes, particularly when exact lower or upper bounds may occur with appreciable probability. We propose a Bayesian reference-based sensitivity analysis framework for bounded longitudinal outcomes, based on a thresholded latent Gaussian working model. Interior observations are mapped one-to-one to the latent scale, whereas boundary observations correspond to latent values below or above prespecified thresholds, yielding a distribution on the observed scale with a continuous interior density and positive mass at the bounds. Reference-based rules are imposed on the latent scale, and posterior predictive imputation proceeds on that scale before mapping back to the original outcome scale. Simulations assessed model calibration, recovery of complete data contrasts, and population treatment effect inference after imputation. For the proposed model, population treatment effect inference showed small bias when the analysis rule matched the truth rule after the trigger, but conservative Rubin intervals, particularly under jump to reference. The method is illustrated using the \textit{BtheB} study, a longitudinal trial with a bounded depression outcome and incomplete follow-up.
\end{abstract}

\vspace{0.6em}
\noindent
\begin{tabularx}{\textwidth}{@{}p{1.7cm}X@{}}
\textbf{Keywords:} &
reference-based sensitivity analysis; bounded longitudinal outcomes; controlled multiple imputation; missing data; latent Gaussian model
\end{tabularx}

\newpage

\section{Introduction}

Missing data arising after treatment discontinuation, dropout, or other deviations after randomization remain a central challenge in the analysis of longitudinal randomized trials. Sensitivity analyses assess how conclusions depend on assumptions about unobserved outcomes that cannot be verified from the observed data alone \citep{ICHE9R1, EMAMissing2010}. Reference-based multiple imputation (RBMI) constructs distributions for missing outcomes in the active arm using information from a reference arm, usually the control group \citep{carpenter2013analysis, cro2020sensitivity}. Under conventional Gaussian formulations, jump to reference (J2R) specifies a marginal mean trajectory that follows the active arm before a specified trigger and the reference arm afterward, with imputation then conditional on the available outcome history. Copy reference (CR) instead uses the reference arm conditional distribution given the available outcome history. Weighted rules interpolate between the conditional means under missing at random (MAR) and J2R. These approaches provide structured alternatives to MAR whose suitability depends on the clinical setting and the treatment effect of interest.

Most reference-based methods have been developed for continuous longitudinal outcomes under Gaussian or multivariate normal working models, closely related to mixed models or analyses of repeated measures under MAR \citep{mallinckrodt2003assessing, mallinckrod2008recommendations, carpenter2013analysis, cro2020sensitivity}. This framework is familiar and convenient, but it is less satisfactory when the outcome is known to lie within a fixed support. Many trial endpoints are bounded, including symptom scores, questionnaire totals, proportions, and related indices. In such settings, direct Gaussian imputation can generate values outside the allowable range, motivating transformations, models that respect the bounds, or adjustments to the imputed values \citep{geraci2018multiple, rodwell2014comparison}.

Several model classes have been proposed for bounded outcomes, including latent variable and coarsening models for bounded scores \citep{lesaffre2007logistic, molas2008comparison, xu2014modeling}, beta regression models with random effects or mixtures \citep{verkuilen2012mixed}, and models that place explicit mass at one or both bounds through inflated beta or related augmented beta rectangular constructions \citep{ospina2010inflated, wang2016augmented, wang2017bayesian}. These approaches are useful for modeling bounded responses, but they were not developed with the simple conditional multivariate normal imputation formulas used in conventional reference-based sensitivity analysis in mind. At the same time, recent work has started to extend reference-based sensitivity analysis beyond the standard Gaussian setting for continuous outcomes, for example, to right-censored time-to-event outcomes and longitudinal binary outcomes \citep{atkinson2021information, lu2020reference, cro2025reference}. To our knowledge, however, no existing reference-based framework combines a continuous interior distribution and explicit mass at the bounds for bounded longitudinal outcomes with Gaussian completion (imputation) on the latent scale.

In this paper, we extend reference-based sensitivity analysis to bounded longitudinal outcomes by using a thresholded latent Gaussian working model. Interior observations are mapped one-to-one to a latent Gaussian scale, whereas observations at the lower and upper bounds correspond to latent values below or above prespecified thresholds. This induces a distribution on the observed scale with a continuous interior density and positive mass at the bounds. The reference-based rules are imposed on the latent scale, where the familiar Gaussian completion formulas remain available. Missing outcomes are completed by posterior predictive sampling on the latent scale and then mapped back to the observed scale, so completed values automatically respect the support and can equal the bounds exactly.

The remainder of the paper is organized as follows. \autoref{sec:motivating_example} presents the motivating \textit{BtheB} example. \autoref{sec:data_structure} introduces the observed data structure and support specification. \autoref{sec:methodology} develops the thresholded latent Gaussian framework and the reference-based completion rules. \autoref{sec:simulation_studies} presents the simulation studies. \autoref{sec:application} presents the application to the \textit{BtheB} data, and \autoref{sec:discussion} concludes.

\section{Motivating data example} \label{sec:motivating_example}

We illustrate the proposed framework using the \textit{BtheB} data, available in the \texttt{HSAUR2} package and originating from a randomized primary care trial comparing the computerized cognitive behavioral therapy program \textit{Beating the Blues} (BtheB) with treatment as usual (TAU) \citep{proudfoot2003computerized, HothornEveritt2025HSAUR2}. The Beck Depression Inventory-II (BDI-II) was measured at baseline and at 2, 3, 5, and 8 months post-baseline. In the methodological development below, the longitudinal outcome vector comprises the four post-baseline measurements, indexed by $t = 1, \dots, 4$ for the 2-, 3-, 5-, and 8-month visits. The BDI-II is a bounded integer score taking values from $0$ to $63$.

\begin{figure}[htbp]
\centering
\includegraphics[width=0.65\textwidth]{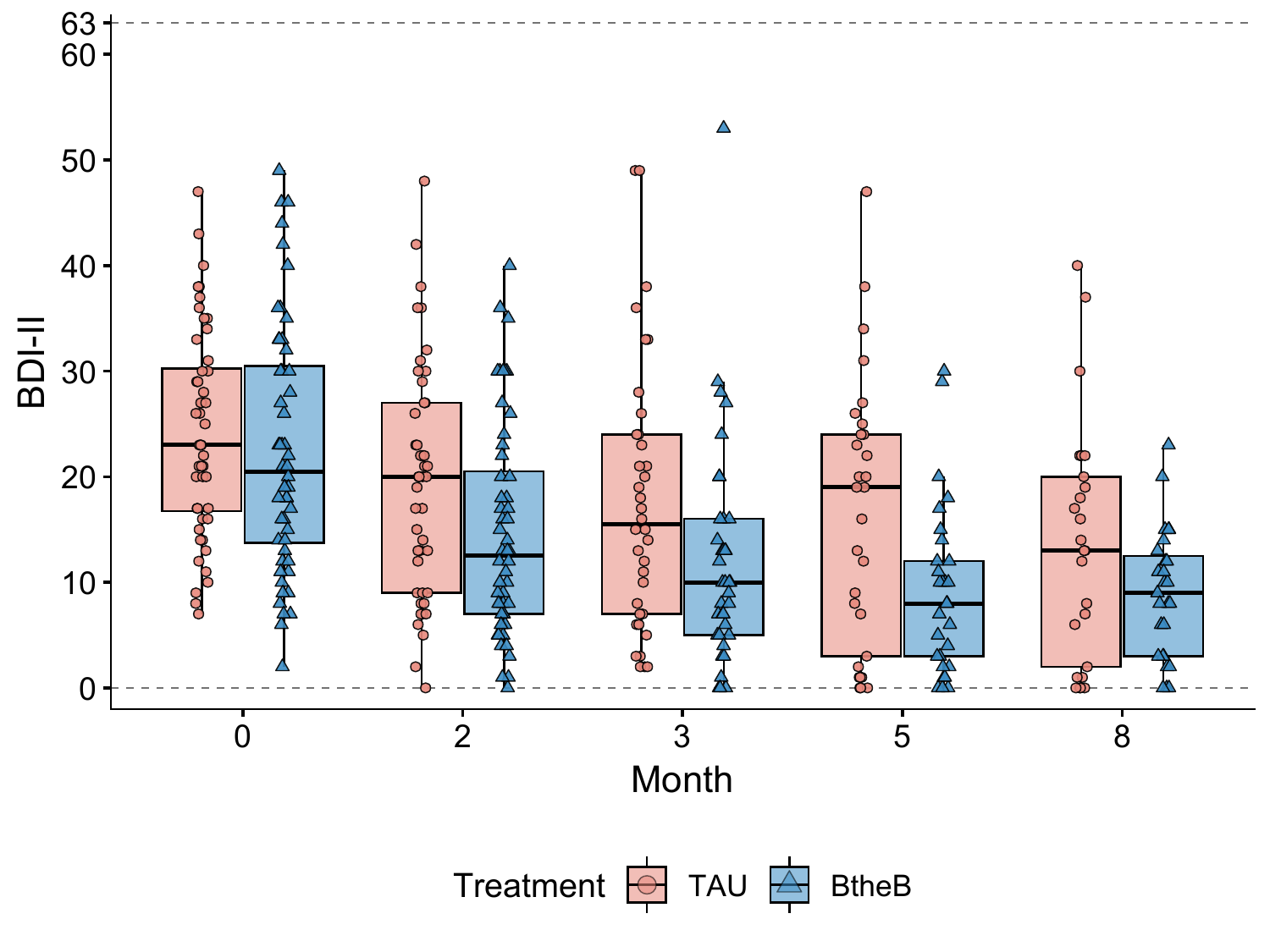}
\caption{Observed BDI-II scores by visit and treatment arm, comparing treatment as usual (TAU) with \textit{Beating the Blues} (BtheB), on the bounded scale from 0 to 63 and shown as boxplots with overlaid individual observed values. The dashed horizontal lines mark the lower and upper bounds of the score.}
\label{fig:btheb-distribution}
\end{figure}

The publicly available \texttt{HSAUR2} dataset contains 100 subjects, of whom 48 were assigned to TAU and 52 to BtheB. For the present application, we excluded 3 TAU subjects with no post-baseline BDI-II measurements, leaving 97 subjects for analysis, with 45 in TAU and 52 in BtheB. Over follow-up, the observed mean profiles suggest improvement in both groups, with a more pronounced decline in the BtheB arm; see \webautoref{fig:btheb-mean-profile} in \webautoref{sec:btheb-descriptive} of the Supporting Information. This same pattern is visible in \autoref{fig:btheb-distribution}, where the BtheB arm shows a progressively greater concentration of lower scores during follow-up. The original trial used a frequentist linear mixed model under MAR, with covariate adjustment and random intercepts and slopes for each subject \citep{proudfoot2003computerized}. The estimated BtheB versus TAU difference, assumed common across visits after treatment, was $-5.21$ BDI-II points (95\% CI: $-8.60$ to $-1.82$). No formal intercurrent event strategy was specified. Our analysis illustrates missing data sensitivity using the publicly available dataset rather than reproducing the original analysis.

\begin{figure}[htbp]
\centering
\includegraphics[width=0.60\textwidth]{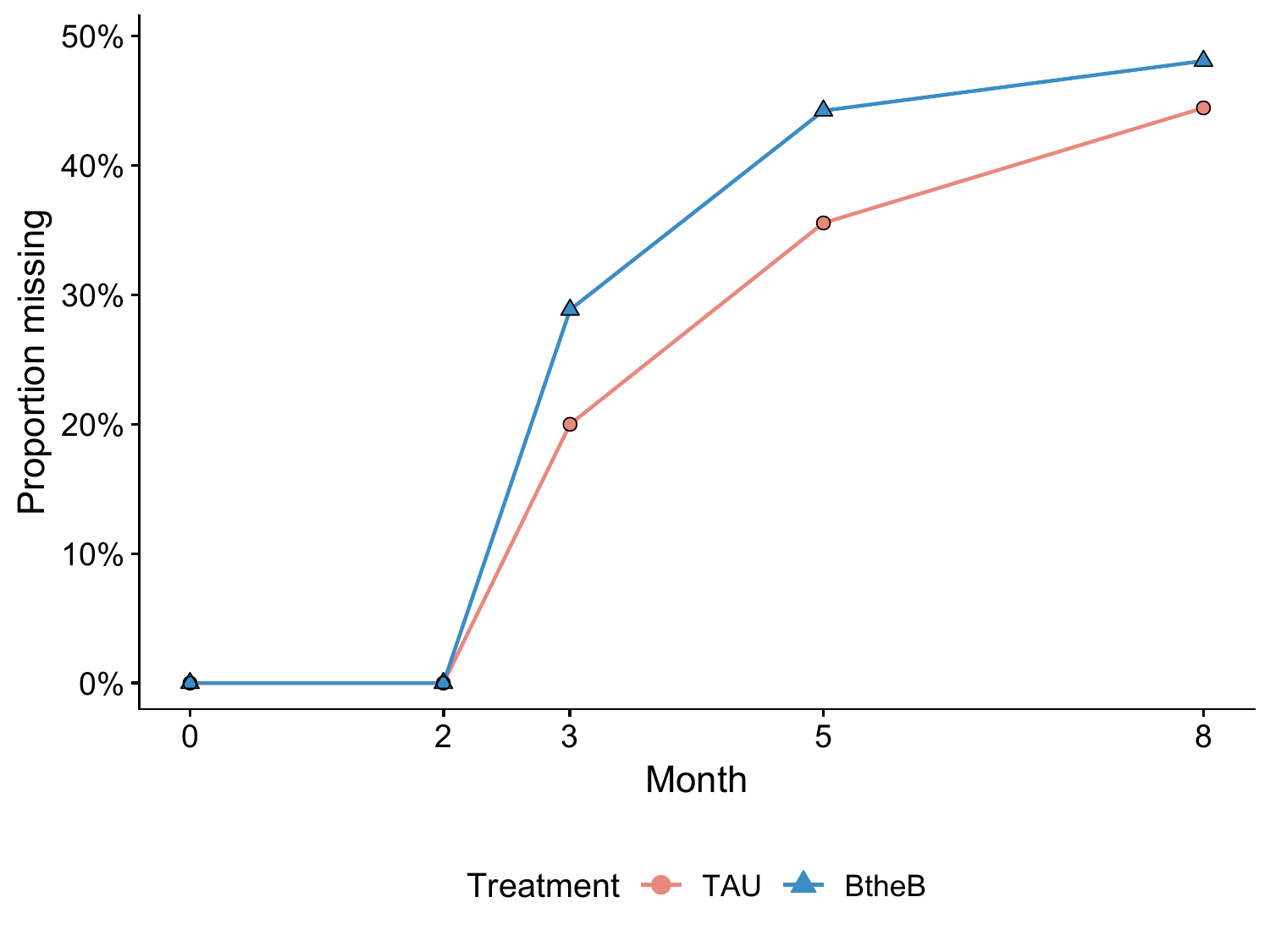}
\caption{Observed proportion of missing BDI-II values by visit and treatment arm, comparing treatment as usual (TAU) with \textit{Beating the Blues} (BtheB). Proportions use the retained analysis set as the denominator within each arm: 45 subjects for TAU and 52 for BtheB. Missingness increases substantially over follow-up in both arms.}
\label{fig:btheb-missingness}
\end{figure}

Two features of these data motivate the proposed methodology. First, the outcome is bounded, and the lower bound is reached in practice. Exact zero values occur with appreciable frequency and become more prominent later in follow-up; see \webautoref{fig:btheb-lower-bound}. Second, longitudinal follow-up is incomplete in the retained analysis set. Missingness is negligible at baseline and increases substantially over the post-baseline visits in both arms, approaching half by 8 months; see \autoref{fig:btheb-missingness}. Subject trajectories likewise show incomplete follow-up for many subjects before the final visit; see \webautoref{fig:btheb-spaghetti}.

Taken together, the \textit{BtheB} data illustrate the central practical problem addressed in this paper: standard Gaussian longitudinal models do not naturally accommodate bounded outcomes together with appreciable probability at the boundary and incomplete follow-up.

\FloatBarrier
\section{Observed data structure and support specification} \label{sec:data_structure}

Consider a longitudinal randomized trial with subjects indexed by $i = 1, \dots, n$ and scheduled post-baseline visits indexed by $t = 1, \dots, T$. Let $Y_{it}$ denote the outcome of interest for subject $i$ at visit $t$, and let $A_i \in \left\{0,1\right\}$ denote randomized treatment assignment, where $A_i = 1$ indicates the active arm and $A_i = 0$ the reference arm. Let $X_{it}$ denote the covariate vector entering the longitudinal mean model, which may include baseline outcome, visit indicators, interactions between baseline outcome and visit, stratification factors, and other prognostic variables. Write $X_i$ for the collection of covariates for subject $i$ across all visits. Define the observation indicator
\begin{equation}
R_{it} =
\begin{cases}
1, & \text{if } Y_{it} \text{ is observed}, \\
0, & \text{if } Y_{it} \text{ is missing}.
\end{cases}
\end{equation}
For subject $i$, let $L_i$ and $U_i$ denote known finite bounds with $L_i < U_i$, such that $L_i \le Y_{it} \le U_i$ for $t = 1, \dots, T$. Although many applications share common bounds across subjects, we allow limits that vary across subjects to ensure generality.

For each subject, define the observed lower bound, interior, upper bound, and missing visit sets by
\begin{equation}
\mathcal{L}_i = \left\{t \in \left\{1, \dots, T\right\} : R_{it} = 1, Y_{it} = L_i\right\},
\end{equation}
\begin{equation}
\mathcal{I}_i = \left\{t \in \left\{1, \dots, T\right\} : R_{it} = 1, L_i < Y_{it} < U_i\right\},
\end{equation}
\begin{equation}
\mathcal{U}_i = \left\{t \in \left\{1, \dots, T\right\} : R_{it} = 1, Y_{it} = U_i\right\},
\end{equation}
and
\begin{equation}
\mathcal{M}_i = \left\{t \in \left\{1, \dots, T\right\} : R_{it} = 0\right\}.
\end{equation}
The full observed visit index set is therefore $\mathcal{O}_i = \mathcal{L}_i \cup \mathcal{I}_i \cup \mathcal{U}_i$.

Let $\tau_i \in \left\{1, \dots, T + 1\right\}$ denote the first visit from which the sensitivity assumption applies for subject $i$, with $\tau_i = T + 1$ if no such visit occurs. We restrict the reference-based formulation to a final missing tail: when $\tau_i \le T$, all outcomes from visit $\tau_i$ onward are missing. An intercurrent event may define the trigger when all outcomes from the first affected visit onward are missing; otherwise, the trigger may be defined from the observed follow-up pattern as the visit immediately after the last observed outcome. Intermittent missing outcomes before the trigger are permitted. We then define the sets of missing visits before the trigger and from the trigger onward by
\begin{equation}
\mathcal{M}_i^{\mathrm{pre}} = \mathcal{M}_i \cap \left\{1, \dots, \tau_i - 1\right\},
\end{equation}
and
\begin{equation}
\mathcal{M}_i^{\mathrm{post}} = \mathcal{M}_i \cap \left\{\tau_i, \dots, T\right\}.
\end{equation}
Here, $\mathcal{M}_i^{\mathrm{pre}}$ is empty when $\tau_i = 1$, and $\mathcal{M}_i^{\mathrm{post}}$ is empty when $\tau_i = T + 1$.

The reference-based sensitivity assumptions introduced below apply only to the predictive distribution of missing values in the active arm from the trigger onward.

In the \textit{BtheB} application considered below, intercurrent events are not recorded for individual subjects, so $\tau_i$ is defined from the observed follow-up pattern rather than from an observed intercurrent event.

\section{Methodology} \label{sec:methodology}

We propose a reference-based sensitivity analysis framework for bounded longitudinal outcomes with possible exact boundary values. The key device is a thresholded latent Gaussian construction that combines a continuous interior distribution with positive mass at the bounds. We first introduce the representation on the observed scale and the model on the latent working scale, and then show how familiar reference-based assumptions can be imposed within that framework.

\subsection{Thresholded Gaussian representation on the latent working scale} \label{sec:thresholded_latent}

Motivated by latent variable threshold formulations for non-Gaussian longitudinal outcomes and related work on bounded outcomes, we introduce the following thresholded latent Gaussian representation \citep{lu2020reference, geraci2018multiple, aguirre2021imputation}.

Let $H_{it} \in \mathbb{R}$ denote a variable on the latent working scale and let $a_i < b_i$ denote fixed lower and upper thresholds. It is convenient to define these through two prespecified constants $p_{L_i}$ and $p_{U_i}$ satisfying $0 < p_{L_i} < p_{U_i} < 1$, with
\begin{equation}
a_i = \Phi^{-1}\left(p_{L_i}\right)
\qquad \text{and} \qquad
b_i = \Phi^{-1}\left(p_{U_i}\right),
\end{equation}
where $\Phi$ denotes the standard normal cumulative distribution function (CDF).

The observed outcome is linked to the latent scale through
\begin{equation}
Y_{it} =
\begin{cases}
L_i, & \text{if } H_{it} \le a_i, \\
L_i + \left(U_i - L_i\right)
\dfrac{\Phi\left(H_{it}\right) - p_{L_i}}{p_{U_i} - p_{L_i}}, & \text{if } a_i < H_{it} < b_i, \\
U_i, & \text{if } H_{it} \ge b_i.
\end{cases}
\end{equation}
This induces positive mass at the lower bound $L_i$ and the upper bound $U_i$, together with a continuous density on the interior $\left(L_i, U_i\right)$. The constants $p_{L_i}$ and $p_{U_i}$ specify fixed latent thresholds using the standard normal CDF; this does not require the latent outcome itself to have a standard normal distribution. If $H_{it}$ were standard normal, the probabilities of the lower and upper bounds would be $p_{L_i}$ and $1 - p_{U_i}$, respectively. More generally, under the Gaussian working model below, the boundary probabilities depend on the latent mean and standard deviation (SD) and can therefore vary even when the thresholds remain fixed.

For interior observations $Y_{it} \in \left(L_i, U_i\right)$, the inverse map to the latent scale is
\begin{equation}
H_{it} = g_i\left(Y_{it}\right)
=
\Phi^{-1}\left(
p_{L_i} + \left(p_{U_i} - p_{L_i}\right)
\frac{Y_{it} - L_i}{U_i - L_i}
\right).
\end{equation}
The corresponding Jacobian factor for a single interior observation is
\begin{equation}
\left| \frac{\partial H_{it}}{\partial Y_{it}} \right|
=
\frac{p_{U_i} - p_{L_i}}{U_i - L_i}
\frac{1}{\phi\left(H_{it}\right)},
\end{equation}
where $\phi$ denotes the standard normal density.

The thresholds $a_i$ and $b_i$ may be chosen to suit the application. For outcomes recorded on a discrete score grid with step size $s_i$, a natural choice is
\begin{equation}
p_{L_i} = \frac{s_i/2}{U_i - L_i + s_i}
\qquad \text{and} \qquad
p_{U_i} = 1 - \frac{s_i/2}{U_i - L_i + s_i}.
\end{equation}
This choice corresponds to extending the score range by half a step at each end when defining the probit rescaling. For integer scores with $s_i = 1$, the interior transformation becomes
\begin{equation}
g_i\left(y\right)
=
\Phi^{-1}\left(
\frac{y - L_i + 0.5}{U_i - L_i + 1}
\right),
\qquad L_i < y < U_i.
\end{equation}
Thus, interior observations use the familiar midpoint correction, while exact boundary observations are represented by latent tail events. As $p_{L_i}$ approaches $0$ and $p_{U_i}$ approaches $1$, the thresholded observation rule reduces in the limit to a purely continuous probit transformation, with no probability mass at either bound.

Let $H_i = \left(H_{i1}, \dots, H_{iT}\right)^{\top}$ denote the vector on the latent working scale for subject $i$. Under the primary analysis model, we assume
\begin{equation}
H_i \left| A_i = a, X_i, \theta\right. \sim \mathcal{N}\left(\mu_{ia}, \Sigma_a\right),
\qquad
a \in \left\{0,1\right\},
\end{equation}
where $\mu_{ia}$ is the mean vector for arm $a$, $\Sigma_a$ is the covariance matrix for repeated outcomes within a subject in that arm, and $\theta$ denotes the full collection of model parameters. Although this notation allows covariance matrices to differ between arms, the reference-based rules, theoretical results, and numerical analyses presented here use a common latent covariance matrix, $\Sigma_0 = \Sigma_1 = \Sigma$.

For any visit index set $\mathcal{J} \subseteq \left\{1, \dots, T\right\}$, let $H_{i,\mathcal{J}}$ denote the subvector of $H_i$ containing the components indexed by $\mathcal{J}$, arranged in increasing visit order. Likewise, $\mu_{ia,\mathcal{J}}$ denotes the corresponding subvector of $\mu_{ia}$. For visit index sets $\mathcal{J}, \mathcal{K} \subseteq \left\{1, \dots, T\right\}$, let $\Sigma_{a,\mathcal{J}\mathcal{K}}$ denote the submatrix of $\Sigma_a$ formed by the rows indexed by $\mathcal{J}$ and the columns indexed by $\mathcal{K}$.

A flexible visit-specific mean specification is
\begin{equation}
\mu_{ia} =
\left(
\mu_{ia1}, \dots, \mu_{iaT}
\right)^{\top},
\qquad
\mu_{iat} = X_{it}^{\top}\beta_t + a W_{it}^{\top}\delta_t,
\qquad
t = 1, \dots, T,
\end{equation}
where $\beta_t$ contains visit-specific prognostic effects and $\delta_t$ contains visit-specific treatment effects. In many applications, $W_{it}$ may simply be a subset of $X_{it}$ corresponding to interactions between treatment and visit.

\begin{figure}[!htbp]
  \centering
  \includegraphics[width=0.85\textwidth]{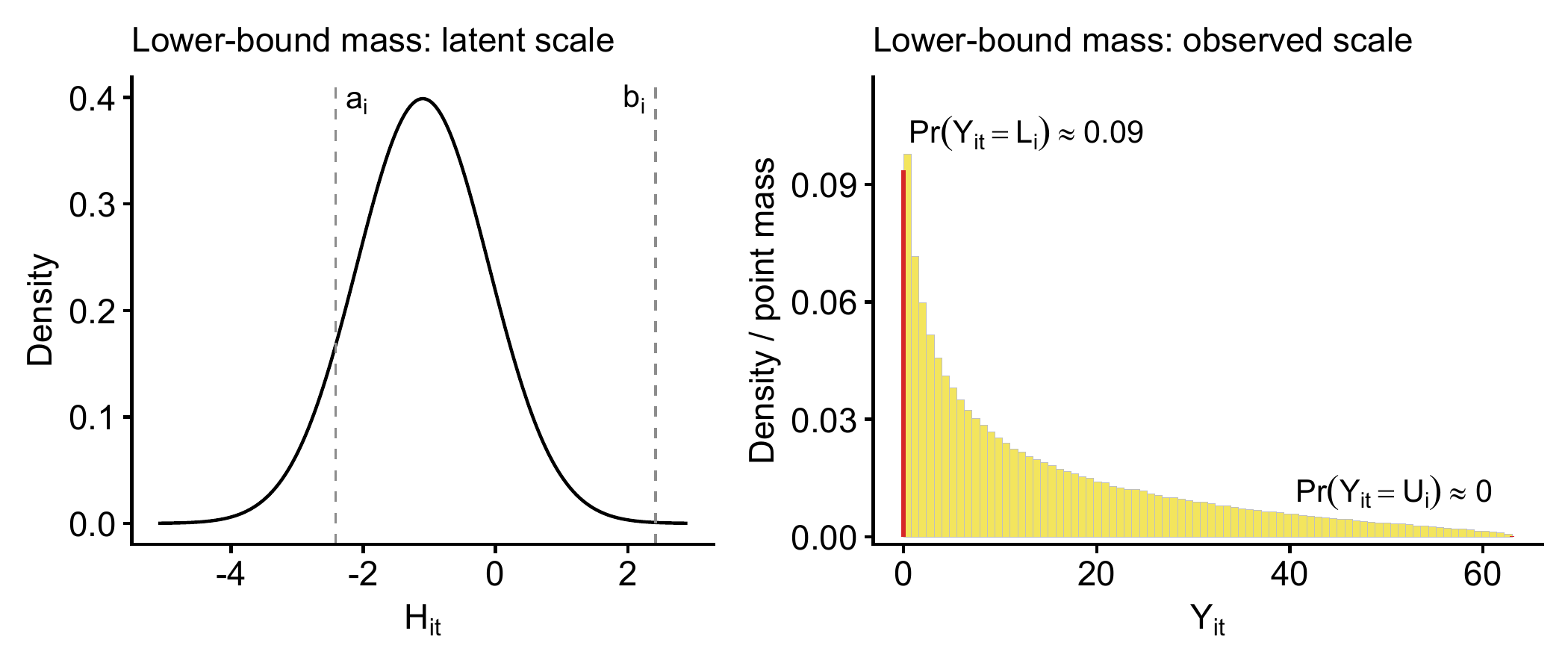}
  \caption{Illustration of the thresholded latent Gaussian construction for a representative setting with mass at the lower bound, $\left(\mu, \sigma\right) = \left(-1.1, 1.0\right)$, which induces approximately $\Pr\left(Y_{it} = L_i\right) = 0.09$ and $\Pr\left(Y_{it} = U_i\right) = 0.00$. The left panel shows the latent Gaussian density with threshold locations $a_i$ and $b_i$. The right panel shows the corresponding distribution on the observed scale, obtained from 1{,}000{,}000 simulated draws under the thresholded observation rule, with the red vertical line indicating the point mass at the lower boundary. The interior histogram is scaled so that its area plus the probabilities at the two bounds equals one.}
  \label{fig:thresholded-lower-mass}
\end{figure}

To illustrate the thresholded construction, consider the special case with common support, $L_i = 0$, $U_i = 63$, and $s_i = 1$ for all subjects, so that the thresholds are also common across subjects. For a single visit, suppose $H_{it} \sim \mathcal{N}\left(\mu, \sigma^2\right)$ with the representative setting $\left(\mu, \sigma\right) = \left(-1.1, 1.0\right)$. Under the thresholded model, this induces approximately $\Pr\left(Y_{it} = L_i\right) = 0.09$ and $\Pr\left(Y_{it} = U_i\right) = 0.00$. \autoref{fig:thresholded-lower-mass} displays the corresponding Gaussian density on the latent scale, the threshold locations, and the induced distribution on the observed scale.

\FloatBarrier
\subsection{Primary posterior under missing at random}

Under the baseline MAR imputation model, outcome missingness is assumed to depend only on observed outcomes, treatment assignment, and observed covariates. The parameters governing missingness are assumed to be distinct from, and a priori independent of, $\theta$, so the missingness mechanism can be ignored in posterior inference. Boundary observations are handled through latent augmentation. Interior observations determine latent values exactly through the inverse map $g_i\left(\cdot\right)$, whereas observations at the lower and upper bounds imply only that the corresponding latent variables satisfy $H_{it} \le a_i$ and $H_{it} \ge b_i$. We therefore adopt a latent data augmentation strategy in Stan \citep{carpenter2017stan}.

For subject $i$, define the augmented observed latent vector $\widetilde{H}_{i,\mathcal{O}_i} = \left(\widetilde{H}_{it} : t \in \mathcal{O}_i\right)^{\top}$, where
\begin{equation}
\widetilde{H}_{it} =
\begin{cases}
H_{it}, & \text{if } t \in \mathcal{I}_i, \\
H_{it}^{L}, & \text{if } t \in \mathcal{L}_i, \\
H_{it}^{U}, & \text{if } t \in \mathcal{U}_i,
\end{cases}
\end{equation}
with constraints $H_{it}^{L} \le a_i$ for $t \in \mathcal{L}_i$ and $H_{it}^{U} \ge b_i$ for $t \in \mathcal{U}_i$. Thus, interior components are known exactly from the observed data, whereas boundary components are latent and sampled jointly with the model parameters.

Up to multiplicative terms that are constant with respect to $\theta$ and the boundary latent variables, the augmented posterior under MAR is
\begin{equation}
\begin{aligned}
&p\left(
\theta,
\left\{H_{it}^{L} : t \in \mathcal{L}_i,\; i = 1, \dots, n\right\},
\left\{H_{it}^{U} : t \in \mathcal{U}_i,\; i = 1, \dots, n\right\}
\middle|
\mathcal{D}_{\mathrm{obs}}
\right)
\\
&\propto
p\left(\theta\right)
\prod_{i=1}^{n}
\left[
p\left(\widetilde{H}_{i,\mathcal{O}_i} \middle| A_i, X_i, \theta\right)
\mathbbm{1}\left(H_{i,\mathcal{L}_i}^{L} \le a_i \mathbf{1}\right)
\mathbbm{1}\left(H_{i,\mathcal{U}_i}^{U} \ge b_i \mathbf{1}\right)
\right].
\end{aligned}
\end{equation}
Here, $\mathcal{D}_{\mathrm{obs}}$ denotes the observed data, and the inequalities inside the indicator functions are understood componentwise. The multivariate normal density is therefore evaluated for the full augmented latent vector at observed visits, while the indicator functions enforce the boundary constraints. This avoids the need for multivariate normal CDFs in model fitting, because the boundary probabilities are represented through constrained latent variables rather than analytically integrated out. Additional details on prior specification, covariance parameterization, and Stan implementation are given in \autoref{app:stan-implementation}.

For later use in posterior predictive imputation, reorder the latent vector and mean vector so that observed visits precede missing visits, retaining increasing visit order within each block, and apply the same permutation to the rows and columns of the covariance matrix. With this reordering understood, write
\begin{equation}
H_i =
\begin{pmatrix}
H_{i,\mathcal{O}_i} \\
H_{i,\mathcal{M}_i}
\end{pmatrix},
\qquad
\mu_{ia} =
\begin{pmatrix}
\mu_{ia,\mathcal{O}_i} \\
\mu_{ia,\mathcal{M}_i}
\end{pmatrix},
\qquad
\Sigma_a =
\begin{pmatrix}
\Sigma_{a,\mathcal{O}_i\mathcal{O}_i} & \Sigma_{a,\mathcal{O}_i\mathcal{M}_i} \\
\Sigma_{a,\mathcal{M}_i\mathcal{O}_i} & \Sigma_{a,\mathcal{M}_i\mathcal{M}_i}
\end{pmatrix},
\end{equation}
where the observed latent subvector is represented by the augmented observed latent vector $\widetilde{H}_{i,\mathcal{O}_i}$. Under MAR within each arm, the conditional distribution of the missing outcomes on the latent working scale is Gaussian:
\begin{equation}
H_{i,\mathcal{M}_i}
\left|\widetilde{H}_{i,\mathcal{O}_i}, A_i = a, X_i, \theta\right.
\sim
\mathcal{N}\left(
m_{ia}^{\mathrm{MAR}}\left(\theta\right),
V_{ia}^{\mathrm{MAR}}\left(\theta\right)
\right),
\end{equation}
where
\begin{equation}
m_{ia}^{\mathrm{MAR}}\left(\theta\right)
=
\mu_{ia,\mathcal{M}_i}
+
\Sigma_{a,\mathcal{M}_i\mathcal{O}_i}
\Sigma_{a,\mathcal{O}_i\mathcal{O}_i}^{-1}
\left(
\widetilde{H}_{i,\mathcal{O}_i} - \mu_{ia,\mathcal{O}_i}
\right),
\end{equation}
and
\begin{equation}
V_{ia}^{\mathrm{MAR}}\left(\theta\right)
=
\Sigma_{a,\mathcal{M}_i\mathcal{M}_i}
-
\Sigma_{a,\mathcal{M}_i\mathcal{O}_i}
\Sigma_{a,\mathcal{O}_i\mathcal{O}_i}^{-1}
\Sigma_{a,\mathcal{O}_i\mathcal{M}_i}.
\end{equation}
Dependence of $m_{ia}^{\mathrm{MAR}}\left(\theta\right)$ on the augmented observed latent vector is suppressed in the notation. The Gaussian conditional distribution applies for fixed model parameters and augmented boundary values; integrating over their posterior uncertainty generally yields a mixture of Gaussian distributions. When the conditioning visit set is empty, the corresponding conditional Gaussian distribution is interpreted as the marginal Gaussian distribution.

\subsection{Reference-based sensitivity rules on the latent scale}

The primary posterior under the arm-specific MAR model is learned from the augmented latent representation described above. The sensitivity rule enters only at the posterior predictive completion stage, through the distribution used to generate missing latent values in the active arm from the trigger onward.

Although the thresholded latent model permits bounds to vary across subjects and, in principle, covariance structures to differ between arms, we restrict attention to the case with a common covariance matrix, $\Sigma_0 = \Sigma_1 = \Sigma$. This keeps the J2R and CR constructions on the latent scale aligned with the conventional Gaussian reference-based framework and yields a shared conditional covariance structure after the trigger, which underlies the variance invariance results presented later. Allowing $\Sigma_0 \neq \Sigma_1$ is conceptually possible but leads to more cumbersome conditioning formulas and is not pursued here.

For subject $i$, let $\mathcal{O}_i^{\star} = \mathcal{O}_i \cup \mathcal{M}_i^{\mathrm{pre}}$ denote the augmented conditioning set obtained after any missing latent values before the trigger have been imputed under the arm-specific MAR model. Under the final missing tail restriction, every visit in $\mathcal{O}_i^{\star}$ precedes $\tau_i$, so the J2R conditioning formula below centers the entire augmented history at the active arm mean before the trigger. Then, $\widetilde{H}_{i,\mathcal{O}_i^{\star}}$ denotes the realized latent history on that set, namely the augmented observed latent vector together with any latent values imputed for missing visits before the trigger, arranged in increasing order of visits, and the only components that remain to be generated under the sensitivity rule are those in $\mathcal{M}_i^{\mathrm{post}}$. Dependence of the conditional means on the displayed conditioning history is suppressed in the notation.

Partition the arm-specific mean vector and the common covariance matrix according to the ordered index blocks $\left(\mathcal{O}_i^{\star}, \mathcal{M}_i^{\mathrm{post}}\right)$, so that
\begin{equation}
\mu_{ia} =
\begin{pmatrix}
\mu_{ia,\mathcal{O}_i^{\star}} \\
\mu_{ia,\mathcal{M}_i^{\mathrm{post}}}
\end{pmatrix},
\qquad
\Sigma =
\begin{pmatrix}
\Sigma_{\mathcal{O}_i^{\star}\mathcal{O}_i^{\star}} &
\Sigma_{\mathcal{O}_i^{\star}\mathcal{M}_i^{\mathrm{post}}} \\
\Sigma_{\mathcal{M}_i^{\mathrm{post}}\mathcal{O}_i^{\star}} &
\Sigma_{\mathcal{M}_i^{\mathrm{post}}\mathcal{M}_i^{\mathrm{post}}}
\end{pmatrix}.
\end{equation}
For $a \in \left\{0,1\right\}$, the MAR completion distribution in arm $a$ for missing latent values from the trigger onward is
\begin{equation}
H_{i,\mathcal{M}_i^{\mathrm{post}}}
\left|
\widetilde{H}_{i,\mathcal{O}_i^{\star}}, A_i = a, X_i, \theta\right.
\sim
\mathcal{N}\left(
m_{ia}^{\mathrm{post}}\left(\theta\right),
V_i^{\mathrm{post}}\left(\theta\right)
\right),
\end{equation}
where
\begin{equation}
m_{ia}^{\mathrm{post}}\left(\theta\right)
=
\mu_{ia,\mathcal{M}_i^{\mathrm{post}}}
+
\Sigma_{\mathcal{M}_i^{\mathrm{post}}\mathcal{O}_i^{\star}}
\Sigma_{\mathcal{O}_i^{\star}\mathcal{O}_i^{\star}}^{-1}
\left(
\widetilde{H}_{i,\mathcal{O}_i^{\star}} - \mu_{ia,\mathcal{O}_i^{\star}}
\right),
\end{equation}
and
\begin{equation}
V_i^{\mathrm{post}}\left(\theta\right)
=
\Sigma_{\mathcal{M}_i^{\mathrm{post}}\mathcal{M}_i^{\mathrm{post}}}
-
\Sigma_{\mathcal{M}_i^{\mathrm{post}}\mathcal{O}_i^{\star}}
\Sigma_{\mathcal{O}_i^{\star}\mathcal{O}_i^{\star}}^{-1}
\Sigma_{\mathcal{O}_i^{\star}\mathcal{M}_i^{\mathrm{post}}}.
\end{equation}
Under MAR in the active arm, we take $a = 1$, so that
\begin{equation}
H_{i,\mathcal{M}_i^{\mathrm{post}}}
\left|
\widetilde{H}_{i,\mathcal{O}_i^{\star}}, A_i = 1, X_i, \theta\right.
\sim
\mathcal{N}\left(
m_{i1}^{\mathrm{post}}\left(\theta\right),
V_i^{\mathrm{post}}\left(\theta\right)
\right).
\end{equation}
Under J2R, we construct the imputation distribution using a marginal latent mean trajectory that follows the active arm before the trigger and the reference arm afterward. Imputation is then conditional on the realized latent history. The completion distribution is
\begin{equation}
H_{i,\mathcal{M}_i^{\mathrm{post}}}
\left|
\widetilde{H}_{i,\mathcal{O}_i^{\star}}, A_i = 1, X_i, \theta\right.
\sim
\mathcal{N}\left(
m_i^{\mathrm{J2R}}\left(\theta\right),
V_i^{\mathrm{post}}\left(\theta\right)
\right),
\end{equation}
where
\begin{equation}
m_i^{\mathrm{J2R}}\left(\theta\right)
=
\mu_{i0,\mathcal{M}_i^{\mathrm{post}}}
+
\Sigma_{\mathcal{M}_i^{\mathrm{post}}\mathcal{O}_i^{\star}}
\Sigma_{\mathcal{O}_i^{\star}\mathcal{O}_i^{\star}}^{-1}
\left(
\widetilde{H}_{i,\mathcal{O}_i^{\star}} - \mu_{i1,\mathcal{O}_i^{\star}}
\right).
\end{equation}
Under CR, missing latent values in the active arm from the trigger onward are drawn from the same conditional distribution that would apply in the reference arm, given the realized augmented latent history:
\begin{equation}
H_{i,\mathcal{M}_i^{\mathrm{post}}}
\left|
\widetilde{H}_{i,\mathcal{O}_i^{\star}}, A_i = 1, X_i, \theta\right.
\sim
\mathcal{N}\left(
m_i^{\mathrm{CR}}\left(\theta\right),
V_i^{\mathrm{post}}\left(\theta\right)
\right),
\end{equation}
where
\begin{equation}
m_i^{\mathrm{CR}}\left(\theta\right)
=
\mu_{i0,\mathcal{M}_i^{\mathrm{post}}}
+
\Sigma_{\mathcal{M}_i^{\mathrm{post}}\mathcal{O}_i^{\star}}
\Sigma_{\mathcal{O}_i^{\star}\mathcal{O}_i^{\star}}^{-1}
\left(
\widetilde{H}_{i,\mathcal{O}_i^{\star}} - \mu_{i0,\mathcal{O}_i^{\star}}
\right).
\end{equation}
The formulation on the latent scale also admits a family of sensitivity rules that interpolates between MAR and J2R through a single parameter. Let $\omega \in \left[0,1\right]$ denote a prespecified sensitivity parameter, and define
\begin{equation}
\ell_{i,\omega}^{\mathrm{post}}\left(\theta\right)
=
\left(1 - \omega\right)m_{i1}^{\mathrm{post}}\left(\theta\right)
+
\omega m_i^{\mathrm{J2R}}\left(\theta\right).
\end{equation}
We then use the completion distribution
\begin{equation}
H_{i,\mathcal{M}_i^{\mathrm{post}}}
\left|
\widetilde{H}_{i,\mathcal{O}_i^{\star}}, A_i = 1, X_i, \theta, \omega\right.
\sim
\mathcal{N}\left(
\ell_{i,\omega}^{\mathrm{post}}\left(\theta\right),
V_i^{\mathrm{post}}\left(\theta\right)
\right).
\end{equation}
This recovers $\omega = 0$ as the MAR law for the active arm and $\omega = 1$ as J2R, while intermediate values represent graded departures from MAR toward the reference arm trajectory.

\subsection{Posterior predictive completion and estimation}
\label{sec:posterior_predictive}

The proposed method proceeds through posterior predictive completion on the latent working scale. Equivalently, for each chosen sensitivity rule, completion is based on the posterior predictive distribution obtained by averaging the corresponding conditional latent completion law over the fitted primary posterior. The primary model is first fitted under the arm-specific MAR assumption, yielding posterior draws
\begin{equation}
\left(\theta^{\left(1\right)}, \widetilde{H}_{\mathcal{L}}^{\left(1\right)}, \widetilde{H}_{\mathcal{U}}^{\left(1\right)}\right), \dots, \left(\theta^{\left(K\right)}, \widetilde{H}_{\mathcal{L}}^{\left(K\right)}, \widetilde{H}_{\mathcal{U}}^{\left(K\right)}\right)
\end{equation}
from the primary posterior under the augmented latent representation, where $\widetilde{H}_{\mathcal{L}}^{\left(k\right)}$ and $\widetilde{H}_{\mathcal{U}}^{\left(k\right)}$ denote the collections of augmented latent values corresponding to observed outcomes at the lower and upper bounds in posterior draw $k$. All model quantities and augmented observed boundary values in the completion formulas are evaluated at posterior draw $k$; the draw index is suppressed where unambiguous. For each retained posterior draw, missing latent outcomes are then completed under the arm-specific MAR law before the trigger time and under the chosen reference-based rule after the trigger in the active arm.

For posterior draw $\theta^{\left(k\right)}$, with $k = 1, \dots, K$, completion proceeds subject by subject. If $\mathcal{M}_i^{\mathrm{pre}} \neq \varnothing$, the missing latent values before the trigger are first sampled under the arm-specific MAR law:
\begin{equation}
H_{i,\mathcal{M}_i^{\mathrm{pre}}}
\left|
\widetilde{H}_{i,\mathcal{O}_i}, A_i, X_i, \theta^{\left(k\right)}\right.
\sim
\mathcal{N}\left(
m_{iA_i}^{\mathrm{pre}}\left(\theta^{\left(k\right)}\right),
V_i^{\mathrm{pre}}\left(\theta^{\left(k\right)}\right)
\right),
\end{equation}
where
\begin{equation}
m_{iA_i}^{\mathrm{pre}}\left(\theta^{\left(k\right)}\right)
=
\mu_{iA_i,\mathcal{M}_i^{\mathrm{pre}}}
+
\Sigma_{A_i,\mathcal{M}_i^{\mathrm{pre}}\mathcal{O}_i}
\Sigma_{A_i,\mathcal{O}_i\mathcal{O}_i}^{-1}
\left(
\widetilde{H}_{i,\mathcal{O}_i} - \mu_{iA_i,\mathcal{O}_i}
\right),
\end{equation}
and
\begin{equation}
V_i^{\mathrm{pre}}\left(\theta^{\left(k\right)}\right)
=
\Sigma_{A_i,\mathcal{M}_i^{\mathrm{pre}}\mathcal{M}_i^{\mathrm{pre}}}
-
\Sigma_{A_i,\mathcal{M}_i^{\mathrm{pre}}\mathcal{O}_i}
\Sigma_{A_i,\mathcal{O}_i\mathcal{O}_i}^{-1}
\Sigma_{A_i,\mathcal{O}_i\mathcal{M}_i^{\mathrm{pre}}}.
\end{equation}
These draws are combined with the observed latent history to form the augmented conditioning set $\mathcal{O}_i^{\star} = \mathcal{O}_i \cup \mathcal{M}_i^{\mathrm{pre}}$. Conditional on $\widetilde{H}_{i,\mathcal{O}_i^{\star}}$, the missing latent values from the trigger onward are then completed. If $\mathcal{M}_i^{\mathrm{post}} = \varnothing$, no further completion is required. Otherwise, the remaining missing latent values are generated under the arm-specific MAR law for subjects in the reference arm and under the chosen reference-based completion rule for subjects in the active arm.

After all missing latent values have been obtained, the completed latent vector
\begin{equation}
H_i^{\left(k\right)} = \left(H_{i1}^{\left(k\right)}, \dots, H_{iT}^{\left(k\right)}\right)^{\top}
\end{equation}
is mapped back to the observed outcome scale through the thresholded observation rule
\begin{equation}
Y_{it}^{\left(k\right)} =
\begin{cases}
L_i, & \text{if } H_{it}^{\left(k\right)} \le a_i, \\
L_i + \left(U_i - L_i\right)
\dfrac{\Phi\left(H_{it}^{\left(k\right)}\right) - p_{L_i}}{p_{U_i} - p_{L_i}}, & \text{if } a_i < H_{it}^{\left(k\right)} < b_i, \\
U_i, & \text{if } H_{it}^{\left(k\right)} \ge b_i.
\end{cases}
\end{equation}
Because this rule is defined on all of $\mathbb{R}$, each draw of a completed dataset automatically respects the required support restriction and may contain values exactly at the lower or upper bound whenever the latent draws fall outside the interior interval $\left(a_i, b_i\right)$.

Inference may target a quantity defined for the completed study cohort. For retained posterior draw $k = 1, \dots, K$, let $Y^{\left(k\right)} = \left\{Y_{it}^{\left(k\right)} : i = 1, \dots, n,\; t = 1, \dots, T\right\}$ denote the completed outcome data on the original scale. Let $Q\left(\,\cdot\,, A, X\right)$ denote such a quantity, for example, an arm mean or a contrast between arm means, and define $Q^{\left(k\right)} = Q\left(Y^{\left(k\right)}, A, X\right)$, where $A = \left(A_1, \dots, A_n\right)$ denotes the treatment assignments and $X$ the collection of covariates for all subjects and visits.

Posterior predictive summaries of quantities defined for the completed study cohort are obtained directly from $Q^{\left(1\right)}, \dots, Q^{\left(K\right)}$ and describe uncertainty about those quantities conditional on the observed data. Population treatment effect inference additionally requires accounting for uncertainty about the treatment effect given a completed dataset. When using a secondary analysis model for this purpose, let $M$ denote the number of completed datasets used for that analysis. The secondary model is fitted to each completed dataset, and the resulting point estimates and within-imputation variance estimates are combined using Rubin's rules \citep{rubin1987multiple}. Thus, $K$ denotes the number of retained post-warmup posterior draws used for direct posterior predictive summaries, whereas $M$ denotes the number of completed datasets used for the secondary analysis.

An important theoretical feature of the proposed construction is variance invariance on the latent scale for the class of post-trigger rules considered here. Conditional on the fitted posterior state, MAR, J2R, CR, and the $\omega$-weighted rules modify only the conditional mean of the vector of missing latent values from the trigger onward and leave its conditional covariance unchanged. This yields an exact conditional variance invariance result for estimands that are affine functions of the completed data on the latent scale. On the observed bounded scale, because the threshold map is nonlinear and is not differentiable at the latent thresholds, we provide only a local first-order heuristic under the conditions stated in \autoref{app:info-anchoring}, rather than a full theorem on information anchoring on the observed scale. Formal statements, proofs, and the heuristic on the observed scale are given in \autoref{app:info-anchoring}. The result pertains to affine estimands computed directly from the completed latent data. It does not, in general, imply exact variance invariance for secondary analyses fitted to the completed datasets on the observed scale and then combined using multiple imputation rules.

\section{Simulation studies}
\label{sec:simulation_studies}

We conducted three complementary simulation studies. Study~1 examined calibration of the bounded outcome working models under incomplete follow-up, and Study~2 assessed recovery of the complete data contrast in each replicate under alternative assumptions after the trigger. Study~3 evaluated population treatment effect inference using imputation followed by a mixed model for repeated measures (MMRM) on the observed scale and Rubin's rules, including a transformed Gaussian comparator and a targeted distributional misspecification scenario.

All three studies used outcomes bounded by $\left[0, 63\right]$ and the half-step thresholds defined using unit score step $s = 1$. Studies~1 and~2 used 50 subjects randomized equally between arms and five post-baseline visits. Study~3 used 100 subjects randomized equally between arms and four post-baseline visits.

In Studies~1 and~2, 300 replicates were attempted per scenario; computational failures were excluded, leaving 297 to 300 replicates per scenario and model in Study~1 and 298 to 300 per scenario in Study~2. Study~3 summaries used 300 replicates per scenario. Study~2 assessed recovery of the contrast between arm means in the complete data for each replicate. Study~3 assessed inference for a fixed population treatment effect by fitting an MMRM to each completed dataset and combining the treatment effect estimates and their variance estimates using Rubin's rules. Exact simulation inputs and supplementary results are provided in \webautoref{sec:simulation-details} of the Supporting Information.

\subsection{Primary model calibration under incomplete follow-up}
\label{sec:simulation_primary_model}

Study~1 was designed to benchmark three working models: a Gaussian model for repeated measures fitted directly on the observed scale, a bounded working model based on a continuous interior transformation without explicit boundary mass, and the proposed mixed-boundary model. We report the summaries most directly tied to calibration of the proposed mixed-boundary formulation.

The four scenarios for data generation were chosen to span a progression from mild to substantial boundary accumulation. Rather than selecting latent means and variances ad hoc, we specified, at each visit, the probabilities of the lower and upper bounds in the control arm on the observed scale, together with a visit-specific standardized treatment shift and an AR(1) correlation parameter. The corresponding latent intercepts and marginal latent SDs were then obtained analytically so that the thresholded latent Gaussian generator reproduced the chosen control arm boundary probabilities exactly. This makes the scenario definitions interpretable on the observed scale while keeping data generation fully aligned with the fitted latent model. The exact inputs are reported in \webautoref{tab:sim-study1-design}.

After generating complete data, we imposed monotone dropout with a hazard that depended only on visit. This missingness mechanism was a special case of MAR. Study~1 was therefore not intended to assess sensitivity analysis assumptions. Its purpose was to assess whether the proposed mixed-boundary model remained well calibrated under realistically incomplete follow-up. The resulting mean missingness pattern by scenario and visit is summarized in \webautoref{tab:sim-primary-missingness}.

The reported Study~1 summaries include recovery of the latent intercepts, latent treatment effects, and implied probabilities of the lower and upper bounds at each visit under the proposed mixed-boundary model, together with recovery of arm means on the observed scale across all three working models. These summaries are reported in \webautoref{tab:sim-study1-alpha-bias-cov}, \webautoref{tab:sim-study1-delta-bias-cov}, \webautoref{tab:sim-study1-lower-boundary-bias-cov}, \webautoref{tab:sim-study1-upper-boundary-bias-cov}, and \webautoref{tab:sim-study1-observed-mean-bias-rmse-cov}.

Posterior mean estimates of boundary probabilities showed upward bias. Empirical interval coverage was generally close to nominal, although some settings showed undercoverage.

Differences among the working models were modest in the milder settings. Under \textit{AggressiveCeiling}, the mixed-boundary model showed smaller absolute bias and coverage closer to nominal than the continuous bounded model for the means in both arms on the observed scale at the final visit. However, performance depended on the scenario, arm, and summary measure; the Gaussian model was not uniformly inferior.

\subsection{Operating characteristics of the reference-based sensitivity analyses}
\label{sec:simulation_rbmi}

Study~2 used the same thresholded latent Gaussian generator but shifted the focus from calibration of the model for the observed data to the distribution of missing outcomes after the trigger. The design crossed two support templates with four truth rules for the active arm after the trigger, yielding eight scenarios in total. The first support template, \textit{ModerateFloor}, represents increasing accumulation at the lower bound with little mass at the upper bound. The second, \textit{AggressiveBothBounds}, represents a deliberately more challenging two-sided boundary setting. Within each support template, the truth in the active arm after the trigger was taken to follow MAR, a weighted J2R rule with $\omega = 0.50$, J2R, or CR. This factorial construction separates the support structure from the truth rule after the trigger and makes clear how the final scenario set was obtained. The exact template specifications and combinations of truth rules are listed in \webautoref{tab:sim-rbmi-design}.

Missingness was introduced through a mechanism for loss to follow-up comprising intermittent gaps before a final monotone missing tail. For each subject, the trigger time was defined pragmatically as the visit immediately after the last observed post-baseline outcome. Missing values before the trigger were generated under MAR within each arm, whereas values in the final missing tail in the active arm followed the truth rule for that scenario. The resulting mean missingness pattern is summarized in \webautoref{tab:sim-rbmi-missingness}.

Each dataset in Study~2 was analyzed only with the proposed mixed-boundary model, holding the fitted model fixed across sensitivity analyses. The analysis rules were MAR, weighted J2R with $\omega = 0.50$, J2R, and CR. The target in each replicate was the difference between the sample means in the active and reference arms at the final scheduled visit on the observed bounded scale. For each replicate, this true contrast was calculated from the complete outcomes generated under the scenario's truth rule, before masking the missing values. It therefore varied across replicates within a scenario. For each analysis rule, we generated one posterior predictive completed dataset per retained post-warmup draw and calculated the corresponding sample contrast. No secondary model was fitted on the observed scale, and no Rubin combination was used.

Bias was the average difference between the posterior mean contrast and the true contrast for that replicate. RMSE was the square root of the average squared difference, and coverage was the proportion of central $95\%$ posterior intervals containing the corresponding true contrast. These summaries assess recovery of the realized complete data contrast rather than inference for a fixed population treatment effect.

Absolute bias was smallest under the matching analysis rule in seven of the eight scenarios, whereas RMSE did not consistently favor the matching rule. The mean contrast estimates under weighted J2R generally lay between those under MAR and J2R. The retained numerical summary is reported in \webautoref{tab:sim-rbmi-final-visit-bias-rmse-cov}.

\subsection{Population treatment effect inference after multiple imputation}
\label{sec:simulation_mi}

Study~3 evaluated the application's procedure: imputation followed by MMRM analysis and pooling using Rubin's rules. Four scenarios crossed two latent error distributions (Gaussian and heavy-tailed multivariate Student~$t$ with five degrees of freedom, denoted $t_5$) with MAR or J2R truth after the trigger. Each replicate included 100 subjects, four post-baseline outcomes, and a fully observed prognostic baseline covariate. Missingness combined intermittent gaps and a final missing tail, with approximately $39.5\%$ missing at the final visit. The target was the fixed population mean difference, active minus reference, on the observed scale at that visit, integrating over the baseline covariate and missingness patterns. Detailed design and computational settings appear in \webautoref{sec:simulation-study3-details}.

We compared mixed-boundary imputation with a Gaussian model for $\log\left(1 + Y\right)$, where $Y$ is the simulated outcome on the original scale. The latter model's imputations were back-transformed and clipped to $\left[0, 63\right]$. Both imputation models and the observed-scale MMRM used visit-specific intercepts, baseline effects, and treatment effects, with an unstructured covariance matrix common to both arms. Each imputation model supplied 4,000 completed datasets per rule under MAR, J2R, CR, and WJ2R(0.50), using all post-warmup draws. This increased the application's 50 imputations per rule to reduce imputation Monte Carlo variability while retaining its analysis and pooling procedure: MMRM analysis of each completed dataset and combination of treatment contrasts using Rubin's rules. The same MMRM fitted before masking outcomes provided a complete-data benchmark. Performance measures were bias, empirical SD, mean estimated standard error (SE), and coverage of nominal $95\%$ intervals.

When the analysis and truth rules matched, the mixed-boundary model showed small bias under both error distributions, including the heavy-tailed scenarios where its Gaussian working distribution was misspecified (\autoref{tab:sim-study3-matched}). Its absolute bias was smaller than the shifted-log comparator's under MAR truth and similar under J2R truth. The comparator had slightly smaller empirical SDs, so neither method was uniformly superior.

Rubin's variance quantifies uncertainty about the fixed population contrast, whereas the empirical SD measures sampling variability of the pooled estimator. Both methods produced conservative J2R intervals, with mean Rubin SEs exceeding empirical SDs. This can arise because Rubin's variance estimator need not equal the pooled estimator's repeated-sampling variance under reference-based imputation \citep{bartlett2023reference}. This assessment is distinct from the latent-scale conditional variance result in \autoref{app:info-anchoring}.

Results for all sensitivity rules appear in \webautoref{tab:sim-study3-all-rules}. Rules differing from the generating truth describe sensitivity to assumptions about missing outcomes and are not expected to yield nominal coverage.

\begingroup
\setlength{\tabcolsep}{5pt}
\begin{table}[ht]
\centering
\small
\caption{Simulation Study~3: population treatment effect inference at the final visit when the imputation and generating truth rules match.}
\label{tab:sim-study3-matched}
\begin{tabular}{@{}llrrrr@{}}
\toprule
\textbf{Method} & \textbf{Rule} & \textbf{Bias} & \textbf{Emp. SD} & \textbf{Mean SE} & \textbf{Cov. (\%)} \\
\midrule
\multicolumn{6}{@{}l}{\textbf{Gaussian; MAR truth} ($\Delta = -4.52$)} \\*
Mixed boundary & MAR & 0.06 & 2.33 & 2.64 & 96.3 \\*
Shifted log & MAR & 0.32 & 2.24 & 2.60 & 97.7 \\*
Complete data & -- & 0.09 & 2.11 & 2.06 & 95.7 \\
\addlinespace[0.8em]
\multicolumn{6}{@{}l}{\textbf{\boldmath Gaussian; J2R truth} ($\Delta = -2.73$)} \\*
Mixed boundary & J2R & 0.08 & 1.58 & 2.75 & 99.7 \\*
Shifted log & J2R & 0.10 & 1.54 & 2.74 & 99.7 \\*
Complete data & -- & 0.06 & 2.24 & 2.21 & 95.7 \\
\addlinespace[0.8em]
\multicolumn{6}{@{}l}{\textbf{\boldmath Student $t_5$; MAR truth} ($\Delta = -4.28$)} \\*
Mixed boundary & MAR & -0.08 & 2.22 & 2.48 & 98.3 \\*
Shifted log & MAR & 0.19 & 2.12 & 2.37 & 97.7 \\*
Complete data & -- & 0.14 & 1.94 & 2.02 & 96.3 \\
\addlinespace[0.8em]
\multicolumn{6}{@{}l}{\textbf{\boldmath Student $t_5$; J2R truth} ($\Delta = -2.59$)} \\*
Mixed boundary & J2R & 0.03 & 1.58 & 2.61 & 100.0 \\*
Shifted log & J2R & 0.03 & 1.55 & 2.53 & 100.0 \\*
Complete data & -- & 0.12 & 2.10 & 2.16 & 95.7 \\
\bottomrule
\end{tabular}
\par\smallskip
\begin{minipage}{\linewidth}
\footnotesize
Each scenario uses 300 replicates of 100 subjects (50 per arm). $\Delta$ is the fixed population mean difference, active minus reference, on the bounded outcome scale at visit~4. Emp. SD is the empirical SD of estimates; Mean SE is the average estimated SE; Cov. is empirical coverage of nominal $95\%$ intervals. Both imputation methods use an observed-scale MMRM and Rubin's rules with 4,000 imputations per replicate. Complete data uses the same MMRM before masking, with the visit~4 estimate and SE forming an interval using a Student~$t$ critical value with 97 residual degrees of freedom. The Monte Carlo SE of coverage at $95\%$ is 1.26 percentage points. Student $t_5$ denotes a multivariate Student~$t$ distribution with five degrees of freedom.
\end{minipage}
\end{table}
\endgroup

\section{Application}
\label{sec:application}

We reanalyzed the 97 retained \textit{BtheB} subjects to estimate differences between BtheB and TAU in mean BDI-II at each post-baseline visit, adjusting for baseline. We focused on the final assessment at 8 months; negative contrasts favor BtheB. MAR served as the baseline analysis, with reference-based rules assessing sensitivity to assumptions about missing data after the trigger. As intercurrent event information was unavailable, the trigger was defined from the follow-up pattern, without assigning a hypothetical or treatment policy interpretation.

There were $T = 4$ post-baseline visits, with visit~$T$ corresponding to the 8-month assessment. For subject $i$, let $t_i^{\mathrm{last}}$ denote the last post-baseline visit at which BDI-II was observed. If $t_i^{\mathrm{last}} < T$, we set $\tau_i = t_i^{\mathrm{last}} + 1$, so that the set of missing visits from the trigger onward corresponds to the final monotone missing tail after the last observed post-baseline value. If the final scheduled visit was observed, we set $\tau_i = T + 1$, and no sensitivity rule was applied.

TAU was the reference arm. Missing outcomes in TAU were imputed under MAR. For the final missing tail in BtheB, MAR retained the active arm predictive distribution, whereas J2R, CR, and WJ2R(0.50) modified the latent conditional mean according to the rules in \autoref{sec:methodology}.

The mixed-boundary imputation model used the thresholded latent Gaussian formulation from \autoref{sec:methodology}. On the latent scale, the visit-specific linear predictor was
\begin{equation}
\mu_{it} = \alpha_t + \beta_t^{\mathrm{base}} \, \text{BDI}_{i,\mathrm{base}} + A_i \delta_t,
\end{equation}
where $\text{BDI}_{i,\mathrm{base}}$ denotes centered baseline BDI-II, $\alpha_t$ is the visit-specific intercept, $\beta_t^{\mathrm{base}}$ is the visit-specific baseline effect, $\delta_t$ is the visit-specific treatment effect, and $A_i$ is the randomized treatment indicator, with $A_i = 1$ for BtheB and $A_i = 0$ for TAU.

\subsection{Predictive model comparison}

We compared the Gaussian, continuous bounded, and mixed-boundary working models using leave-one-subject-out prediction. To place the comparison on a common scale, each model was scored using the joint probability of the intervals $\left[y - 0.5, y + 0.5\right)$ across a subject's observed visits. This avoids comparing boundary probability masses directly with continuous densities. Predictive probabilities were estimated using Pareto-smoothed importance sampling (PSIS), with explicit refits when its diagnostics were inadequate \citep{vehtari2017practical, looPackage, vehtari2024pareto}. \autoref{app:common-comparison} defines the comparison, and \webautoref{sec:loo} of the Supporting Information gives computational details.

The continuous bounded model had the highest expected log predictive density (ELPD), with a modest advantage over the mixed-boundary model (\autoref{tab:btheb-model-comparison}). The mixed-boundary minus continuous bounded difference was $-3.89$ (SE $2.10$). Both bounded models had higher ELPD than the Gaussian model; the mixed-boundary minus Gaussian difference was $24.18$ (SE $15.00$). These SEs account for pairing by subject.

\begin{table}[htbp]
\centering
\caption{Comparison using leave-one-subject-out prediction of common score intervals. Higher ELPD indicates better predictive performance. Differences are relative to the continuous bounded model; their SEs account for pairing by subject.}
\label{tab:btheb-model-comparison}
\begin{tabular}{lrrrr}
\hline
\textbf{Model} & \textbf{ELPD} & \textbf{SE} & \textbf{Difference} & \textbf{SE of difference} \\
\hline
Continuous bounded & $-928.01$ & $45.33$ & $0.00$ & --- \\
Mixed-boundary & $-931.90$ & $45.49$ & $-3.89$ & $2.10$ \\
Gaussian & $-956.08$ & $46.30$ & $-28.07$ & $13.80$ \\
\hline
\end{tabular}
\smallskip
\par\begin{minipage}{0.95\linewidth}
\footnotesize
ELPD: expected log predictive density; SE: standard error.
Scores use joint probabilities of the intervals $[y - 0.5, y + 0.5)$ across each subject's observed visits.
Explicit refits by model: continuous bounded (3); mixed-boundary (4); Gaussian (1).
All reported Markov chain Monte Carlo (MCMC) diagnostic checks passed.
\end{minipage}
\end{table}

We also compared predictions under MAR for subjects with missing BDI-II at month~8 (see \webautoref{tab:btheb-boundary-comparison} in \webautoref{sec:btheb-additional-analyses} of the Supporting Information). The mixed-boundary model assigned positive probability to imputations equal to zero and higher probability to the common interval $\left[-0.5, 0.5\right)$, representing scores that round to zero. For this interval, its predicted probabilities exceeded those of the continuous bounded model by 3.45 percentage points in TAU and 3.77 in BtheB. These differences describe the consequences of model choice, without establishing which model predicts the missing outcomes more accurately.

\subsection{Sensitivity analyses}
    
For the reference-based sensitivity analyses, we used the mixed-boundary model to obtain imputations within the known support that may equal the bounds exactly. The predictive comparison assesses observed outcomes and does not validate the assumptions about missing outcomes after the trigger.

\begin{figure}[htbp]
\centering
\includegraphics[width=0.60\textwidth]{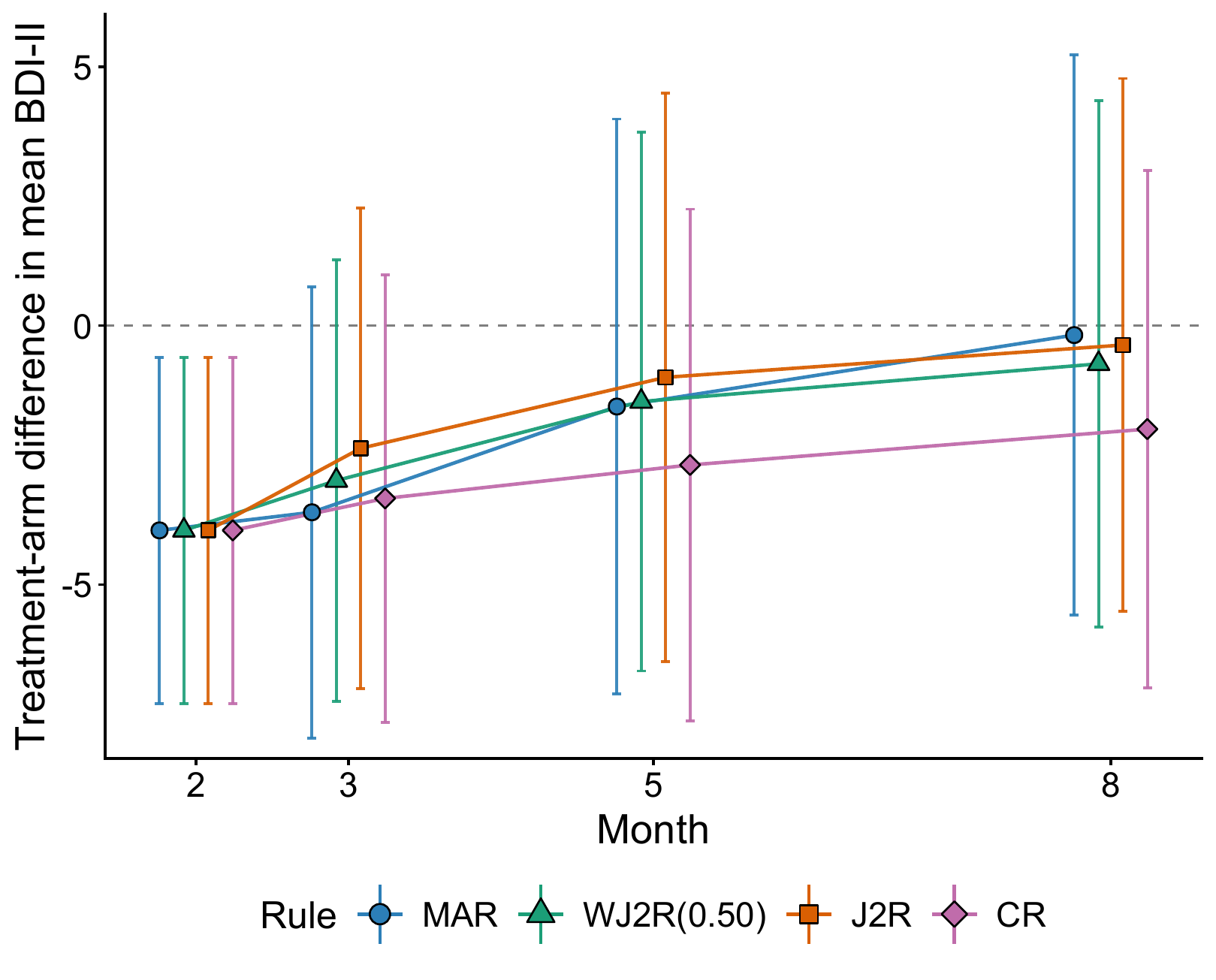}
\caption{Application to the \textit{BtheB} data: visit-specific treatment arm differences in mean BDI-II under the MAR, WJ2R(0.50), J2R, and CR sensitivity rules. Points show estimates pooled using Rubin's rules after fitting an MMRM to each of 50 posterior predictive completed datasets under each rule. Error bars show the corresponding 95\% intervals. Negative values favor \textit{Beating the Blues}, since lower BDI-II indicates fewer depressive symptoms.}
\label{fig:btheb-rbmi-contrast}
\end{figure}

For each completed dataset, we estimated the treatment contrasts on the observed scale using an MMRM with visit-specific intercepts, baseline BDI-II effects, and treatment effects, fitted by restricted maximum likelihood with a common unstructured covariance matrix across treatment arms. The model was fitted using \texttt{nlme::gls} \citep{nlmePackage}, with \texttt{corSymm} for the correlations and \texttt{varIdent} for variances that differ across visits. Let $M = 50$ denote the number of completed datasets used for this downstream multiple imputation analysis. Under each rule, we generated $M$ posterior predictive completed datasets from the fitted mixed-boundary model, fitted an MMRM to each completed dataset, and combined the resulting visit-specific treatment contrasts using Rubin's rules \citep{rubin1987multiple}. \autoref{fig:btheb-rbmi-contrast} shows negative treatment contrasts under all four rules at every visit. At 8 months, CR gave the most negative estimate and MAR the least negative, with substantial overlap between the corresponding 95\% intervals. Thus, the sensitivity assumptions affected the estimated magnitude but did not reverse the direction favoring BtheB, consistent with the original trial. These shifts in point estimates are distinct from interval calibration, which was assessed separately in Study~3. At 8 months, all four 95\% intervals included zero, so none of the sensitivity rules yielded a statistically significant treatment contrast.

\begin{figure}[ht]
\centering
\includegraphics[width=\textwidth]{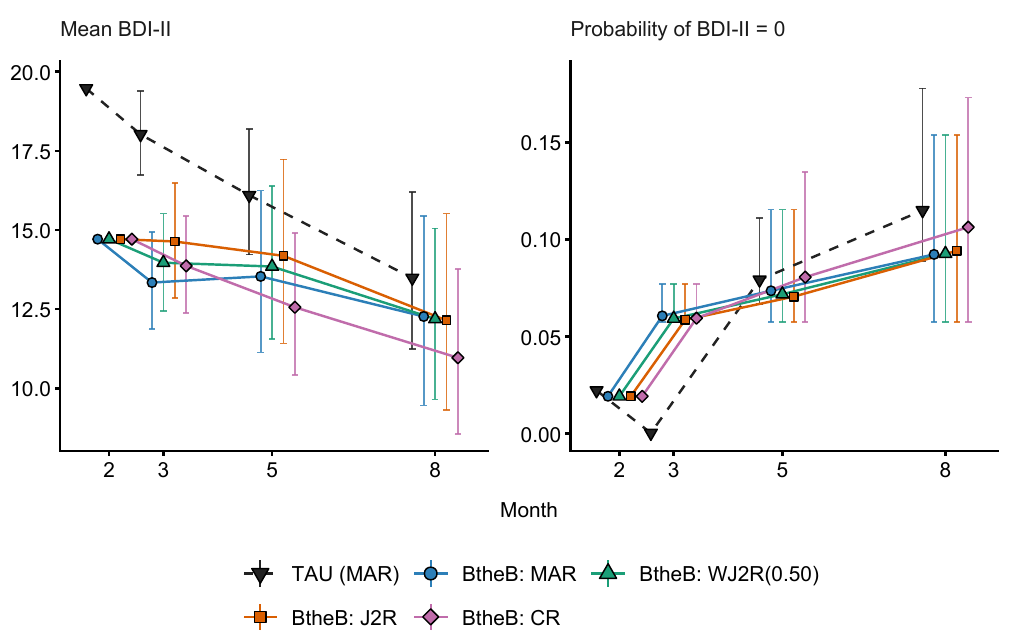}
\caption{Posterior predictive profiles of mean BDI-II and the probability of a score of exactly zero under MAR, WJ2R(0.50), J2R, and CR for BtheB, with TAU under MAR as the reference. Points show posterior means and bars show pointwise 95\% intervals for the completed cohort summaries, with observed outcomes held fixed. All 4,000 post-warmup draws were used. Profiles are offset horizontally for readability. Intervals collapse to points at month 2 because all outcomes at that visit were observed.}
\label{fig:btheb-observed-profiles}
\end{figure}

To interpret the sensitivity rules on the observed scale, \autoref{fig:btheb-observed-profiles} displays mean BDI-II and the probability of a score of exactly zero, including the TAU reference profiles. Under each rule, we generated one completed dataset per retained posterior draw ($K = 4000$), keeping observed outcomes fixed, and calculated the mean and proportion of zeros within each arm at each visit. Points show posterior means; interval limits are the 2.5th and 97.5th percentiles. These intervals describe uncertainty in the completed cohort summaries, separately from the MMRM and Rubin analysis. Differences between the completed cohort arm means need not equal the MMRM treatment contrasts, which adjust for baseline BDI-II.

The profiles in \autoref{fig:btheb-observed-profiles} show that, at 5 and 8 months, CR yielded lower mean BDI-II and higher zero probabilities than the other BtheB rules. J2R did not make the completed BtheB profiles coincide with TAU: the rules apply to missing outcomes conditional on individual histories, while the profiles also reflect retained observations and covariate differences between arms. Moreover, the nonlinear threshold map means that weighting the latent means does not generally produce the same weighting of observed means or boundary probabilities. Here, the WJ2R(0.50) profiles were close to the midpoints of MAR and J2R.

The alternative threshold specifications examined produced little change in the completed cohort means and probabilities of zero at month 8 (\webautoref{tab:btheb-threshold-sensitivity}).

\FloatBarrier
\section{Discussion}
\label{sec:discussion}

This paper extends familiar reference-based sensitivity analysis to bounded longitudinal outcomes by placing the sensitivity assumption on a latent Gaussian working scale and mapping completed values back to the original bounded scale. The resulting framework respects the known support by construction, allows exact boundary values to arise naturally, and retains the conditional Gaussian completion machinery that makes reference-based methods operationally attractive.

The proposed thresholded latent Gaussian model is related to Tobit and coarsening models, which represent boundary observations through latent regions \citep{xu2014modeling}. Reference-based imputation using a Tobit model has also been studied for censored survival outcomes \citep{atkinson2021information}. Here, boundary scores are valid outcomes, and the latent representation does not imply scores beyond the bounds.

Studies~1 and~2 assessed model calibration and recovery of realized complete data contrasts. Study~3 separately evaluated population treatment effect inference after imputation and found small bias when the analysis rule matched the truth rule after the trigger, but conservative Rubin intervals, particularly under J2R. Similar conservatism occurred with the shifted-log comparator. The heavy-tailed scenario provides a limited assessment of distributional robustness; it does not establish robustness to other forms of misspecification. These findings also emphasize that conditional variance invariance on the latent scale does not guarantee nominal frequentist coverage for the downstream MMRM and Rubin procedure. In the \textit{BtheB} application, the estimated treatment contrasts were broadly stable across sensitivity assumptions, although differences at later visits showed the expected dependence on the assumed distribution of missing outcomes after the trigger.

The comparison using common score intervals showed a modest predictive advantage for the continuous bounded model over the mixed-boundary model. The separate comparison for missing outcomes showed that model choice also affected predictions near zero, although these differences do not establish which model better represents the unobserved outcomes.

The framework links boundary probabilities and the interior distribution through a latent Gaussian model with fixed thresholds. A separate process generating boundary outcomes may require a more flexible model. The current development assumes a common latent covariance matrix across arms. Implementation requires MCMC sampling and convergence checks, adding computational effort and requiring familiarity with Bayesian model fitting. The sensitivity assumptions also require interpretation on the observed scale and cannot be verified from the observed data alone. Future work could examine practical guidance for choosing threshold locations, study performance more systematically under sparse versus substantial boundary mass, relax the assumption of a common covariance matrix, and extend the framework to ordinal, semicontinuous, and other coarsened bounded outcomes.

\section*{Acknowledgements}
    
This work is based on research supported by the National Research Foundation (NRF) of South Africa (Grant number 132383). Opinions expressed and conclusions arrived at are those of the author and are not necessarily to be attributed to the NRF.
    
\section*{Conflicts of interest}

The author declares that they have no conflicts of interest.

\section*{Data availability statement}

The data used in the application are publicly available in the \texttt{HSAUR2} package for R as the data frame \texttt{BtheB}. The analysis dataset was derived from this source. Code to reproduce the revised analyses is being prepared and will be provided for reproducible research review and made publicly available upon publication.

\bibliographystyle{apalike}
\bibliography{refs}

\section*{Supporting information}

Additional supporting information can be found online in the Supporting Information section.

\section*{Appendix}

\begin{appendices}

\section{Stan implementation of the primary model}
\label{app:stan-implementation}

The primary model was fitted in Stan by sampling from the augmented posterior described in \autoref{sec:methodology}. In the current implementation, observed interior outcomes enter through their deterministic latent transforms, whereas observed outcomes at the lower and upper bounds are represented by constrained latent variables. Thus, posterior sampling proceeds directly under the augmented latent Gaussian representation, without requiring multivariate normal rectangle probabilities in the fitting step.

In the current implementation, we adopt the formulation with a common covariance matrix described in \autoref{sec:methodology}, so that $\Sigma_0 = \Sigma_1 = \Sigma$. The covariance matrix is parameterized through marginal latent SDs and a correlation matrix:
\begin{equation}
\Sigma = \operatorname{diag}\left(\sigma_1, \dots, \sigma_T\right) R \operatorname{diag}\left(\sigma_1, \dots, \sigma_T\right),
\end{equation}
where $R$ is a $T \times T$ correlation matrix. In Stan, $R$ is represented through its Cholesky factor $L_R$, and we write
\begin{equation}
L_{\Sigma} = \operatorname{diag}\left(\sigma_1, \dots, \sigma_T\right) L_R, \qquad
\Sigma = L_{\Sigma} L_{\Sigma}^{\top}.
\end{equation}
This Cholesky parameterization improves numerical stability and sampling efficiency.

We used the following proper prior distributions. For the visit-specific coefficient vectors entering the mean specification
\begin{equation}
\mu_{iat} = X_{it}^{\top}\beta_t + a W_{it}^{\top}\delta_t,
\qquad
t = 1, \dots, T,
\end{equation}
we used independent normal priors centered at zero, which may be written as
\begin{equation}
\beta_t, \delta_t \sim \mathcal{N}\left(0, 100^2 I\right),
\end{equation}
with identity matrices of conformable dimension.

For the marginal latent SDs, we used independent half-$t$ priors with $3$ degrees of freedom, location $0$, and scale $10$,
\begin{equation}
\sigma_t \sim t_{3}^{+}\left(0, 10\right),
\qquad
t = 1, \dots, T.
\end{equation}
This specification uses the half-$t$ prior family for positive scale parameters discussed by \citet{Gelman2006prior}. For the correlation matrix, we used the Lewandowski-Kurowicka-Joe (LKJ) prior with shape parameter $1$, $R \sim \mathrm{LKJ}\left(1\right)$, which is uniform over correlation matrices \citep{lewandowski2009generating}. Posterior computation was carried out in Stan \citep{carpenter2017stan} using the No-U-Turn Sampler, an adaptive form of Hamiltonian Monte Carlo \citep{hoffman2014nuts}. The number of post-warmup draws and completed datasets used for each analysis are specified in the corresponding sections.

\section{Variance invariance in the latent threshold model}
\label{app:info-anchoring}

Following \citet{cro2019information}, an information-anchored sensitivity analysis preserves the proportion of information about the treatment estimand that is lost due to missing data when moving from the primary analysis to the sensitivity analysis. Theoretical results on information anchoring are available for controlled multiple imputation and RBMI with longitudinal continuous outcomes \citep{cro2019information} and for reference-based sensitivity analysis under imputation models based on truncated normal distributions or Tobit formulations \citep{atkinson2021information}. For the proposed thresholded model for bounded outcomes with discrete and continuous components, we do not establish a full theorem on information anchoring on the observed scale. Instead, we establish exact conditional variance invariance on the latent scale for post-trigger rules that share the same slope matrix for the conditioning history and the same conditional covariance, while allowing the conditional mean intercept to differ. Under the common covariance formulation, this class includes MAR, J2R, CR, and the $\omega$-weighted rules considered in \autoref{sec:methodology}. For quantities that are affine functions of the completed data on the latent scale, this yields exact conditional variance invariance. On the observed bounded scale, the threshold map is nonlinear and is not differentiable at the latent thresholds, so we provide only a local first-order heuristic under the conditions stated below.

Throughout this appendix, let $\vartheta$ denote the full posterior state under the primary model, consisting of the structural parameters together with any augmented latent variables introduced for observed boundary values. Conditional on $\vartheta$ and the observed data $\mathcal{D}_{\mathrm{obs}}$, the sensitivity analysis enters only through the posterior predictive distribution of the missing latent outcomes from the trigger onward; it does not alter the posterior distribution $p\left(\vartheta \middle| \mathcal{D}_{\mathrm{obs}}\right)$. We write $r$ for a generic post-trigger sensitivity rule. In this appendix, the rule $r$ is assumed to belong to the class
\begin{equation}
H_{i,\mathcal{M}_i^{\mathrm{post}}}^{\left(r\right)}
\left|
\widetilde{H}_{i,\mathcal{O}_i^{\star}}, A_i = 1, X_i, \vartheta\right.
\sim
\mathcal{N}\left(
m_i^{\left(r\right)}\left(\vartheta\right),
V_i^{\mathrm{post}}\left(\vartheta\right)
\right),
\end{equation}
with
\begin{equation}
m_i^{\left(r\right)}\left(\vartheta\right)
=
c_i^{\left(r\right)}\left(\vartheta\right)
+
B_i\left(\vartheta\right)\widetilde{H}_{i,\mathcal{O}_i^{\star}},
\end{equation}
and
\begin{equation}
B_i\left(\vartheta\right)
=
\Sigma_{\mathcal{M}_i^{\mathrm{post}}\mathcal{O}_i^{\star}}
\Sigma_{\mathcal{O}_i^{\star}\mathcal{O}_i^{\star}}^{-1}.
\end{equation}
That is, the rules considered here may change the intercept term $c_i^{\left(r\right)}\left(\vartheta\right)$, but they share the same slope matrix $B_i\left(\vartheta\right)$ and the same conditional covariance $V_i^{\mathrm{post}}\left(\vartheta\right)$. Under the formulation with a common covariance matrix described in \autoref{sec:methodology}, this class includes MAR, J2R, CR, and the $\omega$-weighted rules introduced there.

\subsection{Conditional variance invariance on the latent scale}

For subject $i$, recall that the missing latent values before the trigger are imputed under the arm-specific MAR law,
\begin{equation}
H_{i,\mathcal{M}_i^{\mathrm{pre}}}
\left|
\widetilde{H}_{i,\mathcal{O}_i}, A_i, X_i, \vartheta\right.
\sim
\mathcal{N}\left(
m_{iA_i}^{\mathrm{pre}}\left(\vartheta\right),
V_i^{\mathrm{pre}}\left(\vartheta\right)
\right).
\end{equation}
Let $\mathcal{O}_i^{\star} = \mathcal{O}_i \cup \mathcal{M}_i^{\mathrm{pre}}$ denote the augmented conditioning set. Let $B_{i,\mathcal{O}}\left(\vartheta\right)$ and $B_{i,\mathrm{pre}}\left(\vartheta\right)$ denote the columns of $B_i\left(\vartheta\right)$ corresponding to visits in $\mathcal{O}_i$ and $\mathcal{M}_i^{\mathrm{pre}}$, respectively, retaining increasing visit order within each subset. Then
\begin{equation}
\begin{aligned}
B_i\left(\vartheta\right)\widetilde{H}_{i,\mathcal{O}_i^{\star}}
={}&
B_{i,\mathcal{O}}\left(\vartheta\right)\widetilde{H}_{i,\mathcal{O}_i} + B_{i,\mathrm{pre}}\left(\vartheta\right)H_{i,\mathcal{M}_i^{\mathrm{pre}}}.
\end{aligned}
\end{equation}

\begin{proposition}\label{prop:variance_invariance}
For each subject $i$, define the full missing latent vector under rule $r$ by
\begin{equation}
H_{i,\mathcal{M}_i}^{\left(r\right)} = \left(H_{i,\mathcal{M}_i^{\mathrm{pre}}}^{\top}, H_{i,\mathcal{M}_i^{\mathrm{post}}}^{\left(r\right)\top}\right)^{\top}.
\end{equation}
Conditional on $\vartheta$ and $\mathcal{D}_{\mathrm{obs}}$, the covariance matrix of $H_{i,\mathcal{M}_i}^{\left(r\right)}$ is the same for all post-trigger rules $r$ in the class above. Specifically,
\begin{equation}
\mathrm{Cov}\left(
H_{i,\mathcal{M}_i}^{\left(r\right)}
\middle|
\vartheta, \mathcal{D}_{\mathrm{obs}}
\right)
=
\begin{pmatrix}
V_i^{\mathrm{pre}} &
V_i^{\mathrm{pre}} B_{i,\mathrm{pre}}^{\top} \\
B_{i,\mathrm{pre}} V_i^{\mathrm{pre}} &
V_i^{\mathrm{post}} + B_{i,\mathrm{pre}} V_i^{\mathrm{pre}} B_{i,\mathrm{pre}}^{\top}
\end{pmatrix},
\end{equation}
where the dependence on $\vartheta$ has been suppressed for readability.
\end{proposition}

\begin{proof}
By construction, the distribution of the latent vector before the trigger does not depend on the post-trigger rule $r$, so
\begin{equation}
H_{i,\mathcal{M}_i^{\mathrm{pre}}}
=
m_{iA_i}^{\mathrm{pre}} + \varepsilon_i^{\mathrm{pre}},
\qquad
\varepsilon_i^{\mathrm{pre}}
\sim
\mathcal{N}\left(0, V_i^{\mathrm{pre}}\right).
\end{equation}
For the component from the trigger onward, the assumed form of the rule gives
\begin{equation}
H_{i,\mathcal{M}_i^{\mathrm{post}}}^{\left(r\right)}
=
c_i^{\left(r\right)}
+
B_{i,\mathcal{O}} \widetilde{H}_{i,\mathcal{O}_i}
+
B_{i,\mathrm{pre}} H_{i,\mathcal{M}_i^{\mathrm{pre}}}
+
\varepsilon_i^{\mathrm{post}},
\end{equation}
with
\begin{equation}
\varepsilon_i^{\mathrm{post}}
\sim
\mathcal{N}\left(0, V_i^{\mathrm{post}}\right).
\end{equation}
Under the Gaussian conditional representation, $\varepsilon_i^{\mathrm{post}}$ is independent of $\left(\widetilde{H}_{i,\mathcal{O}_i}^{\top}, H_{i,\mathcal{M}_i^{\mathrm{pre}}}^{\top}\right)^{\top}$ given $\vartheta$ and $\mathcal{D}_{\mathrm{obs}}$. Since $\widetilde{H}_{i,\mathcal{O}_i}$ is fixed after conditioning on $\vartheta$ and $\mathcal{D}_{\mathrm{obs}}$, the term $c_i^{\left(r\right)} + B_{i,\mathcal{O}} \widetilde{H}_{i,\mathcal{O}_i}$ affects only the conditional mean and not the conditional covariance. Therefore,
\begin{equation}
\mathrm{Cov}\left(
H_{i,\mathcal{M}_i^{\mathrm{pre}}},
H_{i,\mathcal{M}_i^{\mathrm{post}}}^{\left(r\right)}
\middle|
\vartheta, \mathcal{D}_{\mathrm{obs}}
\right)
=
V_i^{\mathrm{pre}} B_{i,\mathrm{pre}}^{\top},
\end{equation}
and
\begin{equation}
\mathrm{Var}\left(
H_{i,\mathcal{M}_i^{\mathrm{post}}}^{\left(r\right)}
\middle|
\vartheta, \mathcal{D}_{\mathrm{obs}}
\right)
=
B_{i,\mathrm{pre}} V_i^{\mathrm{pre}} B_{i,\mathrm{pre}}^{\top}
+
V_i^{\mathrm{post}},
\end{equation}
both of which are free of $r$. Combining the covariance blocks yields the stated matrix.
\end{proof}

\begin{proposition}\label{prop:affine_latent_invariance}
Let $Q_H^{\left(r\right)} = q_{0}^{\left(r\right)}\left(\vartheta, \mathcal{D}_{\mathrm{obs}}\right) + \sum_{i=1}^{n} d_i^{\top} H_{i,\mathcal{M}_i}^{\left(r\right)}$ be any estimand that is an affine function of the completed data on the latent scale, where the vectors $d_i$ do not depend on the post-trigger rule. Then $\mathrm{Var}\left(Q_H^{\left(r\right)} \middle| \vartheta, \mathcal{D}_{\mathrm{obs}}\right)$ is identical for all rules $r$ in the class above.
\end{proposition}

\begin{proof}
Conditional on $\vartheta$ and $\mathcal{D}_{\mathrm{obs}}$, the only random components in $Q_H^{\left(r\right)}$ are the missing latent vectors $H_{i,\mathcal{M}_i}^{\left(r\right)}$, and these vectors are independent across subjects under the model. Hence
\begin{equation}
\mathrm{Var}\left(
Q_H^{\left(r\right)}
\middle|
\vartheta, \mathcal{D}_{\mathrm{obs}}
\right)
=
\sum_{i=1}^{n}
d_i^{\top}
\mathrm{Cov}\left(
H_{i,\mathcal{M}_i}^{\left(r\right)}
\middle|
\vartheta, \mathcal{D}_{\mathrm{obs}}
\right)
d_i.
\end{equation}
The result now follows immediately from \autoref{prop:variance_invariance}.
\end{proof}

\autoref{prop:affine_latent_invariance} gives an exact conditional variance invariance result on the latent scale for the class of rules considered here. Conditional on the fitted posterior state, the post-trigger sensitivity rule changes only the location of the missing latent Gaussian distribution and does not change its covariance. Thus, for estimands that are affine functions of the completed data on the latent scale, the conditional contribution of missing data to the variance is preserved across sensitivity analyses.

\subsection{Posterior integration and variation across posterior states}

Assume that the completed-data quantities considered below have finite second moments under their posterior predictive distributions.

For any estimand that is an affine function of the completed data on the latent scale, $Q_H^{\left(r\right)}$, the law of total variance gives
\begin{equation}
\mathrm{Var}\left(
Q_H^{\left(r\right)}
\middle|
\mathcal{D}_{\mathrm{obs}}
\right)
=
E\left[
\mathrm{Var}\left(
Q_H^{\left(r\right)}
\middle|
\vartheta, \mathcal{D}_{\mathrm{obs}}
\right)
\middle|
\mathcal{D}_{\mathrm{obs}}
\right]
+
\mathrm{Var}\left(
E\left[
Q_H^{\left(r\right)}
\middle|
\vartheta, \mathcal{D}_{\mathrm{obs}}
\right]
\middle|
\mathcal{D}_{\mathrm{obs}}
\right).
\end{equation}
By \autoref{prop:affine_latent_invariance}, the first term is identical across all rules $r$ in the class above. Therefore, any difference between rules in the posterior variance conditional on $\mathcal{D}_{\mathrm{obs}}$ must arise entirely through the second term, namely variation in the conditional mean across posterior states.

More precisely, for any two rules $r$ and $r^{\prime}$, exact equality of $\mathrm{Var}\left(Q_H^{\left(r\right)} \middle| \mathcal{D}_{\mathrm{obs}}\right)$ and $\mathrm{Var}\left(Q_H^{\left(r^{\prime}\right)} \middle| \mathcal{D}_{\mathrm{obs}}\right)$ continues to hold whenever the difference $E\left[Q_H^{\left(r\right)} \middle| \vartheta, \mathcal{D}_{\mathrm{obs}}\right] - E\left[Q_H^{\left(r^{\prime}\right)} \middle| \vartheta, \mathcal{D}_{\mathrm{obs}}\right]$ is constant in $\vartheta$. This covers the special case in which the rule changes the conditional mean only by an additive constant that does not depend on the posterior state. Otherwise, any remaining discrepancy reflects variation across posterior states rather than a change in the conditional covariance of the missing data.

\subsection{A first-order heuristic on the observed scale}

Let $Q_Y^{\left(r\right)} = \psi\left(Y^{\mathrm{comp},\left(r\right)}, A, X\right)$ denote an estimand evaluated on the completed data on the observed scale. Let $H_{\mathcal{M}}^{\left(r\right)}$ denote the stacked missing latent vector under rule $r$, arranged in a fixed order across subjects and visits. Write $G\left(h\right)$ for the componentwise threshold map that sends a candidate missing latent vector $h$ to the corresponding vector of completed missing observed outcomes under the thresholded observation rule in \autoref{sec:thresholded_latent}. Let $Y^{\mathrm{comp}}\left(h\right)$ denote the full completed observed outcome vector obtained by combining the observed outcomes $Y_{\mathrm{obs}}$ with the missing observed outcomes $G\left(h\right)$. Define $F\left(h\right) = \psi\left(Y^{\mathrm{comp}}\left(h\right), A, X\right)$. For a fixed posterior state $\vartheta$, let $m_r\left(\vartheta\right) = E\left[H_{\mathcal{M}}^{\left(r\right)} \middle| \vartheta, \mathcal{D}_{\mathrm{obs}}\right]$ and $\Sigma_{\mathcal{M}}\left(\vartheta\right) = \mathrm{Cov}\left(H_{\mathcal{M}}^{\left(r\right)} \middle| \vartheta, \mathcal{D}_{\mathrm{obs}}\right)$, where \autoref{prop:variance_invariance} implies that $\Sigma_{\mathcal{M}}\left(\vartheta\right)$ is common across rules in the class above.

The map $G$ is piecewise smooth but is not differentiable at the thresholds. Accordingly, the following argument is only local. For each rule being compared, suppose that there exists a neighborhood of $m_r\left(\vartheta\right)$ on which $F$ is continuously differentiable and that the conditional distribution of $H_{\mathcal{M}}^{\left(r\right)}$ places negligible mass outside that neighborhood. In addition, assume that the first-order Taylor remainder contributes negligibly to the conditional variance, including through its covariance with the linear term. This assumption concerns both curvature within the neighborhood and contributions from draws outside it, including threshold crossings. Under these conditions, a first-order approximation using the delta method gives
\begin{equation}
\mathrm{Var}\left(
Q_Y^{\left(r\right)}
\middle|
\vartheta, \mathcal{D}_{\mathrm{obs}}
\right)
\approx
\nabla F\left(m_r\left(\vartheta\right)\right)^{\top}
\Sigma_{\mathcal{M}}\left(\vartheta\right)
\nabla F\left(m_r\left(\vartheta\right)\right).
\end{equation}
Accordingly, approximate conditional variance invariance on the observed scale is plausible when the local linear approximations are adequate, the gradients $\nabla F\left(m_r\left(\vartheta\right)\right)$ are similar across rules, and the rules do not move appreciable latent probability mass across the thresholds. Smoothness away from the thresholds alone does not ensure similar variances.

This is a heuristic rather than an exact theorem. A rigorous justification of the variance approximation would require quantitative control of the Taylor remainder, including contributions from threshold crossings and the atoms at the bounds. Approximate variance invariance would additionally require control of differences between the gradients across rules. We do not develop those bounds here.

\section{Predictive comparison on a common score scale}
\label{app:common-comparison}

\subsection{Common score intervals}

The mixed-boundary model has a continuous interior density and probability masses at the bounds, whereas the Gaussian and continuous bounded models have continuous densities. Direct comparison of these native predictive ordinates would mix different measures at the bounds. We therefore evaluated all three models through probabilities of the same score intervals. The original likelihoods and priors were retained for fitting; the interval convention was used only for predictive evaluation.

For an integer BDI-II score $y$, define $I\left(y\right) = \left[y - 0.5, y + 0.5\right)$. For model $m$, let $\theta_m$ denote its structural parameters and define
\begin{equation}
q_{im}\left(\theta_m\right)
=
\Pr_m\left(
Y_{it}^{\mathrm{rep}} \in I\left(y_{it}\right)
\text{ for every } t \in \mathcal{O}_i
\middle| A_i, X_i, \theta_m
\right),
\label{eq:common-interval-probability}
\end{equation}
where $Y_i^{\mathrm{rep}}$ is a new outcome vector with the subject's treatment and baseline covariate values. Each subject contributes one joint probability across their observed post-baseline visits, with missing visits marginalized. The resulting predictive mass functions are compared with respect to counting measure on integer score vectors. The comparison therefore uses the logarithmic scoring rule for the predictive distribution of the recorded score categories \citep{gneiting2007strictly}.

For the Gaussian model, the joint interval probability is a normal rectangle probability on the original scale. The intervals for scores $0$ and $63$ are $\left[-0.5, 0.5\right)$ and $\left[62.5, 63.5\right)$, respectively. Gaussian probability outside the valid score categories is retained in the predictive distribution: predictions are neither clipped nor renormalized to the range $0$ to $63$.

In this application, the common latent thresholds are $a = \Phi^{-1}\left(0.5/64\right)$ and $b = \Phi^{-1}\left(63.5/64\right)$. The continuous bounded comparator uses
\begin{equation}
h\left(y\right) = \Phi^{-1}\left(\frac{y + 0.5}{64}\right),
\qquad
Y_{it}^{\mathrm{rep}} = 64\Phi\left(H_{it}^{\mathrm{rep}}\right) - 0.5.
\end{equation}
Its continuous support is $\left(-0.5, 63.5\right)$, and it has no atoms at $0$ or $63$. In its fitting likelihood, both endpoints are evaluated as finite transformed values, $h\left(0\right) = a$ and $h\left(63\right) = b$, with the corresponding Jacobian. The Gaussian fitting likelihood also evaluates densities at these endpoints. The mixed-boundary model instead maps latent values at or below $a$ to $0$, values at or above $b$ to $63$, and interior values through the same inverse transformation. Its fitting likelihood integrates the latent tail regions for exact boundary observations; see \webautoref{sec:loo}.

Under either bounded model, the latent interval for the predictive score at an observed integer $y$ is
\begin{equation}
\left[
\Phi^{-1}\left(\frac{y}{64}\right),
\Phi^{-1}\left(\frac{y + 1}{64}\right)
\right),
\end{equation}
with infinite endpoints interpreted in the usual way. For the mixed-boundary model, the endpoint intervals include the corresponding atoms. The two bounded models thus use identical rectangle limits, but their posterior distributions differ because their fitting likelihoods treat exact bounds differently. This comparison evaluates prediction at the recorded score resolution; it does not distinguish an atom from continuous probability within the same endpoint interval.

\subsection{Prediction with a subject left out}

Let $\mathcal{D}_{-i}$ denote the observed outcomes of the other $n - 1$ subjects. Conditional on treatment and baseline covariates, the contribution for subject $i$ is
\begin{equation}
\ell_{im}
=
\log\left\{
\int q_{im}\left(\theta_m\right)
p_m\left(\theta_m \middle| \mathcal{D}_{-i}\right)
\,d\theta_m
\right\},
\qquad
\widehat{\mathrm{ELPD}}_m = \sum_{i=1}^{n}\widehat{\ell}_{im}.
\end{equation}
All observed outcomes for the held-out subject are excluded from fitting. Baseline centering is held fixed across folds. Higher ELPD indicates better prediction of the observed score intervals.

Let $S$ denote the number of retained posterior draws used for the predictive comparison. Write $L_{im}\left(\theta_m\right)$ for the subject's original observed likelihood contribution, including integration over latent variables associated with observed boundaries. With draws $\theta_m^{\left(s\right)}$ from the posterior fitted to all subjects, the importance ratios are proportional to $1/L_{im}\left(\theta_m^{\left(s\right)}\right)$. After PSIS smoothing and normalization, the weights $\widetilde{w}_{im}^{\left(s\right)}$ give
\begin{equation}
\widehat{\ell}_{im}
=
\log\left\{
\sum_{s=1}^{S}\widetilde{w}_{im}^{\left(s\right)}
q_{im}\left(\theta_m^{\left(s\right)}\right)
\right\},
\qquad
\sum_{s=1}^{S}\widetilde{w}_{im}^{\left(s\right)} = 1.
\end{equation}
The fitting likelihood determines the weights, whereas the common interval probability determines the predictive score. Substituting interval probabilities for the likelihood in the importance ratios would not recover the leave-one-subject-out posterior of the fitted model. When the PSIS diagnostics were inadequate, we refitted the model without subject $i$ and averaged interval probabilities using equal weights over all draws from that refit \citep{vehtari2017practical, looPackage}.

\subsection{Uncertainty in model differences}

For models $m$ and $m'$, define $d_i = \widehat{\ell}_{im} - \widehat{\ell}_{im'}$ and $\bar d = n^{-1}\sum_{i=1}^{n}d_i$. We report the ELPD difference and its approximate SE as
\begin{equation}
\widehat{\Delta}_{m,m'} = \sum_{i=1}^{n}d_i,
\qquad
\widehat{\mathrm{SE}}\left(\widehat{\Delta}_{m,m'}\right)
=
\left\{\frac{n}{n - 1}\sum_{i=1}^{n}\left(d_i - \bar d\right)^2\right\}^{1/2}.
\end{equation}
The SE of an individual model's ELPD is obtained by replacing $d_i$ with $\widehat{\ell}_{im}$ in the same expression. Pairwise SEs use the differences within subjects and cannot be obtained by treating the model estimates as independent \citep{vehtari2017practical, looPackage}. These SEs describe variation across subjects, rather than Monte Carlo or numerical integration error. The comparison concerns observed outcome prediction under the fitted working models and does not test the reference-based assumptions about missing outcomes.

\end{appendices}

\newpage

\markboth{}{}

\setcounter{page}{1}
\setcounter{figure}{0}
\setcounter{table}{0}
\renewcommand\appendixname{Web Appendix}
\renewcommand{\figurename}{Web Figure}
\renewcommand{\tablename}{Web Table}
    
\begin{appendices}
    
\maketitle

\vspace{2cm}

\begin{center}
\noindent
{\Large\bf Supporting Information}
\end{center}

\newpage
\clearpage

\appendix

\section{Descriptive summaries of the motivating example}
\label{sec:btheb-descriptive}

\begin{figure}[!htbp]
\centering
\includegraphics[width=0.6\textwidth]{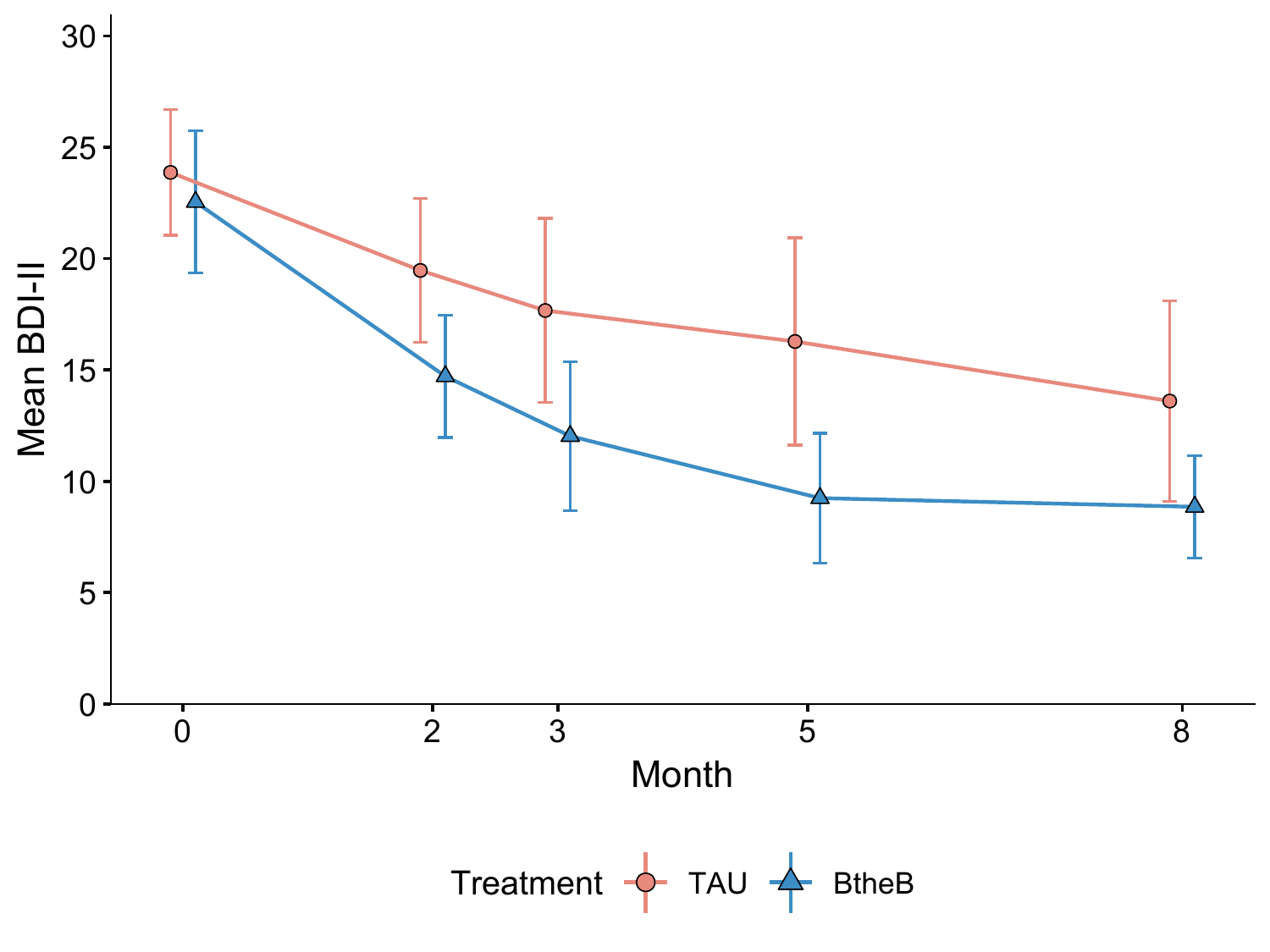}
\caption{Observed mean BDI-II by visit and treatment arm, comparing treatment as usual (TAU) with \textit{Beating the Blues} (BtheB). Error bars show approximate pointwise 95\% confidence intervals computed using normal theory from the available cases at each visit.}
\label{fig:btheb-mean-profile}
\end{figure}

\begin{figure}[!htbp]
\centering
\includegraphics[width=0.6\textwidth]{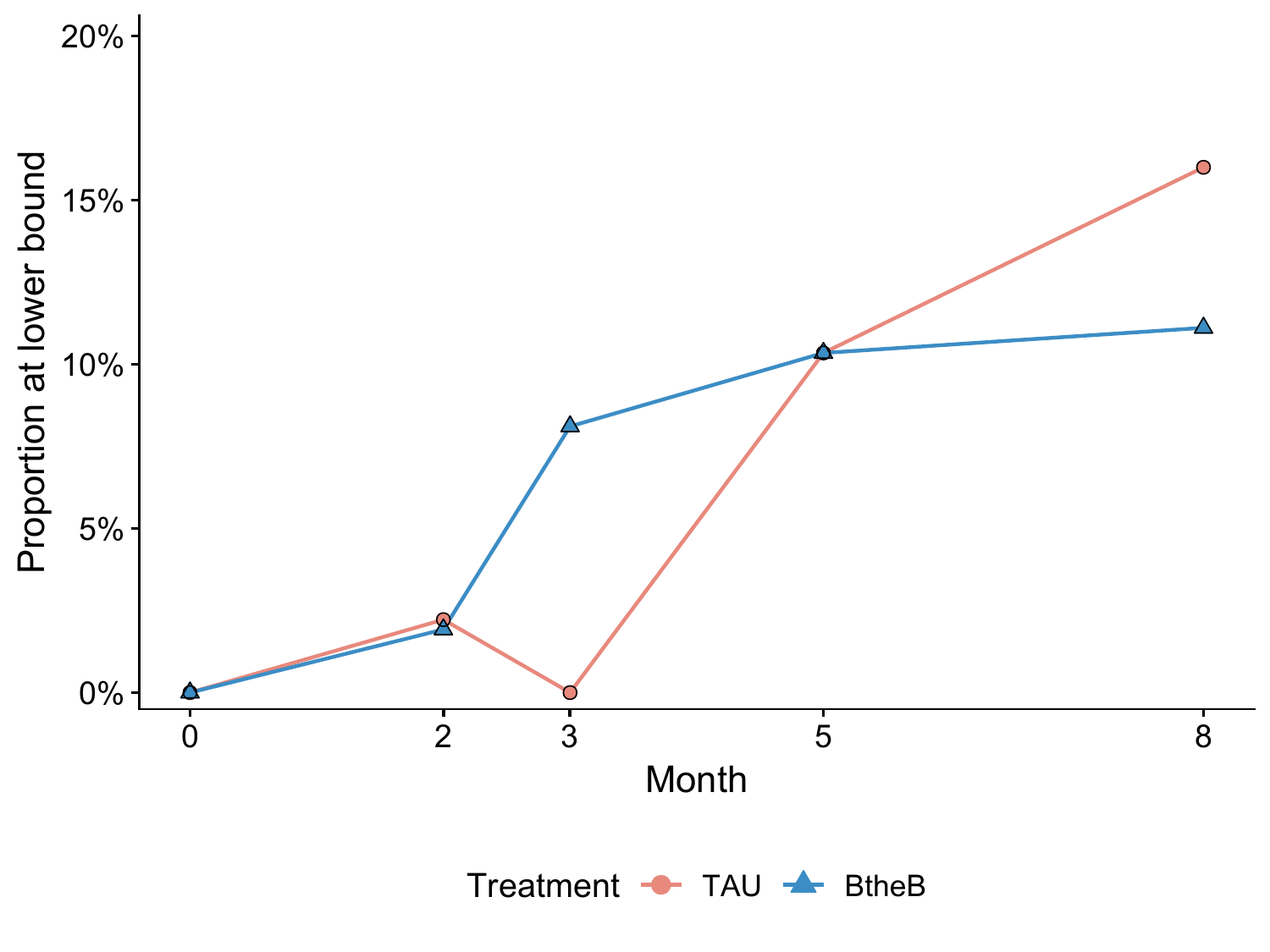}
\caption{Observed proportion of BDI-II scores at the lower bound by visit and treatment arm, comparing treatment as usual (TAU) with \textit{Beating the Blues} (BtheB). Proportions are calculated among subjects with an observed score at each visit within each treatment arm. Values of exactly zero become more prominent later in follow-up.}
\label{fig:btheb-lower-bound}
\end{figure}

\begin{figure}[htbp]
  \centering
  \includegraphics[width=0.8\textwidth]{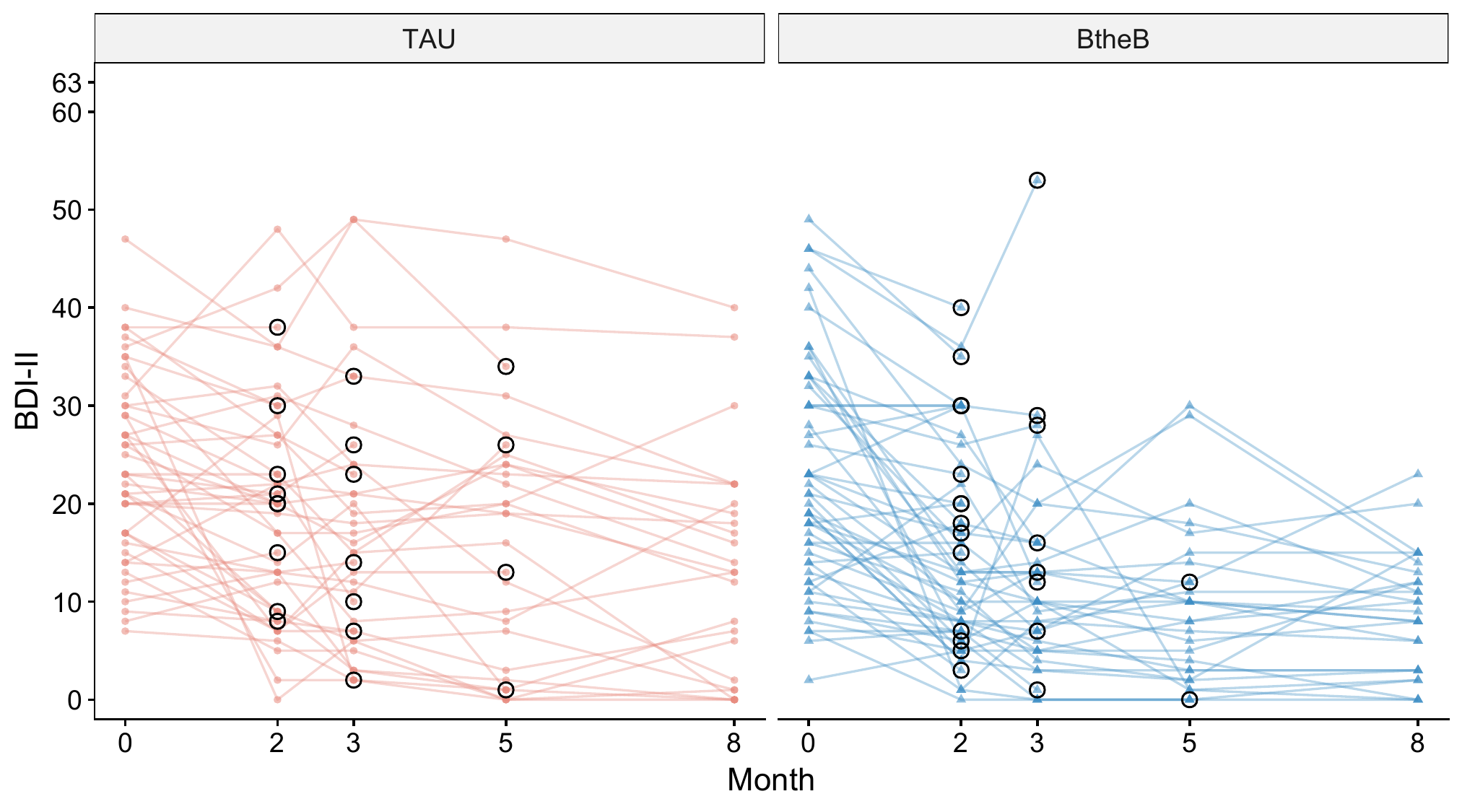}
  \caption{Subject trajectories of BDI-II by treatment arm, comparing treatment as usual (TAU) with \textit{Beating the Blues} (BtheB). The trajectories illustrate heterogeneity across subjects, a general downward trend over time, occasional values of exactly zero, and incomplete follow-up for many subjects before the final visit. Open black circles indicate the last observed BDI-II value for subjects whose final available assessment occurred before the last scheduled visit.}
  \label{fig:btheb-spaghetti}
\end{figure}

\FloatBarrier
\section{Simulation study design and supplementary results}
\label{sec:simulation-details}

This appendix gives the simulation inputs and supplementary numerical results for Studies~1 to 3.

\subsection{Setup for Simulation Studies~1 and~2}

Studies~1 and~2 used the common bounded scale
\begin{equation}
Y_{it} \in \left[0, 63\right],
\end{equation}
with lower bound $L = 0$, upper bound $U = 63$, unit score step $s = 1$, and $T = 5$ post-baseline visits. The unit step size was used only to define the threshold probabilities through the implied half-step boundaries of the underlying score grid.

The half-step construction of \autoref{sec:thresholded_latent} gave $p_L = 0.5/64$ and $p_U = 63.5/64$, with latent thresholds $a = \Phi^{-1}\left(p_L\right)$ and $b = \Phi^{-1}\left(p_U\right)$. Accordingly, the thresholded latent Gaussian generator produced a continuous interior on $\left(0, 63\right)$ together with positive mass at the bounds, rather than outcomes restricted to the full integer lattice.

In Studies~1 and~2, each fitted model used two Markov chains with 1,000 warmup iterations and 1,000 retained post-warmup iterations per chain. All post-warmup draws were retained for inference. Posterior computation was carried out in Stan using dynamic Hamiltonian Monte Carlo with a target acceptance probability of $0.95$ and a maximum tree depth of $10$.

\subsection{Simulation Study~1: construction of the primary model scenarios}

Let $s \in \left\{1, \dots, 4\right\}$ index the Study~1 scenario and let $t \in \left\{1, \dots, T\right\}$ index visit. Scenario $s$ was defined through probabilities of the lower and upper bounds in the control arm on the observed scale, together with a vector of standardized treatment shifts across visits, $\boldsymbol{\kappa}^{\left(s\right)}$, and an AR(1) correlation parameter $\rho^{\left(s\right)}$. Equivalently, each scenario was determined by the vectors
\begin{equation}
\boldsymbol{\pi}_{0}^{L,\left(s\right)} = \left(\pi_{01}^{L,\left(s\right)}, \dots, \pi_{0T}^{L,\left(s\right)}\right)^{\top},
\qquad
\boldsymbol{\pi}_{0}^{U,\left(s\right)} = \left(\pi_{01}^{U,\left(s\right)}, \dots, \pi_{0T}^{U,\left(s\right)}\right)^{\top},
\qquad
\boldsymbol{\kappa}^{\left(s\right)} = \left(\kappa_1^{\left(s\right)}, \dots, \kappa_T^{\left(s\right)}\right)^{\top},
\end{equation}
together with $\rho^{\left(s\right)}$. Here, $\pi_{0t}^{L,\left(s\right)}$ and $\pi_{0t}^{U,\left(s\right)}$ denote the probabilities that the outcome equals the lower and upper bounds, respectively, in the control arm at visit $t$ under scenario $s$.

For scenario $s$, let
\begin{equation}
z_t^{L,\left(s\right)} = \Phi^{-1}\left(\pi_{0t}^{L,\left(s\right)}\right)
\qquad \text{and} \qquad
z_t^{U,\left(s\right)} = \Phi^{-1}\left(1 - \pi_{0t}^{U,\left(s\right)}\right).
\end{equation}
Solving the normal quantile equations $a = \alpha_t^{\left(s\right)} + \sigma_t^{\left(s\right)} z_t^{L,\left(s\right)}$ and $b = \alpha_t^{\left(s\right)} + \sigma_t^{\left(s\right)} z_t^{U,\left(s\right)}$ gave the marginal latent SDs and control arm latent intercepts:
\begin{equation}
\sigma_t^{\left(s\right)} =
\frac{b - a}
{z_t^{U,\left(s\right)} - z_t^{L,\left(s\right)}},
\qquad
\alpha_t^{\left(s\right)} = a - \sigma_t^{\left(s\right)} z_t^{L,\left(s\right)}.
\end{equation}
By construction, this gives the desired control-arm boundary probabilities under the thresholded latent Gaussian model.

Treatment separation was introduced through
\begin{equation}
\delta_t^{\left(s\right)} = \kappa_t^{\left(s\right)} \sigma_t^{\left(s\right)},
\end{equation}
so that the visit-specific latent means were
\begin{equation}
\mu_{0t}^{\left(s\right)} = \alpha_t^{\left(s\right)}
\qquad \text{and} \qquad
\mu_{1t}^{\left(s\right)} = \alpha_t^{\left(s\right)} + \delta_t^{\left(s\right)}.
\end{equation}
Hence, the active arm boundary probabilities were implied quantities rather than primary design inputs.

The within-subject covariance matrix for scenario $s$ was
\begin{equation}
\Sigma^{\left(s\right)} = D^{\left(s\right)} R\left(\rho^{\left(s\right)}\right) D^{\left(s\right)},
\end{equation}
where
\begin{equation}
D^{\left(s\right)} = \operatorname{diag}\left(\sigma_1^{\left(s\right)}, \dots, \sigma_T^{\left(s\right)}\right),
\end{equation}
and
\begin{equation}
R_{tu}\left(\rho^{\left(s\right)}\right) = \left(\rho^{\left(s\right)}\right)^{\left|t - u\right|},
\qquad
t,u = 1, \dots, T.
\end{equation}
The four Study~1 scenarios were chosen for distinct purposes. \textit{SparseBoundary} provides a benchmark with outcomes concentrated in the interior and only mild boundary accumulation. \textit{ModerateFloor} was chosen to mimic progressive accumulation near the lower bound. \textit{AggressiveCeiling} stresses the methods under increasing accumulation at the upper bound. \textit{AggressiveBothBounds} is a deliberately difficult two-sided setting with appreciable mass at both bounds.

After generating complete data, we imposed monotone dropout with a vector of hazards for each scenario,
\begin{equation}
\boldsymbol{h}^{\left(s\right)} = \left(h_1^{\left(s\right)}, \dots, h_T^{\left(s\right)}\right)^{\top}.
\end{equation}
Specifically,
\begin{equation}
\boldsymbol{h}^{\left(\mathrm{SparseBoundary}\right)}
=
\left(0.00, 0.03, 0.05, 0.07, 0.09\right)^{\top},
\end{equation}
whereas
\begin{equation}
\begin{aligned}
&\boldsymbol{h}^{\left(s\right)}
=
\left(0.00, 0.04, 0.06, 0.08, 0.10\right)^{\top},\\
&s
\in
\left\{\mathrm{ModerateFloor}, \mathrm{AggressiveCeiling},
\mathrm{AggressiveBothBounds}\right\}.
\end{aligned}
\end{equation}
Because dropout depended only on visit and not on latent outcomes, treatment arm, or latent random effects, it was a simple special case of the MAR assumption used in the primary model. The first visit was fully observed. At each subsequent visit $t$, subjects who had not previously dropped out were independently assigned dropout with probability $h_t^{\left(s\right)}$. When dropout occurred at visit $t$, the outcome at that visit and all subsequent outcomes were set to missing. Thus, $h_t^{\left(s\right)}$ is a conditional dropout probability among subjects still in follow-up, rather than the marginal probability of a missing outcome at visit $t$.

For the continuous bounded comparator in Study~1, exact scores of $0$ and $63$ were replaced by $0.5$ and $62.5$, respectively, while interior scores were unchanged. Denoting these adjusted scores by $Y^{\dagger}$, we fitted a Gaussian model to $H = \Phi^{-1}\left(Y^{\dagger}/63\right)$ and calculated model-implied means on the observed scale using the inverse map $Y = 63\Phi\left(H\right)$. This comparator differs from the continuous bounded comparator used in the application, whose transformation is defined in \autoref{app:common-comparison}.

Although the latent parameter and boundary probability tables focus on calibration of the proposed mixed-boundary model, the retained Study~1 tables also include recovery of arm means on the observed scale across all three working models, since this comparison more directly shows how modeling that respects the support affects fit on the observed scale. The exact Study~1 inputs are reported in \webautoref{tab:sim-study1-design}, the resulting mean missingness proportions in \webautoref{tab:sim-primary-missingness}, the retained latent and boundary recovery summaries in \webautoref{tab:sim-study1-alpha-bias-cov}, \webautoref{tab:sim-study1-delta-bias-cov}, \webautoref{tab:sim-study1-lower-boundary-bias-cov}, and \webautoref{tab:sim-study1-upper-boundary-bias-cov}, and the comparison of arm means on the observed scale across working models in \webautoref{tab:sim-study1-observed-mean-bias-rmse-cov}.

\begingroup
\setlength{\tabcolsep}{4.5pt}
\fontsize{8pt}{9.0pt}\selectfont

\begin{longtable}{
    ll
    *{5}{S[table-format = 1.3]}
}
\caption{Simulation Study~1: exact scenario inputs on the observed scale. For each scenario and visit, the table reports the probabilities of the lower and upper bounds in the control arm, $\pi_{0t}^{L}$ and $\pi_{0t}^{U}$, the standardized treatment shift $\kappa_t$, visit-specific dropout hazard $h_t$, and the common AR(1) correlation parameter $\rho$ used to construct the latent truth values. Abbreviated scenario labels are used for space: Sparse bdy = \textit{SparseBoundary}, Mod.\ floor = \textit{ModerateFloor}, Agg.\ ceiling = \textit{AggressiveCeiling}, and Two-sided bds = \textit{AggressiveBothBounds}.}
\label{tab:sim-study1-design} \\
\toprule
\textbf{Scenario} & \textbf{Input} &
\multicolumn{1}{c}{\textbf{Visit 1}} &
\multicolumn{1}{c}{\textbf{Visit 2}} &
\multicolumn{1}{c}{\textbf{Visit 3}} &
\multicolumn{1}{c}{\textbf{Visit 4}} &
\multicolumn{1}{c}{\textbf{Visit 5}} \\
\midrule
\endfirsthead

\caption[]{Simulation Study~1: exact scenario inputs on the observed scale (continued).} \\
\toprule
\textbf{Scenario} & \textbf{Input} &
\multicolumn{1}{c}{\textbf{Visit 1}} &
\multicolumn{1}{c}{\textbf{Visit 2}} &
\multicolumn{1}{c}{\textbf{Visit 3}} &
\multicolumn{1}{c}{\textbf{Visit 4}} &
\multicolumn{1}{c}{\textbf{Visit 5}} \\
\midrule
\endhead

\midrule
\multicolumn{7}{r}{Continued on next page} \\
\endfoot

\bottomrule
\endlastfoot

Sparse bdy & $\pi_{0t}^{L}$ & 0.010 & 0.010 & 0.020 & 0.020 & 0.030 \\*
 & $\pi_{0t}^{U}$ & 0.030 & 0.020 & 0.020 & 0.010 & 0.010 \\*
 & $\kappa_t$ & 0.000 & -0.100 & -0.150 & -0.150 & -0.200 \\*
 & $h_t$ & 0.000 & 0.030 & 0.050 & 0.070 & 0.090 \\*
 & AR(1) $\rho$ & 0.600 & 0.600 & 0.600 & 0.600 & 0.600 \\

Mod.\ floor & $\pi_{0t}^{L}$ & 0.010 & 0.030 & 0.050 & 0.080 & 0.120 \\*
 & $\pi_{0t}^{U}$ & 0.080 & 0.050 & 0.030 & 0.020 & 0.010 \\*
 & $\kappa_t$ & 0.000 & -0.150 & -0.220 & -0.280 & -0.350 \\*
 & $h_t$ & 0.000 & 0.040 & 0.060 & 0.080 & 0.100 \\*
 & AR(1) $\rho$ & 0.700 & 0.700 & 0.700 & 0.700 & 0.700 \\

Agg.\ ceiling & $\pi_{0t}^{L}$ & 0.050 & 0.030 & 0.015 & 0.008 & 0.004 \\*
 & $\pi_{0t}^{U}$ & 0.020 & 0.050 & 0.100 & 0.180 & 0.280 \\*
 & $\kappa_t$ & 0.000 & 0.300 & 0.450 & 0.600 & 0.750 \\*
 & $h_t$ & 0.000 & 0.040 & 0.060 & 0.080 & 0.100 \\*
 & AR(1) $\rho$ & 0.750 & 0.750 & 0.750 & 0.750 & 0.750 \\

Two-sided bds & $\pi_{0t}^{L}$ & 0.080 & 0.100 & 0.120 & 0.140 & 0.160 \\*
 & $\pi_{0t}^{U}$ & 0.080 & 0.100 & 0.120 & 0.140 & 0.160 \\*
 & $\kappa_t$ & 0.000 & 0.080 & 0.100 & 0.080 & 0.060 \\*
 & $h_t$ & 0.000 & 0.040 & 0.060 & 0.080 & 0.100 \\*
 & AR(1) $\rho$ & 0.700 & 0.700 & 0.700 & 0.700 & 0.700 \\

\end{longtable}
\endgroup

\begingroup
\setlength{\tabcolsep}{4.5pt}
\fontsize{8pt}{9.6pt}\selectfont

\begin{longtable}{
    l
    *{5}{S[table-format = 1.3]}
}
\caption{Mean proportion of missing values by scenario and visit in Simulation Study~1. Abbreviated scenario labels are used for space: Sparse bdy = \textit{SparseBoundary}, Mod.\ floor = \textit{ModerateFloor}, Agg.\ ceiling = \textit{AggressiveCeiling}, and Two-sided bds = \textit{AggressiveBothBounds}.}
\label{tab:sim-primary-missingness} \\
\toprule
\textbf{Scenario} & \multicolumn{1}{c}{\textbf{Visit 1}} & \multicolumn{1}{c}{\textbf{Visit 2}} & \multicolumn{1}{c}{\textbf{Visit 3}} & \multicolumn{1}{c}{\textbf{Visit 4}} & \multicolumn{1}{c}{\textbf{Visit 5}} \\
\midrule
\endfirsthead

\caption[]{Mean proportion of missing values by scenario and visit in Simulation Study~1 (continued).} \\
\toprule
\textbf{Scenario} & \multicolumn{1}{c}{\textbf{Visit 1}} & \multicolumn{1}{c}{\textbf{Visit 2}} & \multicolumn{1}{c}{\textbf{Visit 3}} & \multicolumn{1}{c}{\textbf{Visit 4}} & \multicolumn{1}{c}{\textbf{Visit 5}} \\
\midrule
\endhead

\midrule
\multicolumn{6}{r}{Continued on next page} \\
\endfoot

\bottomrule
\endlastfoot

Sparse bdy & 0.000 & 0.030 & 0.078 & 0.143 & 0.221 \\* 
  Mod. floor & 0.000 & 0.039 & 0.094 & 0.164 & 0.248 \\* 
  Agg. ceiling & 0.000 & 0.039 & 0.094 & 0.170 & 0.252 \\* 
  Two-sided bds & 0.000 & 0.039 & 0.092 & 0.163 & 0.248 \\ 

\end{longtable}
\endgroup

\begin{landscape}

\begingroup
\setlength{\tabcolsep}{4.5pt}
\fontsize{8pt}{9.6pt}\selectfont

\begin{longtable}{
    l
    *{5}{S[table-format = -1.3] S[table-format = -1.3] S[table-format = 1.3]}
}
\caption{Simulation Study~1: recovery of the visit-specific latent intercepts $\alpha_t$ under the mixed boundary model. For each scenario and visit, the table reports the true value (True), mean bias (Bias), defined as the average posterior mean minus the true value, and empirical coverage (Cov) of the nominal $95\%$ credible interval. Abbreviated scenario labels are used for space: Sparse bdy = \textit{SparseBoundary}, Mod.\ floor = \textit{ModerateFloor}, Agg.\ ceiling = \textit{AggressiveCeiling}, and Two-sided bds = \textit{AggressiveBothBounds}.}
\label{tab:sim-study1-alpha-bias-cov} \\
\toprule
& \multicolumn{3}{c}{\textbf{Visit 1}} & \multicolumn{3}{c}{\textbf{Visit 2}} & \multicolumn{3}{c}{\textbf{Visit 3}} & \multicolumn{3}{c}{\textbf{Visit 4}} & \multicolumn{3}{c}{\textbf{Visit 5}} \\
\cmidrule(lr){2-4}
\cmidrule(lr){5-7}
\cmidrule(lr){8-10}
\cmidrule(lr){11-13}
\cmidrule(lr){14-16}
\textbf{Scenario} &
\multicolumn{1}{c}{\textbf{True}} & \multicolumn{1}{c}{\textbf{Bias}} & \multicolumn{1}{c}{\textbf{Cov}} &
\multicolumn{1}{c}{\textbf{True}} & \multicolumn{1}{c}{\textbf{Bias}} & \multicolumn{1}{c}{\textbf{Cov}} &
\multicolumn{1}{c}{\textbf{True}} & \multicolumn{1}{c}{\textbf{Bias}} & \multicolumn{1}{c}{\textbf{Cov}} &
\multicolumn{1}{c}{\textbf{True}} & \multicolumn{1}{c}{\textbf{Bias}} & \multicolumn{1}{c}{\textbf{Cov}} &
\multicolumn{1}{c}{\textbf{True}} & \multicolumn{1}{c}{\textbf{Bias}} & \multicolumn{1}{c}{\textbf{Cov}} \\
\midrule
\endfirsthead

\caption[]{Simulation Study~1: recovery of the visit-specific latent intercepts $\alpha_t$ under the mixed boundary model (continued).} \\
\toprule
& \multicolumn{3}{c}{\textbf{Visit 1}} & \multicolumn{3}{c}{\textbf{Visit 2}} & \multicolumn{3}{c}{\textbf{Visit 3}} & \multicolumn{3}{c}{\textbf{Visit 4}} & \multicolumn{3}{c}{\textbf{Visit 5}} \\
\cmidrule(lr){2-4}
\cmidrule(lr){5-7}
\cmidrule(lr){8-10}
\cmidrule(lr){11-13}
\cmidrule(lr){14-16}
\textbf{Scenario} &
\multicolumn{1}{c}{\textbf{True}} & \multicolumn{1}{c}{\textbf{Bias}} & \multicolumn{1}{c}{\textbf{Cov}} &
\multicolumn{1}{c}{\textbf{True}} & \multicolumn{1}{c}{\textbf{Bias}} & \multicolumn{1}{c}{\textbf{Cov}} &
\multicolumn{1}{c}{\textbf{True}} & \multicolumn{1}{c}{\textbf{Bias}} & \multicolumn{1}{c}{\textbf{Cov}} &
\multicolumn{1}{c}{\textbf{True}} & \multicolumn{1}{c}{\textbf{Bias}} & \multicolumn{1}{c}{\textbf{Cov}} &
\multicolumn{1}{c}{\textbf{True}} & \multicolumn{1}{c}{\textbf{Bias}} & \multicolumn{1}{c}{\textbf{Cov}} \\
\midrule
\endhead

\midrule
\multicolumn{16}{r}{Continued on next page} \\
\endfoot

\bottomrule
\endlastfoot

Sparse bdy & 0.256 & -0.004 & 0.956 & 0.150 & 0.003 & 0.956 & 0.000 & -0.012 & 0.926 & -0.150 & -0.026 & 0.933 & -0.256 & -0.018 & 0.940 \\* 
  Mod. floor & 0.597 & 0.004 & 0.963 & 0.162 & 0.007 & 0.953 & -0.162 & 0.018 & 0.953 & -0.453 & -0.013 & 0.953 & -0.795 & -0.013 & 0.953 \\* 
  Agg. ceiling & -0.267 & -0.001 & 0.960 & 0.162 & -0.005 & 0.943 & 0.622 & 0.010 & 0.946 & 1.086 & 0.037 & 0.936 & 1.546 & 0.097 & 0.943 \\* 
  Two-sided bds & 0.000 & 0.004 & 0.940 & 0.000 & -0.004 & 0.950 & 0.000 & 0.020 & 0.950 & 0.000 & 0.028 & 0.950 & 0.000 & 0.005 & 0.963 \\ 

\end{longtable}
\endgroup

\begingroup
\setlength{\tabcolsep}{4.5pt}
\fontsize{8pt}{9.6pt}\selectfont

\begin{longtable}{
    l
    *{5}{S[table-format = -1.3] S[table-format = -1.3] S[table-format = 1.3]}
}
\caption{Simulation Study~1: recovery of the visit-specific latent treatment effects $\delta_t$ under the mixed boundary model. For each scenario and visit, the table reports the true value (True), mean bias (Bias), defined as the average posterior mean minus the true value, and empirical coverage (Cov) of the nominal $95\%$ credible interval. Abbreviated scenario labels are used for space: Sparse bdy = \textit{SparseBoundary}, Mod.\ floor = \textit{ModerateFloor}, Agg.\ ceiling = \textit{AggressiveCeiling}, and Two-sided bds = \textit{AggressiveBothBounds}.}
\label{tab:sim-study1-delta-bias-cov} \\
\toprule
& \multicolumn{3}{c}{\textbf{Visit 1}} & \multicolumn{3}{c}{\textbf{Visit 2}} & \multicolumn{3}{c}{\textbf{Visit 3}} & \multicolumn{3}{c}{\textbf{Visit 4}} & \multicolumn{3}{c}{\textbf{Visit 5}} \\
\cmidrule(lr){2-4}
\cmidrule(lr){5-7}
\cmidrule(lr){8-10}
\cmidrule(lr){11-13}
\cmidrule(lr){14-16}
\textbf{Scenario} &
\multicolumn{1}{c}{\textbf{True}} & \multicolumn{1}{c}{\textbf{Bias}} & \multicolumn{1}{c}{\textbf{Cov}} &
\multicolumn{1}{c}{\textbf{True}} & \multicolumn{1}{c}{\textbf{Bias}} & \multicolumn{1}{c}{\textbf{Cov}} &
\multicolumn{1}{c}{\textbf{True}} & \multicolumn{1}{c}{\textbf{Bias}} & \multicolumn{1}{c}{\textbf{Cov}} &
\multicolumn{1}{c}{\textbf{True}} & \multicolumn{1}{c}{\textbf{Bias}} & \multicolumn{1}{c}{\textbf{Cov}} &
\multicolumn{1}{c}{\textbf{True}} & \multicolumn{1}{c}{\textbf{Bias}} & \multicolumn{1}{c}{\textbf{Cov}} \\
\midrule
\endfirsthead

\caption[]{Simulation Study~1: recovery of the visit-specific latent treatment effects $\delta_t$ under the mixed boundary model (continued).} \\
\toprule
& \multicolumn{3}{c}{\textbf{Visit 1}} & \multicolumn{3}{c}{\textbf{Visit 2}} & \multicolumn{3}{c}{\textbf{Visit 3}} & \multicolumn{3}{c}{\textbf{Visit 4}} & \multicolumn{3}{c}{\textbf{Visit 5}} \\
\cmidrule(lr){2-4}
\cmidrule(lr){5-7}
\cmidrule(lr){8-10}
\cmidrule(lr){11-13}
\cmidrule(lr){14-16}
\textbf{Scenario} &
\multicolumn{1}{c}{\textbf{True}} & \multicolumn{1}{c}{\textbf{Bias}} & \multicolumn{1}{c}{\textbf{Cov}} &
\multicolumn{1}{c}{\textbf{True}} & \multicolumn{1}{c}{\textbf{Bias}} & \multicolumn{1}{c}{\textbf{Cov}} &
\multicolumn{1}{c}{\textbf{True}} & \multicolumn{1}{c}{\textbf{Bias}} & \multicolumn{1}{c}{\textbf{Cov}} &
\multicolumn{1}{c}{\textbf{True}} & \multicolumn{1}{c}{\textbf{Bias}} & \multicolumn{1}{c}{\textbf{Cov}} &
\multicolumn{1}{c}{\textbf{True}} & \multicolumn{1}{c}{\textbf{Bias}} & \multicolumn{1}{c}{\textbf{Cov}} \\
\midrule
\endhead

\midrule
\multicolumn{16}{r}{Continued on next page} \\
\endfoot

\bottomrule
\endlastfoot

Sparse bdy & 0.000 & -0.006 & 0.963 & -0.110 & -0.013 & 0.973 & -0.177 & 0.009 & 0.970 & -0.166 & 0.024 & 0.960 & -0.230 & 0.010 & 0.953 \\* 
  Mod. floor & 0.000 & 0.001 & 0.949 & -0.206 & -0.015 & 0.949 & -0.302 & -0.025 & 0.956 & -0.391 & 0.021 & 0.973 & -0.483 & -0.013 & 0.963 \\* 
  Agg. ceiling & 0.000 & 0.014 & 0.963 & 0.411 & 0.037 & 0.956 & 0.630 & 0.022 & 0.943 & 0.873 & 0.027 & 0.943 & 1.121 & 0.077 & 0.953 \\* 
  Two-sided bds & 0.000 & -0.030 & 0.973 & 0.151 & -0.006 & 0.960 & 0.206 & -0.020 & 0.957 & 0.179 & -0.049 & 0.960 & 0.146 & 0.013 & 0.967 \\ 

\end{longtable}
\endgroup

\begingroup
\setlength{\tabcolsep}{4.5pt}
\fontsize{8pt}{9.6pt}\selectfont

\begin{longtable}{
    ll
    *{5}{S[table-format = 1.3] S[table-format = -1.3] S[table-format = 1.3]}
}
\caption{Simulation Study~1: recovery of the probabilities of the lower bound under the mixed boundary model. For each scenario, treatment arm, and visit, the table reports the true value (True), mean bias (Bias), defined as the average posterior mean minus the true value, and empirical coverage (Cov) of the nominal $95\%$ credible interval. Abbreviated scenario labels are used for space: Sparse bdy = \textit{SparseBoundary}, Mod.\ floor = \textit{ModerateFloor}, Agg.\ ceiling = \textit{AggressiveCeiling}, and Two-sided bds = \textit{AggressiveBothBounds}. Treatment abbreviations: Ref = reference and Act = active.}
\label{tab:sim-study1-lower-boundary-bias-cov} \\
\toprule
& & \multicolumn{3}{c}{\textbf{Visit 1}} & \multicolumn{3}{c}{\textbf{Visit 2}} & \multicolumn{3}{c}{\textbf{Visit 3}} & \multicolumn{3}{c}{\textbf{Visit 4}} & \multicolumn{3}{c}{\textbf{Visit 5}} \\
\cmidrule(lr){3-5}
\cmidrule(lr){6-8}
\cmidrule(lr){9-11}
\cmidrule(lr){12-14}
\cmidrule(lr){15-17}
\textbf{Scenario} & \textbf{Treatment} &
\multicolumn{1}{c}{\textbf{True}} & \multicolumn{1}{c}{\textbf{Bias}} & \multicolumn{1}{c}{\textbf{Cov}} &
\multicolumn{1}{c}{\textbf{True}} & \multicolumn{1}{c}{\textbf{Bias}} & \multicolumn{1}{c}{\textbf{Cov}} &
\multicolumn{1}{c}{\textbf{True}} & \multicolumn{1}{c}{\textbf{Bias}} & \multicolumn{1}{c}{\textbf{Cov}} &
\multicolumn{1}{c}{\textbf{True}} & \multicolumn{1}{c}{\textbf{Bias}} & \multicolumn{1}{c}{\textbf{Cov}} &
\multicolumn{1}{c}{\textbf{True}} & \multicolumn{1}{c}{\textbf{Bias}} & \multicolumn{1}{c}{\textbf{Cov}} \\
\midrule
\endfirsthead

\caption[]{Simulation Study~1: recovery of the probabilities of the lower bound under the mixed boundary model (continued).} \\
\toprule
& & \multicolumn{3}{c}{\textbf{Visit 1}} & \multicolumn{3}{c}{\textbf{Visit 2}} & \multicolumn{3}{c}{\textbf{Visit 3}} & \multicolumn{3}{c}{\textbf{Visit 4}} & \multicolumn{3}{c}{\textbf{Visit 5}} \\
\cmidrule(lr){3-5}
\cmidrule(lr){6-8}
\cmidrule(lr){9-11}
\cmidrule(lr){12-14}
\cmidrule(lr){15-17}
\textbf{Scenario} & \textbf{Treatment} &
\multicolumn{1}{c}{\textbf{True}} & \multicolumn{1}{c}{\textbf{Bias}} & \multicolumn{1}{c}{\textbf{Cov}} &
\multicolumn{1}{c}{\textbf{True}} & \multicolumn{1}{c}{\textbf{Bias}} & \multicolumn{1}{c}{\textbf{Cov}} &
\multicolumn{1}{c}{\textbf{True}} & \multicolumn{1}{c}{\textbf{Bias}} & \multicolumn{1}{c}{\textbf{Cov}} &
\multicolumn{1}{c}{\textbf{True}} & \multicolumn{1}{c}{\textbf{Bias}} & \multicolumn{1}{c}{\textbf{Cov}} &
\multicolumn{1}{c}{\textbf{True}} & \multicolumn{1}{c}{\textbf{Bias}} & \multicolumn{1}{c}{\textbf{Cov}} \\
\midrule
\endhead

\midrule
\multicolumn{17}{r}{Continued on next page} \\
\endfoot

\bottomrule
\endlastfoot

Sparse bdy & Ref & 0.010 & 0.008 & 0.950 & 0.010 & 0.007 & 0.960 & 0.020 & 0.009 & 0.960 & 0.020 & 0.011 & 0.943 & 0.030 & 0.017 & 0.946 \\* 
   & Act & 0.010 & 0.008 & 0.946 & 0.013 & 0.008 & 0.963 & 0.028 & 0.010 & 0.966 & 0.028 & 0.011 & 0.950 & 0.046 & 0.019 & 0.936 \\* 
  Mod. floor & Ref & 0.010 & 0.007 & 0.939 & 0.030 & 0.009 & 0.960 & 0.050 & 0.011 & 0.960 & 0.080 & 0.016 & 0.949 & 0.120 & 0.026 & 0.946 \\* 
   & Act & 0.010 & 0.007 & 0.912 & 0.042 & 0.011 & 0.929 & 0.077 & 0.014 & 0.960 & 0.130 & 0.013 & 0.963 & 0.205 & 0.025 & 0.949 \\* 
  Agg. ceiling & Ref & 0.050 & 0.011 & 0.960 & 0.030 & 0.007 & 0.916 & 0.015 & 0.006 & 0.960 & 0.008 & 0.007 & 0.929 & 0.004 & 0.009 & 0.936 \\* 
   & Act & 0.050 & 0.010 & 0.973 & 0.015 & 0.005 & 0.929 & 0.004 & 0.003 & 0.970 & 0.001 & 0.003 & 0.929 & 0.000 & 0.002 & 0.956 \\* 
  Two-sided bds & Ref & 0.080 & 0.017 & 0.960 & 0.100 & 0.017 & 0.943 & 0.120 & 0.012 & 0.950 & 0.140 & 0.014 & 0.957 & 0.160 & 0.025 & 0.953 \\* 
   & Act & 0.080 & 0.019 & 0.960 & 0.087 & 0.018 & 0.950 & 0.101 & 0.014 & 0.960 & 0.123 & 0.019 & 0.960 & 0.146 & 0.025 & 0.950 \\ 

\end{longtable}
\endgroup

\begingroup
\setlength{\tabcolsep}{4.5pt}
\fontsize{8pt}{9.6pt}\selectfont

\begin{longtable}{
    ll
    *{5}{S[table-format = 1.3] S[table-format = -1.3] S[table-format = 1.3]}
}
\caption{Simulation Study~1: recovery of the probabilities of the upper bound under the mixed boundary model. For each scenario, treatment arm, and visit, the table reports the true value (True), mean bias (Bias), defined as the average posterior mean minus the true value, and empirical coverage (Cov) of the nominal $95\%$ credible interval. Abbreviated scenario labels are used for space: Sparse bdy = \textit{SparseBoundary}, Mod.\ floor = \textit{ModerateFloor}, Agg.\ ceiling = \textit{AggressiveCeiling}, and Two-sided bds = \textit{AggressiveBothBounds}. Treatment abbreviations: Ref = reference and Act = active.}
\label{tab:sim-study1-upper-boundary-bias-cov} \\
\toprule
& & \multicolumn{3}{c}{\textbf{Visit 1}} & \multicolumn{3}{c}{\textbf{Visit 2}} & \multicolumn{3}{c}{\textbf{Visit 3}} & \multicolumn{3}{c}{\textbf{Visit 4}} & \multicolumn{3}{c}{\textbf{Visit 5}} \\
\cmidrule(lr){3-5}
\cmidrule(lr){6-8}
\cmidrule(lr){9-11}
\cmidrule(lr){12-14}
\cmidrule(lr){15-17}
\textbf{Scenario} & \textbf{Treatment} &
\multicolumn{1}{c}{\textbf{True}} & \multicolumn{1}{c}{\textbf{Bias}} & \multicolumn{1}{c}{\textbf{Cov}} &
\multicolumn{1}{c}{\textbf{True}} & \multicolumn{1}{c}{\textbf{Bias}} & \multicolumn{1}{c}{\textbf{Cov}} &
\multicolumn{1}{c}{\textbf{True}} & \multicolumn{1}{c}{\textbf{Bias}} & \multicolumn{1}{c}{\textbf{Cov}} &
\multicolumn{1}{c}{\textbf{True}} & \multicolumn{1}{c}{\textbf{Bias}} & \multicolumn{1}{c}{\textbf{Cov}} &
\multicolumn{1}{c}{\textbf{True}} & \multicolumn{1}{c}{\textbf{Bias}} & \multicolumn{1}{c}{\textbf{Cov}} \\
\midrule
\endfirsthead

\caption[]{Simulation Study~1: recovery of the probabilities of the upper bound under the mixed boundary model (continued).} \\
\toprule
& & \multicolumn{3}{c}{\textbf{Visit 1}} & \multicolumn{3}{c}{\textbf{Visit 2}} & \multicolumn{3}{c}{\textbf{Visit 3}} & \multicolumn{3}{c}{\textbf{Visit 4}} & \multicolumn{3}{c}{\textbf{Visit 5}} \\
\cmidrule(lr){3-5}
\cmidrule(lr){6-8}
\cmidrule(lr){9-11}
\cmidrule(lr){12-14}
\cmidrule(lr){15-17}
\textbf{Scenario} & \textbf{Treatment} &
\multicolumn{1}{c}{\textbf{True}} & \multicolumn{1}{c}{\textbf{Bias}} & \multicolumn{1}{c}{\textbf{Cov}} &
\multicolumn{1}{c}{\textbf{True}} & \multicolumn{1}{c}{\textbf{Bias}} & \multicolumn{1}{c}{\textbf{Cov}} &
\multicolumn{1}{c}{\textbf{True}} & \multicolumn{1}{c}{\textbf{Bias}} & \multicolumn{1}{c}{\textbf{Cov}} &
\multicolumn{1}{c}{\textbf{True}} & \multicolumn{1}{c}{\textbf{Bias}} & \multicolumn{1}{c}{\textbf{Cov}} &
\multicolumn{1}{c}{\textbf{True}} & \multicolumn{1}{c}{\textbf{Bias}} & \multicolumn{1}{c}{\textbf{Cov}} \\
\midrule
\endhead

\midrule
\multicolumn{17}{r}{Continued on next page} \\
\endfoot

\bottomrule
\endlastfoot

Sparse bdy & Ref & 0.030 & 0.013 & 0.956 & 0.020 & 0.009 & 0.953 & 0.020 & 0.009 & 0.950 & 0.010 & 0.007 & 0.956 & 0.010 & 0.010 & 0.953 \\* 
   & Act & 0.030 & 0.013 & 0.956 & 0.016 & 0.008 & 0.950 & 0.014 & 0.007 & 0.956 & 0.007 & 0.006 & 0.946 & 0.006 & 0.007 & 0.963 \\* 
  Mod. floor & Ref & 0.080 & 0.014 & 0.956 & 0.050 & 0.012 & 0.939 & 0.030 & 0.011 & 0.939 & 0.020 & 0.010 & 0.956 & 0.010 & 0.012 & 0.953 \\* 
   & Act & 0.080 & 0.015 & 0.946 & 0.036 & 0.009 & 0.936 & 0.018 & 0.008 & 0.939 & 0.010 & 0.008 & 0.956 & 0.004 & 0.007 & 0.943 \\* 
  Agg. ceiling & Ref & 0.020 & 0.008 & 0.946 & 0.050 & 0.008 & 0.929 & 0.100 & 0.010 & 0.939 & 0.180 & 0.019 & 0.929 & 0.280 & 0.032 & 0.939 \\* 
   & Act & 0.020 & 0.009 & 0.943 & 0.089 & 0.011 & 0.943 & 0.203 & 0.013 & 0.923 & 0.376 & 0.017 & 0.936 & 0.566 & 0.018 & 0.943 \\* 
  Two-sided bds & Ref & 0.080 & 0.018 & 0.950 & 0.100 & 0.017 & 0.930 & 0.120 & 0.016 & 0.933 & 0.140 & 0.020 & 0.943 & 0.160 & 0.026 & 0.943 \\* 
   & Act & 0.080 & 0.014 & 0.963 & 0.115 & 0.016 & 0.936 & 0.141 & 0.013 & 0.950 & 0.159 & 0.014 & 0.953 & 0.175 & 0.027 & 0.950 \\ 

\end{longtable}
\endgroup

\begingroup
\setlength{\tabcolsep}{2.0pt}
\fontsize{8pt}{9.5pt}\selectfont

\begin{longtable}{
    l l l
    *{5}{
        S[table-format = 2.3]
        S[table-format = -1.3]
        S[table-format = 1.3]
        S[table-format = 1.3]
    }
}
\caption{Simulation Study~1: recovery of means on the observed scale across working models. For each scenario, model, treatment arm, and visit, the table reports the true value (True), mean bias (Bias), root mean squared error (RMSE), and empirical coverage (Cov) of the nominal $95\%$ credible interval. Abbreviated scenario labels are used for space: Sparse bdy = \textit{SparseBoundary}, Mod.\ floor = \textit{ModerateFloor}, Agg.\ ceiling = \textit{AggressiveCeiling}, and Two-sided bds = \textit{AggressiveBothBounds}. Model abbreviations: Gauss. = Gaussian, CB = continuous bounded, and MB = mixed boundary. Treatment abbreviations: Ref = reference and Act = active.}
\label{tab:sim-study1-observed-mean-bias-rmse-cov} \\
\toprule
& & &
\multicolumn{4}{c}{\textbf{Visit 1}} &
\multicolumn{4}{c}{\textbf{Visit 2}} &
\multicolumn{4}{c}{\textbf{Visit 3}} &
\multicolumn{4}{c}{\textbf{Visit 4}} &
\multicolumn{4}{c}{\textbf{Visit 5}} \\
\cmidrule(lr){4-7}
\cmidrule(lr){8-11}
\cmidrule(lr){12-15}
\cmidrule(lr){16-19}
\cmidrule(lr){20-23}
\textbf{Scenario} & \textbf{Model} & \textbf{Treatment} &
\multicolumn{1}{c}{\textbf{True}} &
\multicolumn{1}{c}{\textbf{Bias}} &
\multicolumn{1}{c}{\textbf{RMSE}} &
\multicolumn{1}{c}{\textbf{Cov}} &
\multicolumn{1}{c}{\textbf{True}} &
\multicolumn{1}{c}{\textbf{Bias}} &
\multicolumn{1}{c}{\textbf{RMSE}} &
\multicolumn{1}{c}{\textbf{Cov}} &
\multicolumn{1}{c}{\textbf{True}} &
\multicolumn{1}{c}{\textbf{Bias}} &
\multicolumn{1}{c}{\textbf{RMSE}} &
\multicolumn{1}{c}{\textbf{Cov}} &
\multicolumn{1}{c}{\textbf{True}} &
\multicolumn{1}{c}{\textbf{Bias}} &
\multicolumn{1}{c}{\textbf{RMSE}} &
\multicolumn{1}{c}{\textbf{Cov}} &
\multicolumn{1}{c}{\textbf{True}} &
\multicolumn{1}{c}{\textbf{Bias}} &
\multicolumn{1}{c}{\textbf{RMSE}} &
\multicolumn{1}{c}{\textbf{Cov}} \\
\midrule
\endfirsthead

\caption[]{Simulation Study~1: recovery of means on the observed scale across working models (continued).} \\
\toprule
& & &
\multicolumn{4}{c}{\textbf{Visit 1}} &
\multicolumn{4}{c}{\textbf{Visit 2}} &
\multicolumn{4}{c}{\textbf{Visit 3}} &
\multicolumn{4}{c}{\textbf{Visit 4}} &
\multicolumn{4}{c}{\textbf{Visit 5}} \\
\cmidrule(lr){4-7}
\cmidrule(lr){8-11}
\cmidrule(lr){12-15}
\cmidrule(lr){16-19}
\cmidrule(lr){20-23}
\textbf{Scenario} & \textbf{Model} & \textbf{Treatment} &
\multicolumn{1}{c}{\textbf{True}} &
\multicolumn{1}{c}{\textbf{Bias}} &
\multicolumn{1}{c}{\textbf{RMSE}} &
\multicolumn{1}{c}{\textbf{Cov}} &
\multicolumn{1}{c}{\textbf{True}} &
\multicolumn{1}{c}{\textbf{Bias}} &
\multicolumn{1}{c}{\textbf{RMSE}} &
\multicolumn{1}{c}{\textbf{Cov}} &
\multicolumn{1}{c}{\textbf{True}} &
\multicolumn{1}{c}{\textbf{Bias}} &
\multicolumn{1}{c}{\textbf{RMSE}} &
\multicolumn{1}{c}{\textbf{Cov}} &
\multicolumn{1}{c}{\textbf{True}} &
\multicolumn{1}{c}{\textbf{Bias}} &
\multicolumn{1}{c}{\textbf{RMSE}} &
\multicolumn{1}{c}{\textbf{Cov}} &
\multicolumn{1}{c}{\textbf{True}} &
\multicolumn{1}{c}{\textbf{Bias}} &
\multicolumn{1}{c}{\textbf{RMSE}} &
\multicolumn{1}{c}{\textbf{Cov}} \\
\midrule
\endhead

\midrule
\multicolumn{23}{r}{Continued on next page} \\
\endfoot

\bottomrule
\endlastfoot

Sparse bdy & Gauss. & Ref & 35.765 & -0.324 & 3.732 & 0.963 & 34.072 & 0.014 & 3.912 & 0.946 & 31.500 & -0.366 & 4.298 & 0.930 & 28.928 & -0.643 & 4.428 & 0.916 & 27.235 & -0.331 & 4.629 & 0.933 \\* 
   &  & Act & 35.765 & -0.226 & 3.919 & 0.967 & 32.186 & -0.204 & 3.641 & 0.960 & 28.592 & -0.090 & 3.796 & 0.970 & 26.129 & 0.082 & 4.128 & 0.950 & 23.505 & 0.043 & 4.173 & 0.963 \\* 
   & CB & Ref & 35.765 & -0.279 & 3.532 & 0.957 & 34.072 & -0.140 & 3.631 & 0.950 & 31.500 & -0.228 & 4.019 & 0.926 & 28.928 & -0.316 & 4.142 & 0.936 & 27.235 & -0.015 & 4.301 & 0.936 \\* 
   &  & Act & 35.765 & -0.387 & 3.655 & 0.957 & 32.186 & -0.158 & 3.428 & 0.977 & 28.592 & 0.053 & 3.578 & 0.970 & 26.129 & 0.264 & 3.830 & 0.950 & 23.505 & 0.430 & 3.806 & 0.963 \\* 
   & MB & Ref & 35.765 & -0.255 & 3.566 & 0.953 & 34.072 & -0.037 & 3.655 & 0.960 & 31.500 & -0.222 & 4.036 & 0.926 & 28.928 & -0.329 & 4.177 & 0.936 & 27.235 & -0.085 & 4.302 & 0.940 \\* 
   &  & Act & 35.765 & -0.363 & 3.657 & 0.956 & 32.186 & -0.177 & 3.433 & 0.970 & 28.592 & 0.034 & 3.589 & 0.970 & 26.129 & 0.150 & 3.859 & 0.946 & 23.505 & 0.288 & 3.845 & 0.963 \\ 
  Mod. floor & Gauss. & Ref & 40.585 & -0.050 & 4.067 & 0.937 & 33.923 & -0.050 & 4.238 & 0.947 & 29.077 & 0.088 & 4.464 & 0.940 & 24.863 & -0.184 & 4.240 & 0.953 & 20.051 & -0.011 & 4.336 & 0.933 \\* 
   &  & Act & 40.585 & 0.030 & 4.044 & 0.943 & 30.841 & -0.165 & 4.497 & 0.913 & 24.634 & 0.023 & 4.251 & 0.957 & 19.481 & 0.351 & 4.119 & 0.957 & 14.083 & -0.052 & 3.756 & 0.970 \\* 
   & CB & Ref & 40.585 & -0.385 & 3.733 & 0.957 & 33.923 & -0.075 & 3.967 & 0.953 & 29.077 & 0.425 & 4.126 & 0.936 & 24.863 & 0.264 & 3.989 & 0.950 & 20.051 & 0.575 & 4.020 & 0.946 \\* 
   &  & Act & 40.585 & -0.428 & 3.825 & 0.936 & 30.841 & -0.099 & 4.148 & 0.933 & 24.634 & 0.353 & 3.972 & 0.957 & 19.481 & 0.956 & 3.753 & 0.960 & 14.083 & 1.138 & 3.636 & 0.950 \\* 
   & MB & Ref & 40.585 & -0.265 & 3.777 & 0.963 & 33.923 & -0.002 & 4.016 & 0.953 & 29.077 & 0.341 & 4.166 & 0.949 & 24.863 & 0.103 & 4.101 & 0.956 & 20.051 & 0.588 & 3.967 & 0.966 \\* 
   &  & Act & 40.585 & -0.252 & 3.846 & 0.943 & 30.841 & -0.091 & 4.239 & 0.926 & 24.634 & 0.128 & 4.063 & 0.956 & 19.481 & 0.565 & 3.712 & 0.960 & 14.083 & 0.757 & 3.587 & 0.966 \\ 
  Agg. ceiling & Gauss. & Ref & 27.382 & 0.054 & 3.887 & 0.963 & 33.923 & -0.208 & 4.248 & 0.947 & 40.505 & -0.046 & 4.348 & 0.897 & 46.211 & 0.129 & 4.190 & 0.887 & 50.918 & 0.113 & 3.869 & 0.830 \\* 
   &  & Act & 27.382 & 0.161 & 4.284 & 0.950 & 39.936 & 0.443 & 4.059 & 0.947 & 48.492 & 0.154 & 3.639 & 0.950 & 54.804 & -0.049 & 2.889 & 0.977 & 58.844 & -0.066 & 2.303 & 0.973 \\* 
   & CB & Ref & 27.382 & 0.231 & 3.619 & 0.963 & 33.923 & -0.161 & 3.949 & 0.950 & 40.505 & -0.159 & 4.080 & 0.919 & 46.211 & 0.309 & 3.990 & 0.916 & 50.918 & 0.700 & 3.730 & 0.826 \\* 
   &  & Act & 27.382 & 0.447 & 3.851 & 0.950 & 39.936 & -0.061 & 3.628 & 0.953 & 48.492 & -0.632 & 3.294 & 0.936 & 54.804 & -1.222 & 2.908 & 0.926 & 58.844 & -1.427 & 2.484 & 0.879 \\* 
   & MB & Ref & 27.382 & 0.102 & 3.698 & 0.963 & 33.923 & -0.131 & 4.004 & 0.943 & 40.505 & -0.079 & 3.957 & 0.933 & 46.211 & -0.193 & 3.827 & 0.936 & 50.918 & -0.552 & 3.498 & 0.960 \\* 
   &  & Act & 27.382 & 0.311 & 3.852 & 0.956 & 39.936 & 0.282 & 3.727 & 0.953 & 48.492 & -0.032 & 3.331 & 0.956 & 54.804 & -0.354 & 2.767 & 0.929 & 58.844 & -0.653 & 2.194 & 0.946 \\ 
  Two-sided bds & Gauss. & Ref & 31.500 & -0.153 & 4.767 & 0.943 & 31.500 & -0.163 & 4.695 & 0.950 & 31.500 & -0.175 & 4.839 & 0.943 & 31.500 & 0.034 & 5.380 & 0.933 & 31.500 & -0.155 & 5.525 & 0.953 \\* 
   &  & Act & 31.500 & -0.198 & 4.338 & 0.950 & 33.292 & -0.114 & 4.876 & 0.937 & 33.777 & -0.016 & 5.060 & 0.930 & 33.348 & -0.370 & 5.580 & 0.937 & 32.903 & 0.116 & 5.721 & 0.967 \\* 
   & CB & Ref & 31.500 & 0.069 & 4.216 & 0.950 & 31.500 & -0.017 & 4.085 & 0.960 & 31.500 & 0.073 & 4.254 & 0.943 & 31.500 & 0.236 & 4.596 & 0.937 & 31.500 & 0.058 & 4.597 & 0.957 \\* 
   &  & Act & 31.500 & -0.208 & 3.766 & 0.970 & 33.292 & -0.286 & 4.227 & 0.940 & 33.777 & -0.293 & 4.254 & 0.953 & 33.348 & -0.528 & 4.668 & 0.957 & 32.903 & -0.118 & 4.738 & 0.967 \\* 
   & MB & Ref & 31.500 & 0.045 & 4.488 & 0.940 & 31.500 & -0.064 & 4.426 & 0.950 & 31.500 & 0.175 & 4.660 & 0.950 & 31.500 & 0.250 & 4.933 & 0.950 & 31.500 & 0.064 & 5.007 & 0.963 \\* 
   &  & Act & 31.500 & -0.285 & 3.903 & 0.970 & 33.292 & -0.232 & 4.399 & 0.953 & 33.777 & -0.125 & 4.530 & 0.946 & 33.348 & -0.339 & 4.944 & 0.953 & 32.903 & 0.026 & 5.083 & 0.946 \\ 

\end{longtable}
\endgroup
\end{landscape}

\subsection{Simulation Study~2: support templates, truth rules, and reported summaries}

Let $s \in \left\{\mathrm{MF}, \mathrm{ABB}\right\}$ index the support template, where $\mathrm{MF}$ denotes \textit{ModerateFloor} and $\mathrm{ABB}$ denotes \textit{AggressiveBothBounds}. Let $r \in \left\{\mathrm{MAR}, \mathrm{WJ2R}_{0.50}, \mathrm{J2R}, \mathrm{CR}\right\}$ index the truth rule for the active arm after the trigger. Each Study~2 scenario was therefore the pair $\left(s, r\right)$.

Support template $s$ was defined by the six inputs
\begin{equation}
\boldsymbol{\pi}_{0}^{L,\left(s\right)},
\qquad
\boldsymbol{\pi}_{0}^{U,\left(s\right)},
\qquad
\boldsymbol{\kappa}^{\left(s\right)},
\qquad
\rho^{\left(s\right)},
\qquad
\boldsymbol{\eta}^{\left(s\right)},
\qquad
\boldsymbol{\lambda}^{\left(s\right)},
\end{equation}
where $\boldsymbol{\pi}_{0}^{L,\left(s\right)} = \left(\pi_{01}^{L,\left(s\right)}, \dots, \pi_{0T}^{L,\left(s\right)}\right)^{\top}$ and $\boldsymbol{\pi}_{0}^{U,\left(s\right)} = \left(\pi_{01}^{U,\left(s\right)}, \dots, \pi_{0T}^{U,\left(s\right)}\right)^{\top}$ are the vectors of probabilities of the lower and upper bounds in the control arm, $\boldsymbol{\kappa}^{\left(s\right)} = \left(\kappa_1^{\left(s\right)}, \dots, \kappa_T^{\left(s\right)}\right)^{\top}$ is the vector of standardized treatment shifts, $\rho^{\left(s\right)}$ is the AR(1) correlation parameter, $\boldsymbol{\eta}^{\left(s\right)} = \left(\eta_1^{\left(s\right)}, \dots, \eta_T^{\left(s\right)}\right)^{\top}$ gives the probabilities of intermittent missingness before the final tail, and $\boldsymbol{\lambda}^{\left(s\right)} = \left(\lambda_1^{\left(s\right)}, \dots, \lambda_T^{\left(s\right)}\right)^{\top}$ gives the dropout hazards for the final missing tail.

The latent mean vectors and covariance matrix under template $s$ were obtained from the same construction as in Study~1, namely through $\delta_t^{\left(s\right)} = \kappa_t^{\left(s\right)} \sigma_t^{\left(s\right)}$ and $\Sigma^{\left(s\right)} = D^{\left(s\right)} R\left(\rho^{\left(s\right)}\right) D^{\left(s\right)}$. The support template, therefore, determined the boundary pattern, structure of treatment shifts, covariance structure, and missingness design, whereas the truth rule $r$ determined only the distribution of missing outcomes in the active arm after the trigger.

The \textit{ModerateFloor} template was intended to represent increasing floor accumulation with limited mass at the upper bound. The \textit{AggressiveBothBounds} template was intended to represent a more difficult two-sided boundary setting and therefore used higher dropout hazards for the final missing tail to increase the influence of the sensitivity rules. The exact template inputs, combinations of truth rules, and scenario labels are listed in \webautoref{tab:sim-rbmi-design}.

The missingness pattern was generated separately for each subject before generating the latent outcomes. We first selected an initial dropout visit sequentially, with $\lambda_t^{\left(s\right)}$ giving the probability of dropout at visit $t$ conditional on no earlier dropout. The initial dropout visit was set to $T + 1$ if no dropout occurred. All outcomes from the initial dropout visit onward were designated missing. Conditional on that visit, each earlier outcome was independently designated missing with probability $\eta_t^{\left(s\right)}$. The same probabilities were used in both treatment arms. After combining these two components, the sensitivity trigger was determined from the last observed visit. Consequently, additional missing values immediately preceding the initially generated dropout tail were included in the final missing tail and could move the trigger earlier. Additional missing values at the end of follow-up could also create a final missing tail when no initial dropout had occurred.

For subject $i$, let
\begin{equation}
\mathcal{O}_i^{\mathrm{obs}} = \left\{t \in \left\{1, \dots, T\right\} : R_{it} = 1\right\}
\end{equation}
denote the observed post-baseline visit set after imposing missingness. If $\mathcal{O}_i^{\mathrm{obs}} = \varnothing$, we set $\tau_i = 1$. Otherwise, let
\begin{equation}
t_i^{\mathrm{last}} = \max \mathcal{O}_i^{\mathrm{obs}},
\end{equation}
and define
\begin{equation}
\tau_i =
\begin{cases}
t_i^{\mathrm{last}} + 1, & \text{if } t_i^{\mathrm{last}} < T, \\
T + 1, & \text{if } t_i^{\mathrm{last}} = T.
\end{cases}
\end{equation}
Thus, missing values at visits $t < \tau_i$, before the trigger, were generated under MAR within each arm, whereas missing values at visits $t \ge \tau_i$ in the final monotone tail of the active arm followed the truth rule for that scenario. The resulting mean missingness proportions are reported in \webautoref{tab:sim-rbmi-missingness}.

Each dataset in Study~2 was analyzed only with the proposed mixed-boundary model. The fitted model was therefore held fixed across sensitivity analyses, and only the posterior predictive completion rule was varied. The analysis rules were MAR, weighted J2R with $\omega = 0.50$, J2R, and CR. The retained Study~2 summary reports the difference between the complete-data arm means on the observed scale at the final scheduled visit; see \webautoref{tab:sim-rbmi-final-visit-bias-rmse-cov}.

\begin{landscape}

\begingroup
\setlength{\tabcolsep}{6pt}
\fontsize{8pt}{9.6pt}\selectfont

\begin{longtable}{@{}p{5.5cm}p{\dimexpr\linewidth-5.5cm-2\tabcolsep\relax}@{}}
\caption{Simulation Study~2: inputs for the support templates and the truth rules used to generate the eight reference-based sensitivity analysis scenarios. In Study~2, each full scenario is defined by the combination of support template and truth rule after the trigger. The two support templates are \textit{ModerateFloor} and \textit{AggressiveBothBounds}; in subsequent tables, these are abbreviated as Mod.\ floor and Two-sided bds, respectively. Abbreviations: CR = copy reference, J2R = jump to reference, MAR = missing at random, and WJ2R(0.50) = weighted jump to reference with $\omega = 0.50$.}
\label{tab:sim-rbmi-design} \\
\toprule
\textbf{Component} & \textbf{Specification} \\
\midrule
\endfirsthead

\caption[]{Simulation Study~2: inputs for the support templates and the truth rules used to generate the eight reference-based sensitivity analysis scenarios (continued).} \\
\toprule
\textbf{Component} & \textbf{Specification} \\
\midrule
\endhead

\midrule
\multicolumn{2}{r}{Continued on next page} \\
\endfoot

\bottomrule
\endlastfoot

Support templates &
\textit{ModerateFloor} and \textit{AggressiveBothBounds} \\*

Probabilities of the lower bound in the control arm $\pi_{0t}^{L}$ &
\textit{ModerateFloor}: $\left(0.01, 0.03, 0.05, 0.08, 0.12\right)$; \newline
\textit{AggressiveBothBounds}: $\left(0.08, 0.10, 0.12, 0.14, 0.16\right)$ \\*

Probabilities of the upper bound in the control arm $\pi_{0t}^{U}$ &
\textit{ModerateFloor}: $\left(0.08, 0.05, 0.03, 0.02, 0.01\right)$; \newline
\textit{AggressiveBothBounds}: $\left(0.08, 0.10, 0.12, 0.14, 0.16\right)$ \\*

Visit-specific standardized treatment shifts $\kappa_t$ &
\textit{ModerateFloor}: $\left(0.00, -0.15, -0.22, -0.28, -0.35\right)$; \newline
\textit{AggressiveBothBounds}: $\left(0.00, 0.15, 0.20, 0.20, 0.18\right)$ \\*

Common AR(1) correlation parameter $\rho$ &
\textit{ModerateFloor}: $0.70$; \textit{AggressiveBothBounds}: $0.70$ \\*

Probabilities of intermittent missingness before the final tail $\eta_t$ &
Common to both templates: $\left(0.01, 0.03, 0.04, 0.05, 0.05\right)$ \\*

Dropout hazards for the final missing tail $\lambda_t$ &
\textit{ModerateFloor}: $\left(0.00, 0.02, 0.04, 0.06, 0.08\right)$; \newline
\textit{AggressiveBothBounds}: $\left(0.00, 0.04, 0.08, 0.12, 0.16\right)$ \\*

Truth rules for the active arm after the trigger &
MAR, WJ2R(0.50), J2R, and CR \\*

Weight for the truth rule &
For WJ2R(0.50), $\omega = 0.50$ \\*

Scenario names &
\textit{ModerateFloorMarTruth}, \textit{ModerateFloorJ2RTruth}, \textit{ModerateFloorCRTruth}, \textit{ModerateFloorWJ2R50Truth} \\*
& \textit{AggressiveBothBoundsMarTruth}, \textit{AggressiveBothBoundsJ2RTruth}, \textit{AggressiveBothBoundsCRTruth}, \textit{AggressiveBothBoundsWJ2R50Truth} \\*

Trigger definition &
If no post-baseline outcome remains observed, $\tau_i = 1$; otherwise, $\tau_i = t_i^{\mathrm{last}} + 1$ when $t_i^{\mathrm{last}} < T$, and $\tau_i = T + 1$ when $t_i^{\mathrm{last}} = T$ \\*

Total scenarios &
$2 \times 4 = 8$ \\

\end{longtable}
\endgroup

\begingroup
\setlength{\tabcolsep}{4.5pt}
\fontsize{8pt}{9.6pt}\selectfont

\begin{longtable}{
    ll
    *{5}{S[table-format = 1.3]}
}
\caption{Mean proportion of missing values by support template, truth rule, and visit in Simulation Study~2. Here, the Support column lists abbreviated labels for the support templates rather than full scenario names: Mod.\ floor = \textit{ModerateFloor} and Two-sided bds = \textit{AggressiveBothBounds}. Each full Study~2 scenario is defined by the combination of support template and truth rule. Abbreviations: WJ2R(0.50) = weighted jump to reference with $\omega = 0.50$.}
\label{tab:sim-rbmi-missingness} \\
\toprule
\textbf{Support} & \textbf{Truth} & \multicolumn{1}{c}{\textbf{V1}} & \multicolumn{1}{c}{\textbf{V2}} & \multicolumn{1}{c}{\textbf{V3}} & \multicolumn{1}{c}{\textbf{V4}} & \multicolumn{1}{c}{\textbf{V5}} \\
\midrule
\endfirsthead

\caption[]{Mean proportion of missing values by support template, truth rule, and visit in Simulation Study~2 (continued).} \\
\toprule
\textbf{Support} & \textbf{Truth} & \multicolumn{1}{c}{\textbf{V1}} & \multicolumn{1}{c}{\textbf{V2}} & \multicolumn{1}{c}{\textbf{V3}} & \multicolumn{1}{c}{\textbf{V4}} & \multicolumn{1}{c}{\textbf{V5}} \\
\midrule
\endhead

\midrule
\multicolumn{7}{r}{Continued on next page} \\
\endfoot

\bottomrule
\endlastfoot

Mod. floor & MAR & 0.011 & 0.053 & 0.095 & 0.164 & 0.231 \\* 
   & WJ2R(0.50) & 0.009 & 0.047 & 0.098 & 0.163 & 0.233 \\* 
   & J2R & 0.011 & 0.051 & 0.098 & 0.159 & 0.229 \\* 
   & CR & 0.010 & 0.051 & 0.097 & 0.160 & 0.228 \\* 
  Two-sided bds & MAR & 0.011 & 0.070 & 0.154 & 0.264 & 0.381 \\* 
   & WJ2R(0.50) & 0.010 & 0.067 & 0.151 & 0.263 & 0.382 \\* 
   & J2R & 0.012 & 0.067 & 0.153 & 0.260 & 0.381 \\* 
   & CR & 0.010 & 0.071 & 0.150 & 0.262 & 0.384 \\ 

\end{longtable}
\endgroup

\begingroup

\setlength{\tabcolsep}{4.5pt}
\fontsize{8pt}{9.6pt}\selectfont

\begin{longtable}{
    ll
    *{4}{S[table-format = -1.3] S[table-format = -1.3] S[table-format = 1.3] S[table-format = 1.3]}
}
\caption{Simulation Study~2: recovery of the sample treatment contrast on the observed scale at the final visit under the RBMI analysis rules. True is the average of the complete data contrasts across replicates. Bias and root mean squared error (RMSE) compare each posterior mean with the true contrast for that replicate; Cov is the proportion of nominal $95\%$ posterior intervals containing that contrast. Support labels: Mod.\ floor = \textit{ModerateFloor} and Two-sided bds = \textit{AggressiveBothBounds}. Each scenario combines a support template and a truth rule. WJ2R(0.50) denotes weighted jump to reference with $\omega = 0.50$.}
\label{tab:sim-rbmi-final-visit-bias-rmse-cov} \\
\toprule
& & \multicolumn{4}{c}{\textbf{MAR}} & \multicolumn{4}{c}{\textbf{WJ2R(0.50)}} & \multicolumn{4}{c}{\textbf{J2R}} & \multicolumn{4}{c}{\textbf{CR}} \\
\cmidrule(lr){3-6}
\cmidrule(lr){7-10}
\cmidrule(lr){11-14}
\cmidrule(lr){15-18}
\textbf{Support} & \textbf{Truth} &
\multicolumn{1}{c}{\textbf{True}} & \multicolumn{1}{c}{\textbf{Bias}} & \multicolumn{1}{c}{\textbf{RMSE}} & \multicolumn{1}{c}{\textbf{Cov}} &
\multicolumn{1}{c}{\textbf{True}} & \multicolumn{1}{c}{\textbf{Bias}} & \multicolumn{1}{c}{\textbf{RMSE}} & \multicolumn{1}{c}{\textbf{Cov}} &
\multicolumn{1}{c}{\textbf{True}} & \multicolumn{1}{c}{\textbf{Bias}} & \multicolumn{1}{c}{\textbf{RMSE}} & \multicolumn{1}{c}{\textbf{Cov}} &
\multicolumn{1}{c}{\textbf{True}} & \multicolumn{1}{c}{\textbf{Bias}} & \multicolumn{1}{c}{\textbf{RMSE}} & \multicolumn{1}{c}{\textbf{Cov}} \\
\midrule
\endfirsthead

\caption[]{Simulation Study~2: recovery of the treatment contrast on the observed scale at the final visit under the RBMI analysis rules (continued).} \\
\toprule
& & \multicolumn{4}{c}{\textbf{MAR}} & \multicolumn{4}{c}{\textbf{WJ2R(0.50)}} & \multicolumn{4}{c}{\textbf{J2R}} & \multicolumn{4}{c}{\textbf{CR}} \\
\cmidrule(lr){3-6}
\cmidrule(lr){7-10}
\cmidrule(lr){11-14}
\cmidrule(lr){15-18}
\textbf{Support} & \textbf{Truth} &
\multicolumn{1}{c}{\textbf{True}} & \multicolumn{1}{c}{\textbf{Bias}} & \multicolumn{1}{c}{\textbf{RMSE}} & \multicolumn{1}{c}{\textbf{Cov}} &
\multicolumn{1}{c}{\textbf{True}} & \multicolumn{1}{c}{\textbf{Bias}} & \multicolumn{1}{c}{\textbf{RMSE}} & \multicolumn{1}{c}{\textbf{Cov}} &
\multicolumn{1}{c}{\textbf{True}} & \multicolumn{1}{c}{\textbf{Bias}} & \multicolumn{1}{c}{\textbf{RMSE}} & \multicolumn{1}{c}{\textbf{Cov}} &
\multicolumn{1}{c}{\textbf{True}} & \multicolumn{1}{c}{\textbf{Bias}} & \multicolumn{1}{c}{\textbf{RMSE}} & \multicolumn{1}{c}{\textbf{Cov}} \\
\midrule
\endhead

\midrule
\multicolumn{18}{r}{Continued on next page} \\
\endfoot

\bottomrule
\endlastfoot

Mod. floor & MAR & -6.083 & 0.033 & 2.535 & 0.946 & -6.083 & 0.632 & 2.463 & 0.943 & -6.083 & 1.381 & 2.773 & 0.913 & -6.083 & 0.939 & 2.525 & 0.936 \\* 
   & WJ2R(0.50) & -5.330 & -0.782 & 2.816 & 0.926 & -5.330 & -0.142 & 2.572 & 0.943 & -5.330 & 0.662 & 2.715 & 0.936 & -5.330 & 0.175 & 2.527 & 0.940 \\* 
   & J2R & -4.437 & -1.534 & 3.098 & 0.887 & -4.437 & -0.894 & 2.668 & 0.927 & -4.437 & -0.101 & 2.525 & 0.960 & -4.437 & -0.597 & 2.497 & 0.937 \\* 
   & CR & -5.580 & -0.817 & 2.870 & 0.903 & -5.580 & -0.188 & 2.548 & 0.943 & -5.580 & 0.591 & 2.577 & 0.950 & -5.580 & 0.087 & 2.416 & 0.957 \\* 
  Two-sided bds & MAR & 4.274 & -0.606 & 5.026 & 0.946 & 4.274 & -1.253 & 4.482 & 0.933 & 4.274 & -1.943 & 4.622 & 0.923 & 4.274 & -1.199 & 4.252 & 0.946 \\* 
   & WJ2R(0.50) & 3.702 & 0.521 & 5.043 & 0.930 & 3.702 & -0.191 & 4.310 & 0.943 & 3.702 & -0.944 & 4.206 & 0.950 & 3.702 & -0.197 & 4.070 & 0.946 \\* 
   & J2R & 1.586 & 2.038 & 5.667 & 0.906 & 1.586 & 1.336 & 4.789 & 0.933 & 1.586 & 0.624 & 4.412 & 0.926 & 1.586 & 1.182 & 4.445 & 0.913 \\* 
   & CR & 3.676 & 0.295 & 4.609 & 0.956 & 3.676 & -0.425 & 4.258 & 0.950 & 3.676 & -1.184 & 4.564 & 0.933 & 3.676 & -0.612 & 4.224 & 0.946 \\ 

\end{longtable}
\endgroup

\end{landscape}

\clearpage
\subsection{Simulation Study~3: design and computational details}
\label{sec:simulation-study3-details}

Each replicate contained 100 subjects randomized equally between arms, four post-baseline visits, and an independent baseline covariate $B_i \sim \mathcal{N}\left(0, 1\right)$. This covariate was fully observed and was not itself a bounded outcome. The MAR latent outcomes were generated as
\begin{equation}
H_{it}^{\mathrm{MAR}} = \alpha_t + \beta_t B_i + A_i\delta_t + \varepsilon_{it},
\end{equation}
with
\begin{equation}
\begin{aligned}
\boldsymbol{\alpha} &= \left(-0.8, -1.1, -1.3, -1.5\right)^{\top},\\
\boldsymbol{\beta} &= \left(0.35, 0.35, 0.30, 0.30\right)^{\top},\\
\boldsymbol{\delta} &= \left(-0.15, -0.30, -0.45, -0.55\right)^{\top}.
\end{aligned}
\end{equation}
The residual covariance had marginal SDs $\left(0.9, 0.95, 1, 1\right)$ and AR(1) correlation parameter $0.6$. Gaussian errors followed $\mathcal{N}_4\left(\mathbf{0}, \Sigma\right)$. Heavy-tailed errors were generated as $\varepsilon_i = \sqrt{3/V_i}\,Z_i$, where $Z_i \sim \mathcal{N}_4\left(\mathbf{0}, \Sigma\right)$ and $V_i \sim \chi^2_5$ were independent. A single scale factor was shared across visits within each subject, giving multivariate $t_5$ errors with covariance $\Sigma$. Outcomes were obtained using the threshold map of \autoref{sec:thresholded_latent}, with $L = 0$, $U = 63$, $p_L = 0.5/64$, and $p_U = 63.5/64$, without rounding interior values.

Missingness patterns were generated independently of baseline and latent errors. The visit-specific hazards for starting a monotone missing tail were $\left(0, 0.12, 0.14, 0.20\right)$; before that tail, intermittent missingness probabilities were $\left(0, 0.05, 0.05, 0\right)$. For each subject, the missingness pattern was sampled from the exact pattern probabilities implied by the sequential dropout and independent gap construction described for Study~2. The sensitivity trigger was the visit after the last observed outcome. J2R truth was obtained by subtracting $A_i\delta_t$ from the MAR latent outcome at each missing visit from the trigger onward. Paired MAR and J2R truths therefore shared the same observed data, allowing posterior fits to be reused. Population targets were calculated by numerical integration over the outcome and baseline distributions and exact averaging over the probabilities of the missingness patterns.

Both imputation models retained Gaussian working distributions, so the $t_5$ scenarios assessed distributional misspecification. Missing values in the reference arm and gaps before the trigger in the active arm were imputed under MAR; the sensitivity rules applied to the final missing tail in the active arm. Each fit used four chains, 500 warmup iterations, and 1,000 post-warmup iterations per chain, with adaptation target $0.95$ and maximum tree depth $12$. All 4,000 post-warmup draws were used. The final 1,200 fits passed the diagnostic checks, with no divergences or hits at the maximum tree depth.

To reduce computation, MMRM estimates and restricted maximum likelihood covariance estimates for complete datasets were calculated using the equivalent multivariate regression formulas, checked against \texttt{nlme::gls}. The regression design matrix had one row per subject and three columns representing the intercept, baseline covariate, and treatment indicator, with separate regression coefficients for each visit. Thus, the complete-data analysis had $100 - 3 = 97$ residual degrees of freedom. Its confidence interval for the treatment contrast at visit~4 used the corresponding estimate and SE with a Student~$t$ critical value based on these degrees of freedom. For the imputation analyses, estimates and variance estimates were combined across completed datasets separately at each visit using Rubin's rules, with the conventional degrees of freedom for multiple imputation. At nominal $95\%$ coverage, 300 replicates give a Monte Carlo SE of approximately 1.26 percentage points.

\begingroup
\setlength{\tabcolsep}{5pt}
\footnotesize
\begin{longtable}{@{}llrrrr@{}}
\caption{Simulation Study~3: inference for the population treatment effect at the final visit under all sensitivity rules.}\label{tab:sim-study3-all-rules} \\
\toprule
\textbf{Method} & \textbf{Rule} & \textbf{Bias} & \textbf{Emp. SD} & \textbf{Mean SE} & \textbf{Cov. (\%)} \\
\midrule
\endfirsthead
\multicolumn{6}{l}{\small Simulation Study~3: all sensitivity rules (continued)} \\
\toprule
\textbf{Method} & \textbf{Rule} & \textbf{Bias} & \textbf{Emp. SD} & \textbf{Mean SE} & \textbf{Cov. (\%)} \\
\midrule
\endhead
\midrule
\multicolumn{6}{r}{\small Continued on next page} \\
\endfoot
\bottomrule
\endlastfoot
\multicolumn{6}{@{}l}{\textbf{\boldmath Gaussian; MAR truth} ($\Delta = -4.52$)} \\*
Mixed boundary & MAR & 0.06 & 2.33 & 2.64 & 96.3 \\*
Mixed boundary & J2R & 1.87 & 1.58 & 2.75 & 95.7 \\*
Mixed boundary & CR & 1.31 & 1.76 & 2.67 & 96.3 \\*
Mixed boundary & WJ2R(0.50) & 0.82 & 2.01 & 2.65 & 96.3 \\*
Shifted log & MAR & 0.32 & 2.24 & 2.60 & 97.7 \\*
Shifted log & J2R & 1.88 & 1.54 & 2.74 & 96.0 \\*
Shifted log & CR & 1.41 & 1.70 & 2.65 & 96.3 \\*
Shifted log & WJ2R(0.50) & 0.97 & 1.94 & 2.62 & 96.3 \\*
Complete data & -- & 0.09 & 2.11 & 2.06 & 95.7 \\
\addlinespace[0.8em]
\multicolumn{6}{@{}l}{\textbf{\boldmath Gaussian; J2R truth} ($\Delta = -2.73$)} \\*
Mixed boundary & MAR & -1.73 & 2.33 & 2.64 & 94.3 \\*
Mixed boundary & J2R & 0.08 & 1.58 & 2.75 & 99.7 \\*
Mixed boundary & CR & -0.48 & 1.76 & 2.67 & 99.3 \\*
Mixed boundary & WJ2R(0.50) & -0.96 & 2.01 & 2.65 & 98.3 \\*
Shifted log & MAR & -1.47 & 2.24 & 2.60 & 95.0 \\*
Shifted log & J2R & 0.10 & 1.54 & 2.74 & 99.7 \\*
Shifted log & CR & -0.38 & 1.70 & 2.65 & 99.3 \\*
Shifted log & WJ2R(0.50) & -0.82 & 1.94 & 2.62 & 99.3 \\*
Complete data & -- & 0.06 & 2.24 & 2.21 & 95.7 \\
\addlinespace[0.8em]
\multicolumn{6}{@{}l}{\textbf{\boldmath Student $t_5$; MAR truth} ($\Delta = -4.28$)} \\*
Mixed boundary & MAR & -0.08 & 2.22 & 2.48 & 98.3 \\*
Mixed boundary & J2R & 1.71 & 1.58 & 2.61 & 95.7 \\*
Mixed boundary & CR & 1.23 & 1.70 & 2.54 & 97.3 \\*
Mixed boundary & WJ2R(0.50) & 0.67 & 1.95 & 2.50 & 98.0 \\*
Shifted log & MAR & 0.19 & 2.12 & 2.37 & 97.7 \\*
Shifted log & J2R & 1.71 & 1.55 & 2.53 & 95.7 \\*
Shifted log & CR & 1.30 & 1.64 & 2.45 & 96.0 \\*
Shifted log & WJ2R(0.50) & 0.81 & 1.88 & 2.40 & 96.3 \\*
Complete data & -- & 0.14 & 1.94 & 2.02 & 96.3 \\
\addlinespace[0.8em]
\multicolumn{6}{@{}l}{\textbf{\boldmath Student $t_5$; J2R truth} ($\Delta = -2.59$)} \\*
Mixed boundary & MAR & -1.76 & 2.22 & 2.48 & 94.3 \\*
Mixed boundary & J2R & 0.03 & 1.58 & 2.61 & 100.0 \\*
Mixed boundary & CR & -0.46 & 1.70 & 2.54 & 100.0 \\*
Mixed boundary & WJ2R(0.50) & -1.02 & 1.95 & 2.50 & 99.0 \\*
Shifted log & MAR & -1.50 & 2.12 & 2.37 & 96.0 \\*
Shifted log & J2R & 0.03 & 1.55 & 2.53 & 100.0 \\*
Shifted log & CR & -0.38 & 1.64 & 2.45 & 100.0 \\*
Shifted log & WJ2R(0.50) & -0.87 & 1.88 & 2.40 & 99.3 \\*
Complete data & -- & 0.12 & 2.10 & 2.16 & 95.7 \\
\end{longtable}
{\footnotesize
Each scenario uses 300 replicates with 100 subjects (50 per arm). $\Delta$ is the fixed population mean difference, active minus reference, on the bounded outcome scale at visit~4. Emp. SD is the empirical standard deviation of the estimates; Mean SE is the average estimated standard error; Cov. is empirical coverage of nominal $95\%$ intervals. Both imputation methods use an MMRM on the observed scale and Rubin's rules with 4,000 imputations per replicate. Complete data denotes the same MMRM fitted before masking outcomes. The interval for the treatment contrast at visit~4 uses the corresponding MMRM estimate and SE with a Student~$t$ critical value based on 97 residual degrees of freedom. The Monte Carlo SE of coverage at $95\%$ is 1.26 percentage points. Coverage under rules that differ from the truth is descriptive and is not expected to equal $95\%$. WJ2R(0.50) uses weight $\omega = 0.50$.
\par}
\endgroup

\section{Predictive comparison: likelihood and computation}
\label{sec:loo}

\autoref{app:common-comparison} defines the common interval score. This appendix describes the original observed likelihood used to obtain the importance weights and the numerical evaluation of both quantities.

\subsection{Observed likelihood for the mixed-boundary model}

For subject $i$, let $\mathcal{I}_i$, $\mathcal{L}_i$, and $\mathcal{U}_i$ denote observed visits in the interior, at the lower bound, and at the upper bound. Write $\mathcal{B}_i = \mathcal{L}_i \cup \mathcal{U}_i$ and let $h_{i,\mathcal{I}_i} = g_i\left(Y_{i,\mathcal{I}_i}\right)$. Define $\mathcal{R}_i$ as the rectangle requiring $H_{it} \le a_i$ for $t \in \mathcal{L}_i$ and $H_{it} \ge b_i$ for $t \in \mathcal{U}_i$. Write $\mu_i = \mu_{iA_i}$ for the mean vector corresponding to subject $i$'s assigned arm. Suppressing the model subscript, the observed likelihood contribution is
\begin{equation}
\begin{aligned}
L_i\left(\theta\right)
={}&
\phi_{\left|\mathcal{I}_i\right|}\left(
h_{i,\mathcal{I}_i};
\mu_{i,\mathcal{I}_i}, \Sigma_{\mathcal{I}_i\mathcal{I}_i}
\right)
\left|J_i\left(h_{i,\mathcal{I}_i}\right)\right|\\
&\times
\Pr\left(
H_{i,\mathcal{B}_i} \in \mathcal{R}_i
\middle|
H_{i,\mathcal{I}_i} = h_{i,\mathcal{I}_i}, A_i, X_i, \theta
\right),
\end{aligned}
\end{equation}
where $\phi_d$ denotes a $d$-dimensional normal density and
\begin{equation}
\left|J_i\left(h_{i,\mathcal{I}_i}\right)\right|
=
\prod_{t \in \mathcal{I}_i}
\frac{p_{U_i} - p_{L_i}}{U_i - L_i}
\frac{1}{\phi\left(h_{it}\right)}.
\end{equation}
The final factor is a conditional normal rectangle probability. If there are no observed boundaries, it equals $1$. If there are no observed interior values, the density and Jacobian factors are omitted and the likelihood is the marginal rectangle probability for the observed boundary events. Missing visits are marginalized in each case. The boundary latent values sampled during fitting are therefore integrated out when computing the importance weights.

For the Gaussian comparator, the observed likelihood is the joint normal density of the observed scores. For the continuous bounded comparator, it is the joint normal density of the transformed observed scores multiplied by the Jacobian, including at scores $0$ and $63$. In the application, the Jacobian contribution per visit is $1/\left\{64\phi\left(h_{it}\right)\right\}$. The same observed visit set is used for the fitting likelihood and the common interval score.

\subsection{Numerical evaluation and diagnostics}

Normal rectangle probabilities were evaluated in R using randomized integration through \texttt{mvtnorm}, targeting relative error $10^{-5}$. When the initial integrations did not meet this criterion, conditional Gaussian quadrature was used, with agreement checked at three successive resolutions and a bound on omitted tail probability. An unresolved integration stopped the calculation. The application requires rectangles of at most four dimensions. Integration was parallelized across posterior draws, and completed calculations were saved for reuse. These integrations are required for the predictive comparison, not for fitting the augmented primary model or generating reference-based imputations.

PSIS weights were obtained with \texttt{loo::psis} from the negative observed log-likelihood contributions, with relative effective sample sizes accounting for MCMC dependence \citep{looPackage}. A model was refitted after removing subject $i$ when the Pareto diagnostic was at least $\min\left\{0.7, 1 - 1/\log_{10}\left(S\right)\right\}$, the PSIS effective sample size was below $100$, or either diagnostic was nonfinite. Each application model and explicit leave-one-subject-out refit used four chains, with 1,000 warmup iterations and 1,000 post-warmup iterations per chain, a target acceptance probability of $0.99$, and a maximum tree depth of $12$. The predictive comparison used all 4,000 post-warmup draws from each fit. The number of explicit refits was $3$ for the continuous bounded model, $4$ for the mixed-boundary model, and $1$ for the Gaussian model. These refits replace the importance approximation for the affected subjects; they are separate from the $50$ completed datasets used for the downstream MMRM analysis.

We assessed agreement between MCMC chains using the rank-normalized split $\widehat{R}$ diagnostic, a refinement of the potential scale reduction factor \citep{vehtari2021rank}. The reported MCMC checks required $\widehat{R} \le 1.01$, bulk and tail effective sample sizes of at least $100$ for the monitored nonconstant structural parameters, no divergent transitions or maximum tree depth hits, and an energy Bayesian fraction of missing information of at least $0.3$ in each chain. All reported checks passed. Subject contributions were also accompanied by Monte Carlo SE estimates accounting for MCMC dependence; these estimates do not include numerical integration error and are distinct from the SEs of ELPD differences across subjects. Approximate Monte Carlo SEs of the overall ELPD differences, accounting for dependence among subject contributions sharing a posterior fit, ranged from $0.31$ to $0.34$. These estimates condition on the estimated PSIS weights and refit decisions and exclude numerical integration error.

\section{Model comparisons and sensitivity analyses}
\label{sec:btheb-additional-analyses}

Under MAR, we compared the mixed-boundary and continuous bounded models for the same subjects with missing BDI-II at month 8: 20 in TAU and 25 in BtheB. For each of all 4,000 retained posterior draws, we calculated each subject's conditional probability of an outcome equal to zero and of an outcome in $\left[-0.5, 0.5\right)$ using normal CDFs, conditional on baseline BDI-II and the observed history. The mixed-boundary calculation included the corresponding posterior draws of latent values for observed boundary scores. Probabilities were averaged across the subjects with missing scores within each arm. We report posterior means and 95\% credible intervals for these average conditional probabilities, rather than intervals for proportions in completed datasets.

The mixed-boundary model assigned positive probability to outcomes equal to zero, whereas the continuous bounded model has no probability mass at zero. Both models assigned positive probability to the common interval representing a rounded zero (\webautoref{tab:btheb-boundary-comparison}). The continuous bounded predictions were not clipped before calculating these probabilities.

\begin{table}[htbp]
\centering
\small
\caption{Predicted probabilities (\%) for subjects with missing BDI-II at month 8 under MAR. Entries are posterior means with 95\% credible intervals for the average conditional probability within each arm's subset of subjects with missing scores.}
\label{tab:btheb-boundary-comparison}
\begin{tabular}{lrlll}
\hline
\textbf{Arm} & \textbf{Missing} & \textbf{Model} & \textbf{Exact zero} & \textbf{Rounds to zero} \\
\hline
TAU & 20 & Mixed-boundary & $5.90$ ($1.89$, $12.16$) & $10.45$ ($4.45$, $18.56$) \\
TAU & 20 & Continuous bounded & $0.00$ ($0.00$, $0.00$) & $7.00$ ($2.79$, $13.14$) \\
BtheB & 25 & Mixed-boundary & $7.23$ ($2.94$, $12.95$) & $11.37$ ($5.83$, $18.20$) \\
BtheB & 25 & Continuous bounded & $0.00$ ($0.00$, $0.00$) & $7.60$ ($3.43$, $13.19$) \\
\hline
\end{tabular}
\smallskip
\par\begin{minipage}{0.98\linewidth}
\footnotesize
Both models predict the same subjects within each arm, conditional on their observed histories and baseline BDI-II. Rounds to zero denotes the common interval $\left[-0.5, 0.5\right)$. The continuous bounded model has no point mass at zero and is not clipped before scoring. Intervals summarize variation in conditional probabilities across posterior states, including augmented boundary values where applicable; they are not intervals for realized proportions in completed datasets. All post-warmup draws were used.
\end{minipage}
\end{table}

We additionally examined sensitivity to the threshold specification using extensions of $c = 0.25$ and $c = 1.00$ points, compared with the original $c = 0.50$. The lower and upper threshold probabilities were $p_L = c/\left(63 + 2c\right)$ and $p_U = 1 - p_L$, with latent thresholds obtained by applying $\Phi^{-1}$. The observed bounds remained 0 and 63, and the recorded score resolution was unchanged. We refitted the mixed-boundary model under each alternative, updating the interior transformation and Jacobian while retaining the same priors and MCMC settings.

For each threshold specification, we used all 4,000 posterior draws after warmup to generate completed datasets under MAR, WJ2R(0.50), J2R, and CR for BtheB, with TAU under MAR. All 97 retained subjects were included, with observed outcomes held fixed. The completed arm means and proportions of zeros at month 8 were summarized as in \autoref{fig:btheb-observed-profiles} and showed little change across the specifications examined (\webautoref{tab:btheb-threshold-sensitivity}). All reported fit and profile diagnostic checks passed.

\begin{table}[htbp]
\centering
\small
\caption{Sensitivity of completed cohort summaries at month 8 to the threshold extension. Entries are posterior means with pointwise 95\% intervals.}
\label{tab:btheb-threshold-sensitivity}
\begin{tabular}{llll}
\hline
\textbf{Profile} & \textbf{0.25 points} & \textbf{0.50 points (original)} & \textbf{1.00 point} \\
\hline
\multicolumn{4}{l}{\textbf{Mean BDI-II}} \\
TAU (MAR) & $13.49$ ($11.12$, $16.31$) & $13.49$ ($11.24$, $16.20$) & $13.49$ ($11.29$, $16.04$) \\
BtheB: MAR & $12.42$ ($9.44$, $15.80$) & $12.27$ ($9.46$, $15.43$) & $12.11$ ($9.55$, $14.84$) \\
BtheB: WJ2R(0.50) & $12.27$ ($9.58$, $15.30$) & $12.19$ ($9.65$, $15.05$) & $12.10$ ($9.76$, $14.62$) \\
BtheB: J2R & $12.16$ ($9.21$, $15.48$) & $12.15$ ($9.32$, $15.52$) & $12.11$ ($9.44$, $15.18$) \\
BtheB: CR & $10.99$ ($8.58$, $13.78$) & $10.97$ ($8.58$, $13.77$) & $10.95$ ($8.74$, $13.48$) \\
\hline
\multicolumn{4}{l}{\textbf{Probability of a score of zero (\%)}} \\
TAU (MAR) & $11.13$ ($8.89$, $15.61$) & $11.50$ ($8.89$, $17.78$) & $11.87$ ($8.89$, $17.78$) \\
BtheB: MAR & $8.88$ ($5.77$, $13.46$) & $9.23$ ($5.77$, $15.38$) & $9.59$ ($5.77$, $15.38$) \\
BtheB: WJ2R(0.50) & $8.96$ ($5.77$, $13.46$) & $9.27$ ($5.77$, $15.38$) & $9.54$ ($5.77$, $15.38$) \\
BtheB: J2R & $9.16$ ($5.77$, $15.38$) & $9.42$ ($5.77$, $15.38$) & $9.61$ ($5.77$, $15.38$) \\
BtheB: CR & $10.23$ ($5.77$, $17.31$) & $10.63$ ($5.77$, $17.31$) & $10.97$ ($5.77$, $17.31$) \\
\hline
\end{tabular}
\smallskip
\par\begin{minipage}{0.98\linewidth}
\footnotesize
For extension $c$, the latent threshold probabilities are $c/\left(63 + 2c\right)$ and $1 - c/\left(63 + 2c\right)$. The observed bounds remain 0 and 63. The mixed-boundary model was refitted for each alternative using the same priors and MCMC settings. Observed scores were held fixed, and all post-warmup draws were used for completion. Intervals describe uncertainty in the completed cohort mean or proportion of zeros; they are not MMRM or Rubin intervals. TAU uses MAR under every specification.
\end{minipage}
\end{table}

\end{appendices}

\end{document}